\documentclass[10pt,oneside,a4paper,onecolumn]{elsarticle}
\usepackage{hyperref}
\pdfoutput = 1

\journal{Computers \& Fluids}

\usepackage{fullpage}
\usepackage{siunitx}
\usepackage{cancel}
\usepackage[usenames,dvipsnames]{xcolor}
\usepackage{multirow}

\usepackage{gensymb}
 \usepackage{tikz}
\usepackage{changepage}
\usepackage{alltt}
\usepackage{titlesec}
\usepackage{upquote}
\usepackage{float}
\usepackage{array}
\usepackage{pgfplots}
\usetikzlibrary{positioning}
\usepackage{graphicx}
\usepackage[toc,page]{appendix}
\usepackage{amsmath,mathtools}
\usetikzlibrary{calc,decorations.markings,fit,shapes,chains,arrows}
\usepackage{amsthm}
\usepackage{amsfonts}
\usepackage{enumitem}
\usepackage{subcaption}
\usepackage{caption}
\usepackage{booktabs}
\usepackage{xcolor}
\usepackage{scalerel}
\usepackage{algorithm}
\usepackage{multicol}
\usepackage{mdframed}
\usepackage{mathtools}
\usepackage{enumerate}
\theoremstyle{remark}

\usepackage{natbib}

\newcommand{\varsdomain}[2]{#1\left(\boldsymbol{x}^{\mathrm{#2}},t\right)}

\pgfdeclarepatternformonly{crossHatchDots}{\pgfqpoint{-1pt}{-1pt}}{\pgfqpoint{5pt}{5pt}}{\pgfqpoint{6pt}{6pt}}%
{
	\pgfpathcircle{\pgfqpoint{0pt}{0pt}}{.5pt}
	\pgfpathcircle{\pgfqpoint{3pt}{3pt}}{.5pt}
	\pgfusepath{fill}
}

\tikzset{
	block/.style={rectangle, draw, rounded corners, text centered,text width = 16em, minimum height = 2em},
	line/.style={draw, -latex'}
}
\tikzset{
	block2/.style={text centered,text width = 22em, minimum height = 2em},
	line/.style={draw, -latex'}
}
\tikzset{
	block3/.style={rectangle, draw, rounded corners, text centered,text width = 19em, minimum height = 1em},
	line/.style={draw, -latex'}
}
\tikzset{
	block4/.style={rectangle, draw, rounded corners, text centered,text width = 15em, minimum height = 1em},
	line/.style={draw, -latex'}
}
\tikzset{
	blockNS/.style={rectangle, draw, fill=black!20, rounded corners, text centered,text width = 20em, minimum height = 2em, label={center:Navier-Stokes solver}},
	line/.style={draw, -latex'}
}
\tikzset{
	blockAC/.style={rectangle, draw, fill=black!20, rounded corners, text centered,text width = 20em, minimum height = 2em, label={center:Allen-Cahn solver}},
	line/.style={draw, -latex'}
}
\tikzset{  
	decision/.style = {diamond, draw, minimum width=4cm, minimum height=0.2cm},
	line/.style={draw, -latex'}
}

\def\@author#1{\g@addto@macro\elsauthors{\normalsize%
		\def\baselinestretch{1}%
		\upshape\authorsep#1\unskip\textsuperscript{%
			\ifx\@fnmark\@empty\else\unskip\sep\@fnmark\let\sep=,\fi
			\ifx\@corref\@empty\else\unskip\sep\@corref\let\sep=,\fi
		}%
		\def\authorsep{\unskip,\space}%
		\global\let\@fnmark\@empty
		\global\let\@corref\@empty
		\global\let\sep\@empty}%
	\@eadauthor={#1}
}

\graphicspath{{figures/}}

\begin{document}
	\begin{frontmatter}
		\title{A finite element framework for fluid-structure interaction of turbulent cavitating flows with flexible structures}
		\author[ubc]{Nihar B. Darbhamulla}
		\ead{nihar.darbhamulla@mail.ubc.ca}
		
		\author[ubc]{Rajeev K. Jaiman\corref{cor1}}
		\ead{rjaiman@mech.ubc.ca}
		\cortext[cor1]{Corresponding author}
		\address[ubc]{Department of Mechanical Engineering, The University of British Columbia, Vancouver, BC V6T 1Z4}
		
\begin{abstract}
We present a finite element framework for the numerical prediction of cavitating turbulent flows interacting with flexible structures. The vapor-fluid phases are captured through a homogeneous mixture model, with a scalar transport equation governing the spatio-temporal evolution of cavitation dynamics. High-density gradients in the two-phase cavitating flow motivate the use of a positivity-preserving Petrov-Galerkin stabilization method in the variational framework. A mass transfer source term introduces local compressibility effects arising as a consequence of phase change. The turbulent fluid flow is modeled through a dynamic subgrid-scale method for large eddy simulations. The flexible structure is represented by a set of eigenmodes, obtained through the modal analysis of the linear elasticity equations. A partitioned iterative approach is adopted to couple the structural dynamics and cavitating fluid flow, where the deforming flow domain is described by an arbitrary Lagrangian-Eulerian frame of reference. Through the numerical validation study, we establish the fidelity of the proposed framework by comparing it against experimental and numerical studies for both rigid and flexible hydrofoils in cavitating flows. Under unstable partial cavitating conditions, we identify specific vortical structures leading to cloud cavity collapse. We further explore features of cavitating flow past a rigid body such as re-entrant jet and turbulence-cavity interactions during cloud cavity collapse. Based on the validation study conducted over a flexible NACA66 rectangular hydrofoil, we elucidate the role of cavity and vortex shedding in governing the structural dynamics. Subsequently, we identify a broad spectrum frequency band whose central peak does not correlate to the frequency content of the cavitation dynamics or the natural frequencies of the structure, indicating the induction of unsteady flow patterns around the hydrofoil.  Finally, we discuss the coupled fluid-structure dynamics during a cavitation cycle and the underlying mechanism associated with the promotion and mitigation of cavitation.
\end{abstract}
		
		
	\end{frontmatter}
	
\section{Introduction}\label{sec:introduction}

Cavitation is a prominent phenomenon in natural and industrial systems involving the liquid phase. In the context of marine applications, the performance of hydrodynamic devices such as propellers, rudders, stabilizers, and fins is largely governed by the extent of cavitation in their operating regimes. Furthermore, cavitation is a major source of noise in the ocean environment, with unsteady vibrations affecting the performance of marine structures, and influencing marine life in surrounding regions. Along with noise, the collapse of cavities is associated with shock waves, whose impact on hydrodynamic devices can lead to material erosion and structural damage \cite{carlton2018marine, kerr1940problems}. To mitigate the effects of cavitation on hydrodynamic structures, it is essential to understand the mechanisms driving the coupled dynamics of cavitating flows interacting with structural response.

The phenomenon of cavitation involves the phase transition of liquid into vapor. Often, the flowing liquid is not a pure substance and contains impurities in the form of suspended solid particles, entrained gas bubbles, or regions of interaction with solid surfaces. The locations of these features are identified as points of weakness in the liquid. \citet{brennen_2013} indicates the ability of liquids to withstand tensile loads within an elastic regime, which leads to vacancies coalescing to form a finite vapor pocket. This rupture of the liquid phase occurs isothermally at the tensile strength of the liquid (defined as $p_v - p_c$, where $p_c$ is the liquid pressure at which vapor pocket forms, and $p_v$ is the vapor pressure). 
The vapor growth is driven by the spatial variation in liquid pressure, with new pockets forming and potentially coalescing in regions where the liquid tensile strength is reached. Furthermore, fluctuations in the local liquid pressure can affect the spatial dimensions of the vapor pockets formed, and they are transported by the flowing liquid until they encounter a region of high pressure, where they have the propensity to undergo a violent collapse.

The flow of liquid water over marine propeller blades generates a low-pressure region in accordance with the mechanism driving thrust generation. Under certain flow conditions, the pressure can drop close to $p_c$, leading to the formation of vapor nuclei and their growth \citep{ross1989mechanics} into different cavity structures. This behavior is termed partial cavitation wherein the cavities grow to occupy a fraction of the blade surface. This phenomenon has received extensive attention in literature, with focus on its inception \citep{arakeri1973CavitationInception, katz1984CavFlowSep, rood1991cavreview, wang2001CavTurbDyn}, classification of partial cavitation regimes \citep{franc1985attachedcavitation, kubota1989cloudcavitationmeasurement, kubota1992numericalcavitation, le1993partialCavities, kawanami1997cloudcavitation} and analysis of the instabilities and mechanisms driving the transition between regimes \citep{kawanami1997cloudcavitation, gopalan2000ModelingClosure, wang2001CavTurbDyn, callenaere2001CavInstab, pelz2017Transition} in the context of both attached and separated flows. Partial cavitation surrounding a propeller blade alters the performance characteristics of the propeller while introducing undesirable effects such as vibrations and noise emissions. Recent studies have made efforts to investigate these phenomenon experimentally \citep{ausoni2007CavVib, benaouicha2009Exp, delatorre2013AddedMass, akcabay2014influence, wu2018transient, smith2020FlexStiff, smith2020FlexComp} and computationally \citep{akcabay2014influence,ji2015les, wu2015numericalFlexHydrofoil, chen2019LES, suraj2023cavviv}. However, in the context of fluid-structure interaction (FSI), the approaches primarily involve modeling the flexible structure as a two-dimensional elastically mounted body. In this context, it becomes important to model the strongly coupled fluid-structure dynamics using either a full-order or reduced-order representation of the three-dimensional structure, to elucidate the features of cavity dynamics governing the structural response, and vice-versa.

\subsection{Review of FSI Effects for Hydrofoils and Marine Propellers}	

As aforementioned above, marine propellers can exhibit complex vibration characteristics owing to unsteady fluid forces associated with turbulent flow with or without cavitation effects. In this regard, propeller blades are represented as a hydrofoil where-in the flow characteristics can involve different partial cavitation regimes \citep{young2008FSIcompositeProp}. Within these regimes, a flexible hydrofoil is found to experience attenuation and periodic fluctuations of loads leading to marked differences in structural response relative to the flexible hydrofoil in non-cavitating conditions \citep{franc1988oschydrofoil, kato2006Pitching,ausoni2007CavVib, akcabay2014influence, smith2017FSICloudCav}.

In a cavitating environment, experimental investigations of the FSI effects of hydrofoils have illustrated a rich and complex interplay of the cavitating flow with the structure dynamics. \citet{franc1988oschydrofoil} demonstrated the incompatibility of attached cavities with a turbulent boundary layer in the flow past an oscillating hydrofoil, and attached cavities were found to be swept away by an upstream turbulent front. Furthermore, the leading edge roughness and the frequency of hydrofoil oscillations influence the thickness of the attached cavity \citep{caron2000physical}.

In the context of free-vibration responses of cantilevered hydrofoils in a cavitating flow, \citet{ausoni2007CavVib} studied the von-K\'arm\'an vortex street in the wake of a cantilevered rectangular hydrofoil, with a truncated trailing edge. In the non-cavitating regime, the von-K\'arm\'an vortex street was observed to lock in with the hydrofoil eigenfrequency corresponding to the first torsional mode. Within the cavitating regime, the vibration amplitudes were observed to be reduced accompanied by the synchronization of vortex-shedding with the structural frequency. 
The authors hypothesized that structural characteristics govern the unsteady flow field and vortex-shedding. The presence of cavitation alters the added mass effects owing to the variable fluid-solid density ratios observed due to the phase change. \citet{delatorre2013AddedMass} investigated the role of stable partial cavities on the added mass effects experienced by a flexible hydrofoil. They found that added-mass effects decreased with a decreasing cavitation number, owing to a larger cavity on the suction surface.

\citet{benaouicha2009Exp} and \citet{akcabay2014influence} experimentally investigated the interaction of sheet and cloud cavitation with a flexible hydrofoil. They observed excitations in subharmonics of the foil's wetted natural frequencies and the broadening of frequency content in the cloud cavitation regime. \citet{wu2018transient} further explored the hydroelastic response of flexible hydrofoils in a cavitating environment and delineated the behavior of vibration velocity with the state of surface cavitation in the cloud cavitation regime. They indicate the cloud cavity's collapse leads to a substantial increase in the velocity fluctuations. Further, they indicate variations observed with respect to the cavitation development in the case of a highly compliant hydrofoil, particularly with regard to the collapse of the large-scale cloud cavity into smaller structures. Simultaneously, \citet{smith2017FSICloudCav} investigated the effects of cloud cavitation on a flexible trapezoidal hydrofoil, with a dual objective of correlating hydrofoil flexibility with the frequency content of the force fluctuations and establishing three-dimensional flow regimes. They observed bending deformations to damp the higher frequency fluctuations in the normal force measurements. Furthermore, they attributed the cavity-shedding primarily to the re-entrant jet mechanism, with the thin cavity leading to the shedding of small-scale vapor pockets.

In subsequent investigations by \citet{smith2020FlexStiff},\citep{smith2020FlexComp}, the authors characterized the instabilities driving cavity shedding in the vicinity of a stiff and compliant hydrofoil and they identified a regime of cavity shedding associated with lock-in phenomenon. The authors illustrated that compliant hydrofoils exhibit a prominent bend-twist coupling, which caused an early transition from sheet to cloud cavitation and led to the formation of a longer sheet cavity. The time scale of sheet cavity formation leads to a reduced shedding frequency. The compliant hydrofoil attenuates the high-frequency oscillations in the force, while simultaneously operating at lock-in conditions with the first bending mode. The aforementioned studies illustrated the intricacies of flow physics associated with flexible hydrofoils in a turbulent cavitating flow.

In the light of physics observed in experiments on the response of flexible hydrofoils in a turbulent-cavitating environment, it becomes essential to develop high-fidelity models for understanding  complex physical phenomena. It is equally crucial to develop stable, accurate and robust numerical discretization strategies to ensure the fidelity of numerical results to the underlying model and the corresponding physical observations. Within the context of numerical FSI studies for marine propellers and hydrofoils, the primary numerical challenge is to accurately capture the cavity and flow patterns, in order to predict the forces experienced by the structure.
\subsection{Numerical Studies of FSI with Cavitating Flows}
Approaches based on potential flow theories have been extensively used to evaluate the hydroelastic response of flexible marine propellers. Works by \citet{lee2014hydro, maljaars2018boundary} use a coupled potential theory-based boundary element method (BEM) and finite element method (FEM) for investigating the hydroelastic response of flexible marine propellers. The FSI of fluid and propeller blades has also been investigated with the use of panel methods, through a loosely coupled interaction between the hull wake and propeller \citep{jiang2018ship}. While inviscid flow models are effective in making general design decisions for propellers, they are unable to capture the range of vortical structures that are associated with the unsteady loads experienced by the propeller. In the context of viscous flow modeling, \citep{lee2017fluid} employed a tightly coupled CFD-FEM solver to study high Reynolds number flows over a flexible blade undergoing flow-induced vibrations. The authors illustrated a requirement to deploy tightly coupled FSI solvers for studying large amplitude vibrations of flexible blades in the flow regimes considered.

The numerical modeling of cavitating flows has received extensive attention in the literature, particularly in the domain of flows around propellers. A majority of these studies have been aimed at capturing the turbulence-cavitation interaction, and an accurate estimate of the loads on hydrofoils feasible within the available scope of cavitation and turbulence models. \citet{dang2001cavFlows} presents a comprehensive review of the early approaches towards numerically simulating cavitating flows, and adopted a potential flow theory approach in conjunction with defining a cavity detachment point within a panel method framework. \citet{schnerr2008numerical} further investigated the cloud cavitation and collapse-induced shock dynamics through a compressible formulation of the governing equations, with thermodynamic closure. They indicated the need to have a high temporal resolution within this framework, owing to wave dynamics associated with acoustic cavitation, and to capture regions of instantaneous high-pressure loads. Further studies by \citet{seo2009cloudcav} used a density-based mixture-theory model for cavitation in conjunction with the Spalart-Allmaras turbulence model. Within this framework, they captured the unsteady mode of the unstable partial sheet and cloud cavitation and illustrated the RANS turbulence model's capability to capture the shear-layer instability formed by the re-entrant jet mechanism.

\citet{ji2015les} numerically studied the cavitation structures and vortex shedding dynamics using large eddy simulations (LES) and a homogeneous mixture theory-based model to compute the pressure, velocity and vapor volume fraction in the vicinity of a NACA66 hydrofoil. They investigated the turbulence-cavitation interaction through an analysis of the vorticity transport equation for variable density flows. \citet{chen2019LES} extended this work further and presented a comprehensive evaluation of the forces and moments experienced by the hydrofoil and presented a preliminary analysis of the turbulent structures generated in the hydrofoil's vicinity. \citet{sedlar2016cavInv} investigated the interactions between re-entrant flow and cavitation structures as well as cavitation excited pressure fluctuations. They compared the performance of three different turbulence models to evaluate the fidelity of numerically captured flow features to physical data. They observed the best description of vortical structures around the hydrofoil is predicted by LES, at the cost of overestimating the dominant frequencies of cavity oscillation. These frequencies were observed to be better captured by the SAS-SST turbulence model and Detached Eddy Simulations (DES).

While the above studies have primarily aimed at capturing the re-entrant jet dynamics and their role in driving the cavity breakdown process, they also establish a comparison between numerical approaches for evaluating the forces experienced by a structure, which forms the backbone of studying the response of flexible structures within a cavitating environment. \citet{akcabay2014influence} employed a loose hybrid coupling to couple a commercial two-dimensional URANS solver with a 2-DOF hydrofoil model to study the effect of cavitation on the hydroelastic stability of hydrofoils. Further work by \citet{wu2015numericalFlexHydrofoil} investigated both numerically and experimentally the cavitating flow past a NACA66 hydrofoil. They attributed the hydrofoil's displacement response to flow-induced flutter, a hypothesis that warrants closer investigation. They solved the URANS equations with a $k-\omega$ SST turbulence model, with compressibility-based turbulent-viscosity modifications. For the structure, they used a 2-DOF hydrofoil model to represent the bending and torsional modes of deformation. The studies however depict significant deviation quantitatively from experimental observations, and thus reinforce the need for a robust and accurate unified framework for the strongly-coupled FSI studies of cavitating flows over propellers.

\citet{suraj2021femcav} established a novel variational framework for cavitating flows with moving fluid-structure interfaces, with appropriate stabilization and the use of a homogeneous mixture model theory for capturing cavitation, and the use of DES with the SA turbulence model. The framework depicted high fidelity with the studies of \citet{leroux2004experimental, senocak2001numerical} and demonstrated the method's robustness in capturing flows with highly unsteady cavitation dynamics. The stabilization involves a positivity-preserving variational (PPV) scheme proposed by \citet{joshi2018positivity} for capturing the two-phase flow of immiscible fluids. The use of a local discrete upwind operator in an interface region with oscillations acts to diffuse the overshoots in solution and ensures the positivity of underlying element-level matrices. The scheme has been shown to be effective in damping spurious pressure oscillations in multiphase flow with fluids having a high-density ratio. Thus, this approach lends itself well to transport-equation modeling (TEM) of cavitating flows. The current work extends this framework by introducing an eigenmode representation of a three-dimensional flexible structure, allowing for accurate numerical capture of the FSI of multi-modal structures in cavitating flows.
\subsection{Numerical Modeling of Cavitating Flows}

In the context of marine propellers, cavitation manifests in the form of structures evolving over multiple orders of spatial and temporal scales \citep{arndt2015singing,brennen_2013}. In this regard, an appropriate cavitation model requires the identification of spatio-temporal scales of significance. At macroscopic spatial scales, an extensively used strategy is the treatment of fluid as a homogeneous mixture of liquid and vapor phases, which differ primarily in the treatment of mixture density. One approach to compute density involves using the equations of state under the assumption of equilibrium flow theory\citep{schnerr2008numerical}. For isothermal flows, the density has a barotropic equation of state \citep{deshpande1994cavity,chen1996modeling}. An advantage of this approach is the absence of empirical coefficients for modeling, and the use of well-established equations of state. These models are generally difficult to deploy in an incompressible fluid framework. Furthermore, in an isothermal flow, the gradients of density and pressure are parallel, leading to the baroclinic torque ($\nabla \rho \times \nabla p$) being zero. The baroclinic torque plays an important role in cavitating flows for vorticity generation owing to the high gradient of density, particularly in the closure region of attached cavities\cite{gopalan2000flow}. An alternate approach assumes the individual phases of the fluid to be incompressible. A phase indicator variable is used to interpolate the density. A transport equation model (TEM) has to be constructed for the phase indicator variable, the source term of which captures the finite mass transfer rate between phases\citep{merkle1998computational,schnerr2001physical,singhal2002mathematical,zwart2004two}.

Owing to the phenomenological nature of the source terms for the phase indicator TEM, different approaches have been employed for modeling the source term. \citet{merkle1998computational} related the source term to the local pressure and phase fraction of the liquid based on dimensional arguments for bubble clusters. \citet{schnerr2001physical}, \citet{singhal2002mathematical}, \citet{zwart2004two} developed models, which assume the cavities to be present in the form of clusters of small spherical bubbles, and directly used a simplified form of the Rayleigh-Plesset equation \citep{brennen_2013} for spherical bubble dynamics to model the mass transfer rate. The phenomenological arguments for modeling the bubble-bubble interaction in these models are different, which is what differentiates them. Despite the applicability of the TEM approach for capturing cavitation, they are constrained by problem-dependent semi-empirical coefficients in the source term. Furthermore, the phase indicator variable can have large gradients, which can force unphysical spikes in pressure near the liquid-vapor interface. In incompressible flow simulations, these spikes can diffuse through the computational domain leading to numerical instabilities. \citet{senocak2002pressure} indicated the necessity of handling these spurious pressure spikes as a requirement for robustly capturing cavitating flows.
However, the versatility of the TEMs allows for their use in modeling cavitation across different regimes of cavitation, and have been employed extensively for studying hydrodynamic cavitation over geometries of marine interest such as hydrofoils and circular cylinders \citep{senocak2001numerical,gnanaskandan2015numerical,brandao2020CavCylinder,ji2015les,chen2019LES,akcabay2014influence}.
\subsection{Contributions and Organization}
In the current work, we extend our finite element formulation for the numerical modeling of cavitating flows presented in \citet{suraj2021femcav} to couple with a linear modal structural representation of the structure under consideration. The cavitation transport equation model by \citet{schnerr2001physical} based on a homogeneous mixture assumption of the liquid-vapor phases is used. Large Eddy Simulations (LES) are conducted to capture the turbulent vortical and wake structures around the hydrofoil. The structural deformation, fluid flow and cavitation solvers are coupled in a staggered partitioned manner, with predictor-corrector iterations deployed for convergence and stability. A fully-implicit generalized-$\alpha$ time integration scheme \cite{jansen2000generalized} is used to advance the solution in time for the fluid flow and cavitation equation solvers, while the trapezoidal rule is employed to advance the structural modes in time. Thus, the following requirements are desired from the framework developed: (i) accurate prediction of the turbulent flow field, (ii) robust capturing of cavity collapse phenomenon, and (iii) coupled numerical stability under highly transient structural loading.

The layout of the paper is as follows. Section \ref{sec:methodology} presents the governing equations for the coupled fluid-structure-cavitation system and a brief overview of the LES framework used. Variational formulations of the governing equations and the corresponding stabilizations are presented in Section \ref{sec:varForm}. The LES validation of the cavitating rigid hydrofoil is presented in Section \ref{sec:rigidhydrofoil}, and the LES validation of the cavitating rigid and flexible hydrofoil is presented in Section \ref{sec:flexhydrofoil}. The numerical studies conducted are summarized and concluded in Section \ref{sec:conclusions}.
	
\section{Governing Equations}\label{sec:methodology}
In this section, we present a three-dimensional framework for the coupled fluid-structure solver based on the Navier-Stokes equations for fluid flow, the scalar transport model for cavitation transport, and the linear structural equation for the flexible body.

\subsection{Fluid Properties in Multiphase Flow}\label{sec:formMP}
In the current problem, we consider a physical domain occupied by the fluid ($\Omega^{\mathrm{f}}\left(\mathbf{x}^{\mathrm{f}},t\right)$) confined within the fluid boundary ($\Gamma^{\mathrm{f}}\left(t\right)$), where $\mathbf{x}^{\mathrm{f}}$ and $t$ denote the spatial and temporal co-ordinates respectively. The fluid within the domain is assumed to be a homogeneous mixture of liquid and vapor phases. A phase indicator variable $\phi\left(\mathbf{x}^{\mathrm{f}},t\right)$ is employed to represent the phase fraction of the liquid phase within the fluid domain. The fluid density ($\rho^{\mathrm{f}}$) and viscosity ($\mu^{\mathrm{f}}$) are treated as linear functions of the phase indicator variable as follows:
\begin{align}
	\rho^{\mathrm{f}} &= \rho_l\phi + \rho_v\left(1 - \phi\right),\\
	\mu^{\mathrm{f}} &= \mu_l\phi + \mu_v\left(1 - \phi\right),	
\end{align}
where $\rho_l$ and $\rho_v$ are the densities of the pure liquid and vapor phases, and $\mu_l$ and $\mu_v$ are the dynamic viscosities of the liquid and vapor phases respectively.
	
\subsection{Navier-Stokes Equations in the Deforming Domain}\label{sec:formNS}
Owing to the requirement of simulating the interaction of the fluid flow with a compliant structure, a body-fitted moving boundary-based approach is adopted. The influence of the deforming domain on the fluid kinematics is captured through formulating the equations governing the fluid flow and cavitation transport in an Arbitrary Lagrangian-Eulerian framework (ALE) \citep{hughes1981ale}. The Navier-Stokes equations in the ALE frame are:
\begin{align}
	\left.\rho^{\mathrm{f}} \frac{\partial \boldsymbol{u}^{\mathrm{f}}}{\partial t}\right|_{\boldsymbol{\chi}}
	+\rho^{\mathrm{f}}\left(\left(\boldsymbol{u}^{\mathrm{f}}-\boldsymbol{u}^{\mathrm{m}}\right) \cdot \nabla\right) \boldsymbol{u}^{\mathrm{f}}
	-\nabla \cdot \boldsymbol{\sigma}
	=\boldsymbol{f}^{\mathrm{f}},&&\mathrm{on}\ (\boldsymbol{x}^{\mathrm{f}},t)\in \Omega^{\mathrm{f}}, \label{NS_mom}\\
	\left. \frac{\partial \rho^{\mathrm{f}}}{\partial t}\right|_{\boldsymbol{\chi}}+ \rho^{\mathrm{f}}\nabla \cdot \boldsymbol{u}^{\mathrm{f}} + \left(\boldsymbol{u}^{\mathrm{f}}-\boldsymbol{u}^{\mathrm{m}}\right)\cdot\nabla\rho^{\mathrm{f}} = 0,&&\mathrm{on}\ (\boldsymbol{x}^{\mathrm{f}},t)\in \Omega^{\mathrm{f}}, \label{NS_mass}	
\end{align} where $\boldsymbol{\chi}$ denotes the referential coordinate system, ${\boldsymbol{u}^{\mathrm{f}}} = \varsdomain{\mathbf{u}}{f}$ is the fluid velocity at each spatial location $\boldsymbol{x}^{\mathrm{f}} \in \Omega^{\mathrm{f}}$ and $\boldsymbol{u}^{\mathrm{m}}$ is the relative velocity of the spatial coordinates $\boldsymbol{x}^{\mathrm{f}}$ with respect to the referential coordinate system $\boldsymbol{\chi}$. $\boldsymbol{f}^{\mathrm{f}}$ is the body force applied on the fluid and $\boldsymbol{\sigma}$ is the stress tensor. The stress tensor $\boldsymbol{\sigma}$ is further decomposed as
\begin{equation}
	\boldsymbol{\sigma} = \boldsymbol{\sigma}^{\mathrm{f}} + \boldsymbol{\sigma}^{\mathrm{sgs}}, \label{eq:stressDec}
\end{equation}
where ${\boldsymbol{\sigma}^{\mathrm{f}}}$ and $\boldsymbol{\sigma}^{\mathrm{sgs}}$ denote the Cauchy stress tensor for a Newtonian fluid and the subgrid-scale turbulent stress tensor respectively. The Cauchy stress tensor for the Newtonian fluid is written as
\begin{align}
	{\boldsymbol{\sigma}^{\mathrm{f}}} 
	&= -{p^{\mathrm{f}}}\boldsymbol{I} + \mu^{\mathrm{f}}( \nabla{\boldsymbol{u}^{\mathrm{f}}}+ (\nabla{\boldsymbol{u}^{\mathrm{f}}})^T),
\end{align}
where ${p^{\mathrm{f}}}$ denotes the fluid pressure. The subgrid-scale turbulent stresses are captured using a spatial-filtering based LES approach. The details of the LES approach adopted are briefly discussed in Sec. \ref{sec:formSGS}.
	
\subsection{Cavitation Transport Equation}\label{sec:formCavitation}
The phase indicator $\phi$ is obtained as the solution of a scalar transport equation, which in the ALE framework is:
\begin{align} \label{TEM}
	\left. \frac{\partial \phi}{\partial t}\right|_{\boldsymbol{\chi}}
	+ \phi \nabla \cdot \boldsymbol{u}^{\mathrm{f}} 
	+ \left(\left(\boldsymbol{u}^{\mathrm{f}}-\boldsymbol{u}^{\mathrm{m}}\right)\cdot\nabla\right)\phi 
	= \dfrac{\dot{m}}{\rho_{l}},&&\mathrm{on}\ (\boldsymbol{x}^{\mathrm{f}},t)\in \Omega^{\mathrm{f}}.
\end{align}
The source term $\dot{m}$ in the transport equation models the finite mass transfer rate which governs the production and destruction of vapor under conditions conducive to cavitation. In the current study, we adopt the mass transfer term proposed by \citet{schnerr2001physical}, which has a non-linear dependence on $\phi$ and $p^{\mathrm{f}}$, given by:
\begin{align}
		\dot{m}  
		=  \frac{3 \rho_{l} \rho_{v}}{\rho^{\mathrm{f}} R_{B}} \sqrt{\frac{2}{3 \rho_{l}\left|p^{\mathrm{f}}-p_{v}\right|}}
		\bigg[ C_{c} \phi&(1-\phi) \operatorname{max}\left(p^{\mathrm{f}}-p_{v}, 0\right) \nonumber \\
		&+ C_{v} \phi(1 + \phi_{nuc} - \phi) \operatorname{min}\left(p^{\mathrm{f}}-p_{v}, 0\right)
		\bigg] \label{eq:schnerrSauer}.
\end{align} The model relates the finite mass transfer rate to the rate of growth/collapse of an equivalent spherical bubble under an external pressure field. Cavitation is assumed to initiate from nucleation sites present in the flow \citep{brennen_2013}. The concentration of nuclei per unit volume $\left(n_0\right)$ and the nucleus diameter $\left(d_{nuc}\right)$ are taken to be constant parameters in the model. $R_B\left(\mathbf{x}^{\mathrm{f}},t\right)$ in Eq.\ref{eq:schnerrSauer} represents the equivalent spherical radius of the vapor volume at a given spatial location. $\phi_{nuc}$ denotes the phase fraction of the initial nucleation sites in a unit volume. These are calculated as:
\begin{equation}
	R_B = \left( \frac{3}{4\pi n_0} \frac{1+\phi_{nuc}-\phi}{\phi} \right)^{1/3} \quad\mathrm{and}\quad 
	\phi_{nuc} = \frac{\dfrac{\pi n_0 d^3_{nuc}}{6}}{1+\dfrac{\pi n_0 d^3_{nuc}}{6}}.	
\end{equation}
To maintain numerical control over the model, the parameters $C_c$ and $C_v$ were introduced, and termed as the condensation and evaporation coefficients respectively \citep{cazzoli2016cavModels, ghahramani2019cavModels}. The cavitation model under consideration has been applied extensively to the study of cavitating flow over hydrofoils \citep{ji2015les,chen2019LES,suraj2023cavviv}.
	
\subsection{Modeling of Subgrid-Scale Stresses}\label{sec:formSGS}
For modeling the turbulent structures in the fluid flow at high Reynolds numbers, a dynamic subgrid-scale (SGS) model is utilized to model the subgrid-scale stresses $\boldsymbol{\sigma}^{\mathrm{sgs}}$ as described in Eq. \ref{eq:stressDec}. The dynamic subgrid model applies a filter to the incompressible Navier-Stokes equations to model the scales smaller than the spatial resolution $\Delta$. The subgrid-scale stress is defined as $\sigma_{ij}^{\mathrm{sgs}} = \overline{u_i^\mathrm{f}u_j^\mathrm{f}} - \overline{u}_i^\mathrm{f}\overline{u}_j^\mathrm{f}$ where the overline indicates the filtered flow variables. The need for modeling arises as $\overline{u}_i^\mathrm{f}\overline{u}_j^\mathrm{f}$ involves the unknown SGS quantities $u_i^\mathrm{f}$ and $u_j^\mathrm{f}$. The nonlinear SGS stress tensor \citep{gatski1993stressmodels} is expressed as:
\begin{align}
\sigma_{ij}^\mathrm{sgs} - \frac{\delta_{ij}}{3}\sigma_{kk}^\mathrm{sgs} \approx -2\mu_t\overline{S_{ij}} - 6C_{NL}\frac{\mu_t^2}{\sigma_{kk}^{sgs}}\left(\overline{S}_{ik}\overline{\Omega}_{kj} + \overline{S}_{jk}\overline{\Omega}_{ki} - 2 \overline{S}_{ik}\overline{S}_{kj} + \frac{2}{3}\overline{S}_{nk}\overline{S}_{kn}\delta_{ij}\right),
\end{align}
where $\mu_t$ is the dynamic eddy viscosity given by $\mu_t = \rho^{\mathrm{f}}\left(C_s\overline{\Delta}\right)^2\left|\overline{S}\right|$, the resolved strain-rate tensor $\overline{S}_{ij} = \frac{1}{2}\left(\frac{\partial \overline{u}_i^\mathrm{f}}{\partial x_j} + \frac{\partial \overline{u}_j^\mathrm{f}}{\partial x_i}\right)$. The assumption of local equilibrium between production and dissipation of energy allows for deriving the value of $C_s$ in the Smagorinsky model \citep{smagorinsky1963turb}. 

In the dynamic SGS model\citep{germano1991dynamicsgs}, two filters are defined: the grid filter with scale dimension $\Delta$ and the test filter with scale dimension $\widehat{\Delta}$. The first filter is dependent on the grid resolution and the test filter, denoted by the operator $\widehat{\left(\bar{   }\right)}$, which can be any coarser level filter. An identity relating the subgrid-scale stresses generated by different filters is given by
\begin{align}
L_{ij} = T_{ij} - \sigma_{ij}^\mathrm{sgs},
\end{align}
where the Leonard tensor, $L_{ij}$, is the stress generated by performing the test filter on the grid filtered data, given by
\begin{align}
L_{ij} = \widehat{\overline{u}_i^\mathrm{f}\overline{u}_j^\mathrm{f}} - \widehat{\overline{u_i}}^\mathrm{f}\widehat{\overline{u_j}}^\mathrm{f},\label{eq:Leonard}
\end{align}
and the stress at test level $T_{ij}$ is 
\begin{align}
T_{ij} = \widehat{\overline{u_i^\mathrm{f} u_j^\mathrm{f}}} - \widehat{\overline{u_i}^\mathrm{f}}\widehat{\overline{u_j}^\mathrm{f}}.\label{eq:TestFilter}
\end{align}

As no assumption is considered with regard to modeling the turbulent stresses in Eqs. \ref{eq:Leonard} and \ref{eq:TestFilter}, any dynamic procedure can be adopted for computing the subgrid-scale stresses. Through using the Smagorinsky eddy-viscosity model for the unknown stress tensors $\boldsymbol{\sigma}^\mathrm{sgs}$ and $\mathbf{T}$, the following relation for the Leonard stress is developed:
\begin{align}
L_{ij} = -2C_s^2\left(\widehat{\Delta}^2\widehat{|\overline{S}|}\widehat{\overline{S}_{ij}} - \Delta^2|\overline{S}|\overline{S}_{ij}\right).
\end{align}

In order to aid the resolution of the viscous sublayer, an algebraic eddy viscosity model is employed to ensure reasonable behavior along the wall: $\frac{\mu_t}{\mu_f} = \kappa y_w^+\left(1 - e^-y_w^+/A\right)$, where $y^+ = y_w u_\tau/\nu$ is the distance to the wall in wall units based on the friction velocity $u_\tau$. $\kappa$ is the model coefficient and $A = 19$ is a constant \citep{balaras1996algebraicturb}. Further details of the implementation can be found in \citet{jaiman2016les}. The resolution of the turbulent stresses, in conjunction with the cavitation transport model and the Navier-Stokes equations, allows for investigating the behavior of the fluid flow. Subsequently, we develop the evolution of the structure and briefly discuss the coupling of the fluid-structure interface.

\subsection{Modeling of Structural Deformation}\label{sec:formStruct}
\noindent In the current study, the flexible cantilevered hydrofoil primarily experiences temporally fluctuating pressure forces. The hydrofoil is modeled using eigenmodes obtained from solving the eigenvalue problem derived from the linear elastic structural model. The structural domain $\Omega^{\mathrm{s}}$ is geometrically defined by the co-ordinates $\mathbf{x}^{\mathrm{s}}$, with displacements $\mathbf{w}^{\mathrm{s}}(\mathbf{x}^{\mathrm{s}},t)$. Neglecting the effects of damping and body forces, the equations governing the linear elastic deformation of a structure are:
\begin{equation}
\rho^\mathrm{s}\frac{\partial^2 \mathbf{w}^{\mathrm{s}}}{\partial t^2} = \nabla \cdot\boldsymbol{\sigma}^{\mathrm{s}} + \rho^\mathrm{s}\mathbf{b}, \label{eq:solidstress}
\end{equation}
where $\rho^s$ is the solid density, $\boldsymbol{\sigma}^{\mathrm{s}}$ is the Cauchy stress tensor of the solid and $\mathbf{b}$ is the body force experienced by the solid per unit volume, and for a linear elastic material, is defined as:
\begin{equation}
\boldsymbol{\sigma}^{\mathrm{s}} = \lambda^\mathrm{s} tr(\boldsymbol{\epsilon}^\mathrm{s}) + 2\mu^\mathrm{s}\boldsymbol{\epsilon}^\mathrm{s},
\end{equation}
where $\lambda^{\mathrm{s}}$ and $\mu^\mathrm{s}$ are the Lame's parameters and $\boldsymbol{\epsilon}$ is the infinitesimal strain tensor, which is defined as
\begin{equation}
\boldsymbol{\epsilon}^\mathrm{s} = \frac{1}{2}\left(\nabla \mathbf{w}^{\mathrm{s}} + \left(\nabla \mathbf{w}^{\mathrm{s}}\right)^T\right)\label{eq:strn}.
\end{equation} 

To obtain the eigenvalue problem, we separate the spatial and temporal components of the displacement response as:
\begin{equation}
\mathbf{w}^\mathrm{s}(\mathbf{x}^\mathrm{s},t) = \sum_{n=1}^{\infty}\hat{\mathbf{w}}_n^\mathrm{s}(\mathbf{x}^\mathrm{s})e^{i\omega_n t}, \label{eq:modaldec}
\end{equation}
where $\hat{\mathbf{w}}_n^\mathrm{s}(\mathbf{x}^\mathrm{s})$ denotes an eigenvector or mode shape of the structure, and $\omega_n$ represents the corresponding eigenvalue or modal frequency. Substituting Eq.\ref{eq:modaldec} in Eq.\ref{eq:solidstress} and neglecting body forces, we obtain the following representation:
\begin{equation}
-\rho^s\omega_n^2\hat{\mathbf{w}}_n^\mathrm{s}e^{i\omega_n t} = \nabla \cdot \boldsymbol{\sigma}^{s}(\hat{\mathbf{w}}^\mathrm{s})e^{i\omega_n t},
\end{equation}
which owing to linear elastic modeling can be simplified to:
\begin{equation}
\left(\mathcal{L} + \rho_s\omega_n^2\mathbf{I}\right)\hat{\mathbf{w}}_n^\mathrm{s} = 0, \label{eq:eigvalue}
\end{equation}
where $\mathcal{L}$ denotes the linear differential operator associated with Eqs.\ref{eq:solidstress} and \ref{eq:strn}, completing the definition of the structural eigenvalue problem. The boundary conditions of the cantilevered beam arrangement are given by:
\begin{align}
\left. \mathbf{w}^{\mathrm{s}}(\mathbf{x}^\mathrm{s},t)\right|_{z = 0} &= 0, \label{eq:fixed}\\
\left. \boldsymbol{\sigma}^\mathrm{s} \mathbf{n}^\mathrm{s}\right|_{z\neq 0} & = 0. \label{eq:free}
\end{align}
The solution to Eq.\ref{eq:eigvalue} is obtained through a finite element discretization of the structural domain, and the mode shapes and modal frequencies are obtained from the finite-element software CalculiX. The equivalent finite-element problem is briefly discussed in section\ref{sec:numStruct}.
	
\subsection{Exchange of Information at Fluid-Structure Interface}\label{sec:formFSI}
The fluid and structure are coupled through the imposition of continuity in velocity and traction at their interface. Let $\Gamma^{\mathrm{fs}}(0) = \partial \Omega^{\mathrm{f}}(0) \cap \partial \Omega^{\mathrm{s}}$ denote the fluid-structure interface at $t = 0$ and $\Gamma^{\mathrm{fs}}(t) = \boldsymbol{\psi}^{\mathrm{s}}(\Gamma^{\mathrm{fs}}(0),t)$ be the interface at time t. The interface boundary conditions are given by:
\begin{align}
	\mathbf{u^\mathrm{f}\left(\boldsymbol{\psi}^s(\mathbf{x_0}^{\mathrm{s}},t)\right)} &= \mathbf{\dot{w}^{\mathrm{s}}(\mathbf{x_0}^{\mathrm{s}},t)},\\
	\int_{\boldsymbol{\psi}^{\mathrm{s}}(\gamma, t)} \boldsymbol{\sigma}^{\mathrm{f}} \cdot \mathbf{n}^{\mathrm{f}} d \Gamma &= -\int_{\gamma} \boldsymbol{\sigma}^{\mathrm{s}} \cdot \mathbf{n}^{\mathrm{s}} d \Gamma,		
\end{align} where $\boldsymbol{\psi}^{\mathrm{s}}$ denotes the position vector mapping the initial position $\mathbf{x}_0^\mathrm{s}$ of the flexible body to its position at time t, i.e. $\boldsymbol{\psi}^s(\mathbf{x_0}^{\mathrm{s}},t) = \mathbf{x_0}^{\mathrm{s}} + \mathbf{w}^s(\mathbf{x}^{\mathrm{s}},t)$. $\mathbf{t}^{\mathrm{s}}$ is the fluid traction which relates to the fluid load as $\mathbf{f}^{\mathrm{s}}_{ext} (z,t) = \int_{\Gamma^{\mathrm{fs}}}\mathbf{t}^{\mathrm{s}} d\Gamma$. $\mathbf{\dot{w}}^{\mathrm{s}}$ is the structural velocity at time $t$ given by $\mathbf{\dot{w}}^{\mathrm{s}} = \partial \boldsymbol{\psi}^{\mathrm{s}}/\partial t$. $\mathbf{n}^{\mathrm{f}}$ and $\mathbf{n}^{\mathrm{s}}$ denote the outward normals to the fluid-structure interface and $\gamma$ is any part of the interface $\Gamma^\mathrm{fs}$ in the reference configuration.

\section{Numerical Formulation and Implementation}\label{sec:varForm}
In this section, we present the framework within which the equations governing the FSI problem are numerically formulated. We start with a description of the temporal discretization adopted for the fluid-flow and cavitation transport equations in Sec.\ref{sec:numTemporal}, and subsequently, we discuss the space of finite elements within which the variational statements for the governing equations are formulated in Sec.\ref{sec:numSpatial}. We conclude the section with a discussion of the spatio-temporal evolution of the linear structure in Sec.\ref{sec:numStruct}.

\subsection{Temporal Discretization} \label{sec:numTemporal}
The generalized-$\alpha$ method follows a predictor-corrector approach to advance the solution in time. It has been shown to be second-order accurate and unconditionally stable for linear problems. Further, a single parameter $\rho_{\infty}$ allows the user to achieve high-frequency error damping. For a given spatio-temporally evolving set of variables $\mathbf{\Phi}$, the generalized-$\alpha$ method progresses according to the following equations:=
\begin{align}
	\mathbf{\Phi}^{n+1} &= \mathbf{\Phi}^n + \Delta t\partial_t\mathbf{\Phi}^n + \gamma\Delta t\left(\partial_t\mathbf{\Phi}^{n+1} - \partial_t\mathbf{\Phi}^n\right),\nonumber\\
	\partial_t\mathbf{\Phi}^{n+\alpha_m} &= \partial_t\mathbf{\Phi}^{n} + \alpha_m\left(\partial_t \mathbf{\Phi}^{n+1} - \partial_t \mathbf{\Phi}^n\right),\label{eq:genAlpha}\\
	\mathbf{\Phi}^{n+\alpha} &= \mathbf{\Phi}^n + \alpha\left(\mathbf{\Phi}^{n+1} - \mathbf{\Phi}{n}\right),\nonumber
\end{align}
where $\Delta t$ is the time step size and $\partial_t\mathbf{\Phi}^{n+\alpha_m}$ is the temporal derivative of $\mathbf{\Phi}$ at the $n+\alpha_m$ time level such that $t^{n+\alpha_m} \in \left[t^n, t^{n+1}\right]$. $\alpha_m,\:\alpha$ and $\gamma$ are generalized-$\alpha$ parameters based on $\rho_{\infty}$. For a non-linear PDE system $G(\mathbf{\Phi}) = 0$, the predictor-corrector approach solves the following for $\Delta\mathbf{\Phi}^{n+\alpha}$:
\begin{align}
	\mathbf{G}\left(\mathbf{\dot{\Phi}}^{n+\alpha_m},\mathbf{\Phi}^{n+\alpha}\right) + \frac{\partial \mathbf{G}\left(\mathbf{\dot{\Phi}}^{n+\alpha_m},\mathbf{\Phi}^{n+\alpha}\right)}{\partial \mathbf{\Phi}^{n+\alpha}}\Delta\mathbf{\Phi}^{n+\alpha} =  0. \label{eq:residualForm}
\end{align}
In Eq.\ref{eq:residualForm}, the first term is the residual evaluated at $n+\alpha_m$ and the derivative represents the Jacobian of the system, obtained after linearization of the PDE system. Based on Eqs.\ref{eq:genAlpha} and \ref{eq:residualForm}, we construct the semi-discrete form of the Navier-Stokes equations as
\begin{align}
	\left.\rho^{\mathrm{f}} \partial_{t} \boldsymbol{u}^{\mathrm{f}, \mathrm{n}+\alpha_{\mathrm{m}}}\right|_{\chi}
	+\rho^{\mathrm{f}}\left(\left(\boldsymbol{u}^{\mathrm{f}, \mathrm{n}+\alpha}-\boldsymbol{u}^{\mathrm{m}, \mathrm{n}+\alpha}\right) \cdot \nabla\right)\boldsymbol{u}^{\mathrm{f}, \mathrm{n}+\alpha}
	-\nabla \cdot \boldsymbol{\sigma}^{\mathrm{n}+\alpha} 
	- \boldsymbol{f}^{\mathrm{n}+\alpha} 
	&= 0,\label{NS_momSemiDisc}\\
	\nabla \cdot \boldsymbol{u}^{\mathrm{f},\mathrm{n}+\alpha}
	-\left(\frac{1}{\rho_{l}} - \frac{1}{\rho_{v}} \right)\dot{m}
	&=0,
\end{align}
and the cavitation transport equation as
\begin{align}
	\partial_t{\phi}^{\mathrm{f}, \mathrm{n}+\alpha_\mathrm{m}} 
	+ \left(\left(\boldsymbol{u}^{\mathrm{f}}-\boldsymbol{u}^{\mathrm{m}}\right)\cdot\nabla\right)\phi^{\mathrm{f}, \mathrm{n}+\alpha} 
	+ s\phi^{\mathrm{f}, \mathrm{n}+\alpha} = 0,\label{eq:TEM_CDR}
\end{align}
where the reaction coefficient $s$ for the Schnerr-Sauer model is given by:
\begin{align}
	s &= -\frac{3 }{ R_{B}} \sqrt{\frac{2}{ 3\rho_{l}\left|p^{\mathrm{f}}-p_{v}\right|}} 
	\bigg[ C_{c} (1-\boldsymbol{\phi}^{\mathrm{f}}) \operatorname{max}\left(p^{\mathrm{f}}-p_{v}, 0\right) 
	+ C_{v} (1 + \phi_{nuc} -\boldsymbol{\phi}^{\mathrm{f}}) \operatorname{min}\left(p^{\mathrm{f}}-p_{v}, 0\right)
	\bigg].\\
\end{align}
	
\subsection{Spatial Discretization and Variational Statement}\label{sec:numSpatial}
For the spatial discretization of the flow and cavitation equations, we use a stabilized Petrov-Galerkin finite element framework. The domain $\Omega^\mathrm{f}$ is discretized into $N$ elements such that $\Omega^\mathrm{f} = \cup_{e=1}^N\Omega_e^\mathrm{f}$, and $\emptyset = \cap_{e=1}^N\Omega_e^\mathrm{f}$. For the finite element framework, we allow $\mathcal{S}^h$ to be the space of trial solutions, which satisfy the Dirichlet boundary conditions, and $\mathcal{V}^h$ to be the space of test functions which vanish on the Dirichlet boundaries. The stabilized variational form for the Navier-Stokes equations has been discussed extensively in the works of \citet{jaiman2016les,joshi2017positivity,joshi2017variationally, suraj2021femcav}, and is discussed here briefly. The variational statement for the fluid flow equations involves finding $[\mathbf{u}_h^{n+\alpha},p_h^{n+\alpha}]\in\mathcal{S}^h$, such that $\forall[\boldsymbol{\Psi}_h,q_h] \in \mathcal{V}^h$, 	
\begin{align}
	&\int_{\Omega^{\mathrm{f}}} \boldsymbol{\Psi}_{h} \cdot\left(\left.\rho^{\mathrm{f}} \partial_{t} \boldsymbol{u}_{h}^{\mathrm{f}, \mathrm{n}+\alpha_{\mathrm{m}}}\right|_{\boldsymbol{\chi}}+\rho^{\mathrm{f}}\left(\left(\boldsymbol{u}_{h}^{\mathrm{f}, \mathrm{n}+\alpha}-\boldsymbol{u}^{\mathrm{m}}\right) \cdot \nabla\right) \boldsymbol{u}_{h}^{\mathrm{f}, \mathrm{n}+\alpha}\right) d \Omega\nonumber \\		
	&+ \int_{\Omega^{\mathrm{f}}} \nabla \boldsymbol{\Psi}_{h}: \boldsymbol{\sigma}_{h}^{\mathrm{n}+\alpha} d \Omega 	
	+ \int_{\Omega^{\mathrm{f}}\left(t^{\mathrm{n}+1}\right)} q_{h}\left(\nabla \cdot \boldsymbol{u}_{h}^{\mathrm{f}, \mathrm{n}+\alpha}\right) d \Omega\nonumber \\
	&+ \sum_{e=1}^{N} \int_{\Omega^{\mathrm{f}}} \frac{\tau_{m}}{\rho^{\mathrm{f}}}\left(\rho^{\mathrm{f}}\left(\left(\boldsymbol{u}_{h}^{\mathrm{f}, \mathrm{n}+\alpha}-\boldsymbol{u}^{\mathrm{m}}\right) \cdot \nabla\right) \boldsymbol{\Psi}_{h}+\nabla q_{h}\right) \cdot \boldsymbol{\mathcal{R}}_{m} d \Omega^{\mathrm{e}}
	+ \sum_{e=1}^{N} \int_{\Omega^{e}} \nabla \cdot \boldsymbol{\Psi}_{h} \tau_{c} \rho^{\mathrm{f}} \boldsymbol{\mathcal{R}}_{c} d \Omega^{e}\\
	&-\displaystyle\sum_\mathrm{e=1}^\mathrm{N}\int_{\Omega^\mathrm{e}} \tau_\mathrm{m} \boldsymbol{\Psi}_\mathrm{h}\cdot (\left(\boldsymbol{\mathcal{R}}_\mathrm{m} \cdot \nabla\right) {\boldsymbol{u}}_\mathrm{h}^\mathrm{n+\alpha}) \mathrm{d\Omega^e} 
	-\displaystyle\sum_\mathrm{e=1}^\mathrm{N}\int_{\Omega^\mathrm{e}} \frac{\nabla \boldsymbol{\Psi}_\mathrm{h}}{\rho(\phi)}:(\tau_\mathrm{m}\boldsymbol{\mathcal{R}}_\mathrm{m} \otimes \tau_\mathrm{m}\boldsymbol{\mathcal{R}}_\mathrm{m}) \mathrm{d\Omega^e}\nonumber \\
	&= \int_{\Omega^{\mathrm{f}}} \boldsymbol{\Psi}_{h} \cdot \boldsymbol{f}^{\mathrm{f}, \mathrm{n}+\alpha} \; d \Omega
	+\int_{\Gamma_{N}^{\mathrm{f}}} \boldsymbol{\Psi}_{h} \cdot \boldsymbol{h}^{\mathrm{f}, \mathrm{n}+\alpha} \; d \Gamma
	+ \int_{\Omega^{\mathrm{f}}} q_{h}\left(\frac{1}{\rho_{l}} - \frac{1}{\rho_{v}} \right) \dot{m}^{\mathrm{f}, \mathrm{n}+\alpha} \; d \Omega\nonumber,	
\end{align}
where the first and second lines contain the standard Galerkin finite element terms of the momentum and continuity equations. The third line contains the Galerkin Least Squares stabilization terms for the momentum and mass continuity equations. The fourth line contains stabilization terms based on the multi-scale argument \citep{hughes2005conservation,hsu2010improving}. The fifth line contains the Galerkin terms for the body force and the Neumann boundary in the momentum equation. The element-wise residuals of the momentum and continuity equations are denoted by $\boldsymbol{\mathcal{R}}_{m}$ and $\boldsymbol{\mathcal{R}}_{c}$ respectively, and are given by	
\begin{align}		
		\boldsymbol{\mathcal{R}}_\mathrm{m}({\boldsymbol{u}^{\mathrm{f}}},{p}^{\mathrm{f}}) 
		&= \rho^{\mathrm{f}}\partial_t{\boldsymbol{u}}_\mathrm{h}^\mathrm{n+\alpha_m} 
		+ \rho^{\mathrm{f}}\left(\boldsymbol{u}^{\mathrm{f}, \mathrm{n}+\alpha}-\boldsymbol{u}^{\mathrm{m}}\right) \cdot \nabla{\boldsymbol{u}}_\mathrm{h}^{\mathrm{f}, \mathrm{n}+\alpha} - \nabla \cdot {\boldsymbol{\sigma}}_\mathrm{h}^{\mathrm{f}, \mathrm{n}+\alpha} - \boldsymbol{f}^{\mathrm{f}, \mathrm{n}+\alpha}_\mathrm{h}, \\
		\boldsymbol{\mathcal{R}}_\mathrm{c}({\boldsymbol{u}^{\mathrm{f}}},{p}^{\mathrm{f}}, \phi^{\mathrm{f}}) &= \nabla\cdot\boldsymbol{u}^{\mathrm{f}, \mathrm{n}+\alpha}_\mathrm{h}-\left(\frac{1}{\rho_{l}} - \frac{1}{\rho_{v}} \right) \dot{m}^{\mathrm{f}, \mathrm{n}+\alpha}.
\end{align}

Furthermore, $\tau_m$ and $\tau_c$ are stabilization parameters \citep{brooks1982streamline,tezduyar1992incompressible,franca1992stabilized} defined as
\begin{align}
	\tau_\mathrm{m} &= \bigg[ \bigg( \frac{2}{\Delta t}\bigg)^2 + {\boldsymbol{u}}_\mathrm{h}\cdot \boldsymbol{G}{\boldsymbol{u}}_\mathrm{h} + C_I \bigg(\frac{\mu(\phi)}{\rho(\phi)}\bigg)^2 \boldsymbol{G}:\boldsymbol{G}\bigg] ^{-1/2},\\
	\tau_\mathrm{c} &= \frac{1}{\mathrm{tr}(\boldsymbol{G})\tau_\mathrm{m}}, \label{tau_c}	
\end{align}
where $C_I$ is a constant derived from the element-wise inverse estimate \cite{harari1992c} and $\mathrm{tr(\boldsymbol{G})}$  denotes the trace of the element contravariant metric tensor $\boldsymbol{G}$, which is defined as
\begin{align}	
	\boldsymbol{G} = \frac{\partial \boldsymbol{\xi}^T}{\partial \boldsymbol{x}^{\mathrm{f}}}\frac{\partial \boldsymbol{\xi}}{\partial \boldsymbol{x}^{\mathrm{f}}},\label{eq:defG}
\end{align}
where $\mathbf{x}^f$ and $\boldsymbol{\xi}$ are the physical and parametric co-ordinates respectively. The stabilization in the variational form provides stability to the velocity field in the convection-dominated regimes of the fluid domain and circumvents the Babu\v{s}ka-Brezzi condition which is required to satisfy any mixed Galerkin method \citep{johnson2009fem}.

In a similar manner, the variational statement for the cavitation transport equation involves finding $\phi_h^{n+\alpha} \in \mathcal{S}^h$ such that $\forall w_h\in \mathcal{V}^h$,	\begin{align} \label{eq:TEMvariational}
	&\int_{\Omega^{\mathrm{f}}}  w_h\bigg( \partial_t{\phi}^{{n+\alpha_m}}_{h} 
	+ \left(\left(\boldsymbol{u}^{\mathrm{f}, n+\alpha}-\boldsymbol{u}^{\mathrm{m}, n+\alpha}\right)\cdot\nabla\right)\phi^{n+\alpha}_{h} 
	+ s\phi^{n+\alpha}_\mathrm{h} \bigg) \mathrm{d}\Omega^{\mathrm{f}} \nonumber \\
	&+ \displaystyle\sum_\mathrm{e=1}^\mathrm{N}\int_{\Omega^{{\mathrm{f}},\mathrm{e}}} \left(\left(\boldsymbol{u}^{\mathrm{f}, \mathrm{n}+\alpha}-\boldsymbol{u}^{\mathrm{m}}\right)\cdot \nabla w_h\right) \boldsymbol{\mathcal{R}}_\phi\nonumber\\
	&+ \displaystyle\sum_\mathrm{e=1}^\mathrm{N}\int_{\Omega^{{\mathrm{f}},\mathrm{e}}} \zeta \frac{|\boldsymbol{\mathcal{R}}_\phi|}{|\nabla\phi_h^{n+\alpha}|}k_s^\mathrm{add} \nabla w_{h}\cdot \bigg( \frac{\left(\boldsymbol{u}^{\mathrm{f}, n+\alpha}-\boldsymbol{u}^{\mathrm{m}}\right)\otimes \left(\boldsymbol{u}^{\mathrm{f}, n+\alpha}-\boldsymbol{u}^{\mathrm{m}}\right)}{|\left(\boldsymbol{u}^{\mathrm{f}, n+\alpha}-\boldsymbol{u}^{\mathrm{m}}\right)|^2} \bigg) \cdot \nabla\phi^{n+\alpha}_{h} \mathrm{d}\Omega^{{\mathrm{f}},\mathrm{e}}\\
	&+ \sum_\mathrm{e=1}^\mathrm{N} \int_{\Omega^{{\mathrm{f}},\mathrm{e}}}\zeta \frac{|\boldsymbol{\mathcal{R}}_\phi|}{|\nabla \phi^{n+\alpha}_{h}|} k^\mathrm{add}_{c} \nabla w_{h} \cdot \bigg( \mathbf{I} - \frac{\left(\boldsymbol{u}^{\mathrm{f}, n+\alpha}-\boldsymbol{u}^{\mathrm{m}}\right)\otimes \left(\boldsymbol{u}^{\mathrm{f}, n+\alpha}-\boldsymbol{u}^{\mathrm{m}}\right)}{|\left(\boldsymbol{u}^{\mathrm{f},n+\alpha}-\boldsymbol{u}^{\mathrm{m}}\right)|^2} \bigg) \cdot \nabla\phi^{\mathrm{f}, n+\alpha}_\mathrm{h} \mathrm{d}\Omega^{{\mathrm{f}},\mathrm{e}} \nonumber \\
	&= 0,\nonumber 	
\end{align}	where the first line represents the standard Galerkin finite element terms. The second line consists of linear stabilization terms with the stabilization parameter $\tau_\phi$ given by \cite{shakib1991new} \begin{align}
	\tau_\phi &= \bigg[ \bigg( \frac{2}{\Delta t}\bigg)^2 +\left(\boldsymbol{u}^{\mathrm{f}}-\boldsymbol{u}^{\mathrm{m}}\right)\cdot \boldsymbol{G}\left(\boldsymbol{u}^{\mathrm{f}}-\boldsymbol{u}^{\mathrm{m}}\right) + s^2\bigg] ^{-1/2}.
\end{align} $\mathcal{R}_\phi$ is the residual of the phase indicator transport equation given by \begin{align}
	\boldsymbol{\mathcal{R}}_\phi = \partial_t\phi^{\mathrm{f}, n+\alpha}_h + \left(\left(\boldsymbol{u}^{\mathrm{f}, n+\alpha}-\boldsymbol{u}^{\mathrm{m}}\right)\cdot\nabla\right)\phi^{\mathrm{f}, \mathrm{n}+\alpha}_\mathrm{h} + s\phi^{\mathrm{f}, \mathrm{n}+\alpha}_\mathrm{h}. 		
\end{align}	The phase indicator variable $\phi_h$, takes values between $[0,1]$, and is associated with high gradients in regions of phase transition. Consequently, additional treatment of the variational form is essential to address oscillations in the solution near regions of high gradients \citep{joshi2018positivity}. The positivity-preserving property is essential to prevent the introduction of numerical instability through non-physical interpolations of density and viscosity. Additional non-linear stabilization terms are added to impart the positivity property to the underlying element level matrix. The third and fourth lines of Eq.\ref{eq:TEMvariational} contain the positivity preserving nonlinear stabilization terms in the streamwise and crosswind directions respectively \citep{joshi2018positivity}. The PPV parameters $\zeta$,$k_s^{\mathrm{add}}$ and ,$k_c^{\mathrm{add}}$ for the phase indicator transport equation are obtained as:	\begin{align}
	\zeta &= \frac{2}{|s|h + 2|\left(\boldsymbol{u}^{\mathrm{f}}-\boldsymbol{u}^{\mathrm{m}}\right)|},\\
	k_s^\mathrm{add} &= \mathrm{max} \bigg\{ \frac{||\left(\boldsymbol{u}^{\mathrm{f}}-\boldsymbol{u}^{\mathrm{m}}\right)| - \tau_\phi|\left(\boldsymbol{u}^{\mathrm{f}}-\boldsymbol{u}^{\mathrm{m}}\right)|s|h}{2} - \tau_\phi|\left(\boldsymbol{u}^{\mathrm{f}}-\boldsymbol{u}^{\mathrm{m}}\right)|^2 + \frac{sh^2}{6}, 0 \bigg\},\\
	k_c^\mathrm{add} &= \mathrm{max} \bigg\{ \frac{|\left(\boldsymbol{u}^{\mathrm{f}}-\boldsymbol{u}^{\mathrm{m}}\right)|h}{2} + \frac{sh^2}{6}, 0 \bigg\},	
\end{align} where $|\left(\boldsymbol{u}^{\mathrm{f}}-\boldsymbol{u}^{\mathrm{m}}\right)|$ is the magnitude of the convection velocity and $h$ is the characteristic element length \cite{joshi2017positivity}. 
	
\subsection{Modal Decomposition and Spatio-Temporal Discretization of the Linear Structure} \label{sec:numStruct}
We define the discrete solution to the linear-elastic problem as $\mathbf{w}_h^\mathrm{s} \in \mathcal{S}^h_s$, where $\mathcal{S}^h_s$ is the space of trial solutions for the structure and satisfy the Dirichlet boundary conditions, and $\mathcal{V}^h_s$ is the space of test functions for the structure which vanish on the Dirichlet boundaries. The variational statement involves finding $\mathbf{w}_h^\mathrm{s} \in \mathcal{S}^h_s$, such that $\forall ~ \mathbf{N}^h_s\in \mathcal{V}^h_s$:\begin{equation}
\int_{\Omega^{\mathrm{s}}}  \mathbf{N}_h^\mathrm{s}\frac{\partial^2 \mathbf{w}_h^\mathrm{s}}{\partial t^2} \mathrm{d}\Omega^{\mathrm{s}} + \int_{\Omega^\mathrm{s}}\nabla \mathbf{N}_h^\mathrm{s} : \boldsymbol{\sigma}_h^\mathrm{s} \mathrm{d}\Omega^\mathrm{s} = 0.
\end{equation}
Using the decomposition indicated in Eq.\ref{eq:modaldec} and simplifying we can show the discrete eigenvalue problem takes the form:\begin{equation}
\left(\mathbf{K}^s_h - \omega_n^2 \mathbf{M}^s_h\right)\hat{\mathbf{w}}_{nh}^\mathrm{s} = 0,
\end{equation} where $\mathbf{M}^\mathrm{s}_h$ and $\mathbf{K}^\mathrm{s}_h$ are the mass and stiffness matrices obtained from the finite element discretization, and $\hat{\mathbf{w}}_{nh}^\mathrm{s}$ is the discrete mode shape. The modal frequencies $\omega_n$ and mode shapes $\hat{\mathbf{w}}_{nh}^\mathrm{s}$ of the structure are computed in the pre-processing step. In the current study, we consider $N_m$ modes whose effective modal mass cumulatively constitutes $90\%$ of the total structural mass. 

From Eq.\ref{eq:solidstress}, we can consider the equation for structural motion to take the form: \begin{equation}
\mathbf{M}^\mathrm{s} \ddot{\mathbf{w}}^\mathrm{s} + \mathbf{K}^\mathrm{s}\mathbf{w}^\mathrm{s} = \mathbf{f}_{ext}^\mathrm{s}, \label{eq:structSD}
\end{equation} where $\mathbf{M}^\mathrm{s}$ and $\mathbf{K}^\mathrm{s}$ are the mass and stiffness matrices per unit length respectively. $\ddot{\mathbf{w}}^\mathrm{s}$ denotes the acceleration of the structure, and $\mathbf{f}_{ext}^\mathrm{s}$ is the external force vector per unit length. The force vector $\mathbf{f}_{ext}^\mathrm{s}$ is evaluated by computing the hydrodynamic forces generated on the structure as a consequence of the fluid pressure and viscous stresses. As the structure is assumed to be linear, its motion is assumed to be a linear combination of the vibration modes. Therefore, the modal decomposition of the structure at the discrete level can be expressed as \begin{equation}
\mathbf{w}_h^\mathrm{s} = \sum_{i = 1}^{N_m} \xi^\mathrm{s}_i{\hat{\mathbf{w}}}_{ih}^\mathrm{s}, \label{eq:modalDecomposition}
\end{equation} where $\hat{\mathbf{w}}_{ih}$ is the mode shape, $N_m$ is the number of modes and ${\xi}_i^\mathrm{s}$ is the modal displacement of the $i^{th}$ mode. The transformation given in Eq.\ref{eq:modalDecomposition} allows for decomposing the equation of structural motion into $N_m$ ordinary differential equations in time given by: \begin{equation}
\mathbf{M_\xi}^\mathrm{s} \ddot{\boldsymbol{\xi}}^\mathrm{s} + \mathbf{K_\xi}^\mathrm{s}\boldsymbol{\xi}^\mathrm{s} = \mathbf{F_\xi}_{ext}^\mathrm{s}. \label{eq:structSD1}
\end{equation} Here, $\mathbf{M_\xi}^\mathrm{s} = \hat{\mathbf{W}}_h^T\mathbf{M}^\mathrm{s}\hat{\mathbf{W}}$ and $\mathbf{K_\xi}^\mathrm{s} = \hat{\mathbf{W}}_h^T\mathbf{K}^\mathrm{s}\hat{\mathbf{W}}_h$ denote the projected mass and stiffness matrices in the eigenspace respectively. The force vector is similarly projected into the eigenspace as $\mathbf{F_\xi}_{ext}^\mathrm{s} = \hat{\mathbf{W}}_h^T\mathbf{f}_{ext}^\mathrm{s}$. The system of decoupled ODE's thus obtained is advanced in time using the generalized-$\alpha$ method as described in Sec.(\ref{sec:numTemporal}).
	
\subsection{Unified ALE FSI framework}
The coupling between the fluid and structure is carried out through a partitioned iterative approach using the Nonlinear Iterative Force Correction (NIFC) scheme described in \citep{jaiman2016les,jaiman2016partitioned}. The displacements obtained through solving the structural equations have to be incorporated into the fluid-flow framework, which is accomplished by controlling the motion of each node in the mesh while maintaining the kinematic consistency of the discretized interface. Assuming the fluid mesh to represent a hyperelastic solid model, and using a Lagrangian finite element framework, allows for adapting the fluid mesh to the deformed geometry of the structure.

To advance the solution from $t^n$ to $t^{n+1}$, a sequence of predictor-corrector steps are taken, with the dual objectives of enforcing convergence of the individual numerical methods and coupling the multiphysics phenomena. Using the flow information $\left(\mathbf{u}^\mathrm{f}(\mathbf{x}^\mathrm{f},t^n),p\right)$, the structural displacement is predicted $[A]$. Next, ALE compatibility and velocity continuity at the fluid-structure interface $\Gamma^\mathrm{fs}$ is maintained by equating the computed structural displacements to the mesh displacement on the wetted boundaries, as well as imposing the mesh velocity on the fluid velocity at the same spatial locations at time $t^{n+\alpha}$ $[B]$. These conditions can be expressed as: \begin{align}
\mathbf{d}_m^{n+1} &= \mathbf{w}^\mathrm{s} \quad \text{on} \quad \Gamma^\mathrm{fs},\label{eq:kinematicComp}\\
\mathbf{u}^{\mathrm{f},n+\alpha} &= \dot{\mathbf{d}}^{n+\alpha}_m \quad \text{on} \quad \Gamma^\mathrm{fs}. \label{eq:fluidkinematicComp}
\end{align}
In Eqs.(\ref{eq:kinematicComp}) and (\ref{eq:fluidkinematicComp}), $\mathbf{d}_m^{n+1}$ denotes the mesh displacement at time $t^{n+1}$ and $\dot{\mathbf{d}}^{n+\alpha}_m$ represents the mesh velocity at time $t^{n+\alpha}$, and is defined as: \begin{equation}
\dot{\mathbf{d}}^{n+\alpha}_m = \frac{\mathbf{d}_m^{n+1} - \mathbf{d}_m^{n}}{\Delta t} \quad \text{on} \quad \Gamma^\mathrm{fs}.
\end{equation}
Once the fluid mesh and structure satisfy the compatibility constraints, the fluid-flow equations are solved first $[C]$, followed by the cavitation transport equation $[D]$ to a reasonable level of convergence. This allows for the computation of the forces acting on the structure, and the NIFC filter further filters the forces to provide stability to the overall partitioned fluid-structure coupling. The NIFC scheme acts to construct the cross-coupling effect of strong fluid-structure interaction along the interface without forming the off-diagonal Jacobian term in an explicit manner \citep{jaiman2016partitioned}. The sequence of equations $[A],[B],[C],[D]$ are presented in matrix-vector form as : \begin{align}
[A] &\quad: \left[\boldsymbol{K^s}_\xi\right]\{\Delta \xi\}=\{{\boldsymbol{\mathcal { R }}}_{\xi}\} \label{eq:structmatrices},\\
[B] &\quad: \left[\boldsymbol{K^s}_d\right]\{\mathbf{d}^{n+1}\}=\{{\boldsymbol{0}}\} \label{eq:alematrices},\\
[C] &\quad: \left[\begin{array}{cc}
\boldsymbol{K}_{\Omega^{\mathrm{f}}} & \boldsymbol{G}_{\Omega^{\mathrm{f}}} \\
-\boldsymbol{G}_{\Omega^{\mathrm{f}}}^{T} & \boldsymbol{C}_{\Omega^{\mathrm{f}}}
\end{array}\right]\left\{\begin{array}{c}
\Delta \boldsymbol{u}^{\mathrm{f}, n+\alpha} \\
\Delta p^{\mathrm{f}, n+1}
\end{array}\right\}=\left\{\begin{array}{l}
{\boldsymbol{\mathcal { R }}}_{\mathrm{m}} \\
{\boldsymbol{\mathcal { R }}}_{\mathrm{c}}
\end{array}\right\} \label{eq:NSmatrices},\\
[D] &\quad: \left[\boldsymbol{K}_{\phi}\right]\{\Delta \phi^{\mathrm{f}, n+\alpha}\}=\{{\boldsymbol{\mathcal { R }}}_{\phi}\} \label{eq:phimatrices}
\end{align}

In this numerical study, the linearized systems for incremental pressure, velocity and cavitation fraction arising from the finite element discretization are solved using the Generalized Minimum RESidual (GMRES) \citep{saad1986gmres} with a diagonal preconditioner. For all computations presented, a Krylov space of 25 orthonormal vectors has been used. In the current implementation, only the block matrix is explicitly constructed, and the residual for the linear system is constructed through matrix-vector products. The mesh motion is solved for using a conjugate gradient method owing to the symmetric structure of the left-hand side matrix. For the flow and cavitation fields, equal-order interpolations are considered, with linearly interpolated hexahedral and prism elements being used at the required precision. The solver is written in a combination of C and Fortran languages for scalability to high-performance computing. The solver uses communication protocols based on standard message passing interface \cite{MPI} for parallel computing on distributed memory clusters.

\begin{figure}[!h]
	\includegraphics[width=1\textwidth]{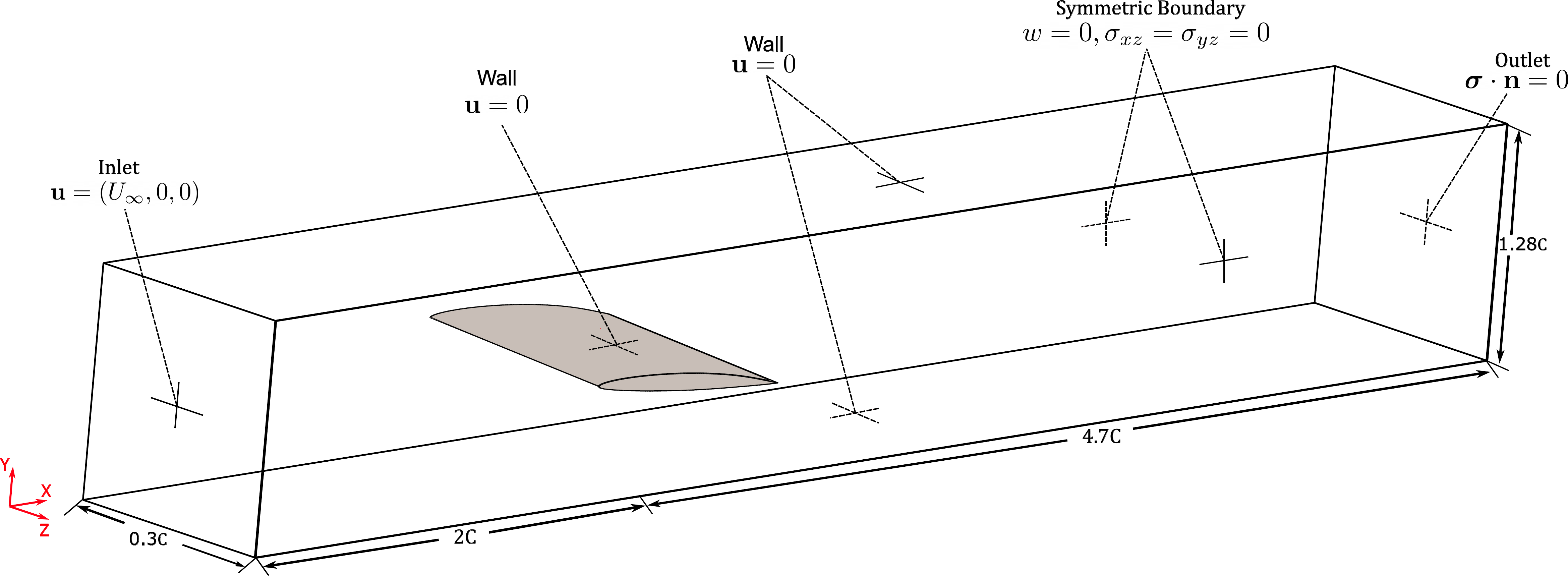}      
	\caption{A schematic of the flow past the NACA66 hydrofoil at $Re = 750000$. The computational setup and boundary conditions are shown for the filtered Navier–Stokes equations. Here, $\mathbf{u} = (u , v , w)$ denotes the components of the fluid velocity and $C$ denotes the chord length.}
	\label{fig:Domain_CavVal}
\end{figure}
	
\section{Validation of LES for Cavitating flow past rigid NACA66 hydrofoil} \label{sec:rigidhydrofoil}
In the current section, we present the validation of the cavitating flow past a rigid NACA66 hydrofoil validated against the experimental studies of \citet{leroux2004experimental} and the large eddy simulations presented in \citet{chen2019LES}. The success of LES depends highly on the resolution achieved by the mesh. Accordingly, the principle objective of the current verification study is to establish the mesh characteristics necessary to capture the complex turbulent-cavitation structures emerging in the flow field and evaluate the flow statistics. For all the cases considered in the current study, we set $\rho_{l} = 998.1\mathrm{kg m}^{-3}$, $\rho_{v} = 0.023\mathrm{kg m}^{-3}$, $\mu_l = 0.0011 \mathrm{kg m}^{-1}\mathrm{s}^{-1}$, $\mu_v = 9.95\times10^{-6} \mathrm{kg m}^{-1}\mathrm{s}^{-1}$ for the liquid and vapor phase properties. For the cavitation model, the parameters are $C_c = 10^{-3}$, $C_v = 5\times10^{-3}$, $d_{nuc} = 2.5\times10^{-6}$, $n_0 = 10^{13}$. The freestream velocity $U_\infty$ is calculated based on the liquid phase density and viscosity, in conjunction with the chord-based Reynolds number. 
\subsection{Problem Description and Mesh Characteristics}\label{sec:rigidmesh}
\noindent We validate the cavitating flow past the rigid NACA66 hydrofoil at $Re = 7.50\times10^5$, $\sigma = 1.25$ and $\alpha = 6^\circ$. At these conditions, the flow experiences a laminar to turbulent transition over the hydrofoil surface. Further, the fluid experiences highly unsteady cavitation, with periods of cavity separation and complete collapse, owing to a combination of re-entrant jet instability and the interfacial instabilities induced on the upper surface of the cavity. Upon collapse, an intricate vortex-cavity field develops over the suction side of the hydrofoil, and subsequently vortex-cavities convect downstream into the wake. \citep{ji2015les,wu2015numericalFlexHydrofoil, chen2019LES, suraj2023cavviv}

To investigate the flow as described above, we define the hydrofoil by coordinates prescribed in \citet{leroux2004experimental}. We denote the hydrofoil's chord length by $C$, and truncate the hydrofoil $0.02C$ from the trailing edge. Fig. \ref{fig:Domain_CavVal} illustrates the computational domain used for conducting the numerical validation. The inlet and outlet boundaries are positioned $2C$ and $4.7C$ upstream and downstream of the leading edge (LE) respectively. The channel's width is set to be $1.28C$ in accordance with the channel height prescribed in \citet{leroux2004experimental}. The span $S$ of the hydrofoil is taken to be $0.3C$, and symmetric boundary conditions are imposed along the Z-axis domain limits. A uniform flow field is imposed at the inlet, and the freestream velocity is set as the initial condition for the non-cavitating flow. The non-cavitating flow is allowed to fully develop up to the non-dimensional time $t^* = tU_\infty/C = 18$, after which the flow is allowed to enter the cavitating regime. A uniform time-step of $\Delta t^* = \Delta tU_\infty/C = 9.434\times10^{-3}$ is used for the rigid hydrofoil study.

\begin{table}[!h]	
	\centering
	\begin{tabular}{|c c c c|}
		\hline
		Property & M1 & M2 & M3 \\
		\hline
		$y^+$ & $0.6$ & $0.3$ & $0.15$\\
		$\left(\Delta x^+,\Delta z^+\right)$ & $(65,75)$ & $(45,60)$ & $(35,45)$\\
		$\left(N_{chord},N_{span}\right)$ & $(200,90)$ & $(300,120)$ & $(400,150)$\\
		$\left(N_{BL}, r_{BL}\right)$ & $(51,1.136)$ & $(66,1.103)$ & $(81,1.096)$\\
		$N^{nodes}$ & $3.59 M$ & $8.03 M$ & $15.25M$\\
		$N^{elem.}$ & $3.68 M$ & $8.18 M$ & $15.51M$\\
		\hline
	\end{tabular}
	\caption{Mesh parameters for the LES validation of flow past NACA66 hydrofoil. Here $N_{chord}$ and $N_{span}$ denote the number of points in the chord-wise and span-wise directions on the hydrofoil surface. $N_{BL}$ denotes the number of points in the boundary layer and $r_{BL}$ denotes the growth ratio of element sizes normal to the boundary. $N^{nodes}$ and $N^{elem.}$ denote the total number of nodes and elements in the mesh respectively. For the current study, only 8 node tri-linear hexahedral elements are used.}
	\label{tab:MeshParameters}
\end{table}

\begin{figure}[!ht]	
	\begin{subfigure}[h]{0.5\textwidth}
		\centering
		\includegraphics[width=1\textwidth]{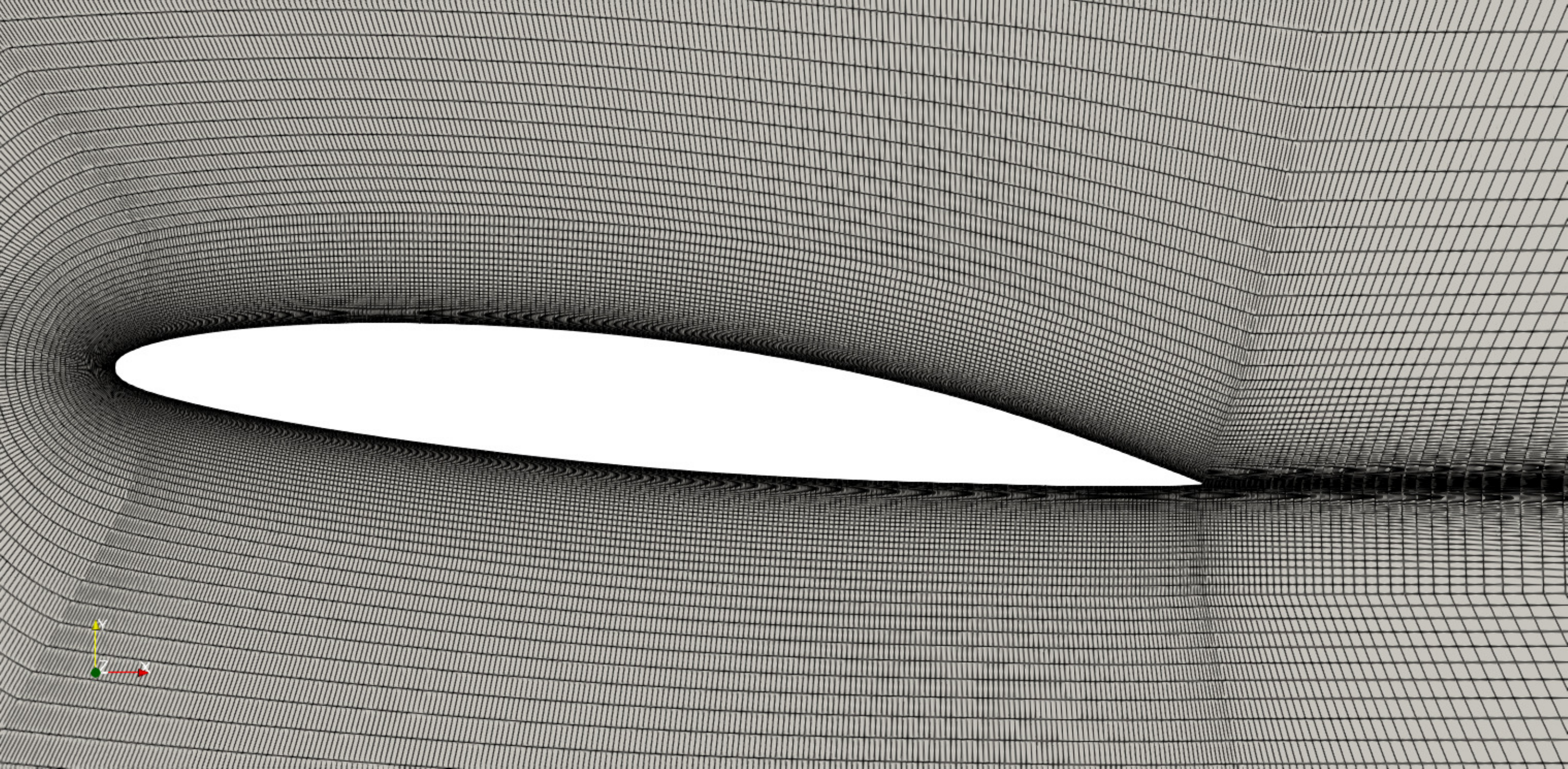}
		\caption{}
		\label{fig:MeshCross}
	\end{subfigure}
	\begin{subfigure}[h]{0.5\textwidth}
		\centering
		\includegraphics[width=1\textwidth]{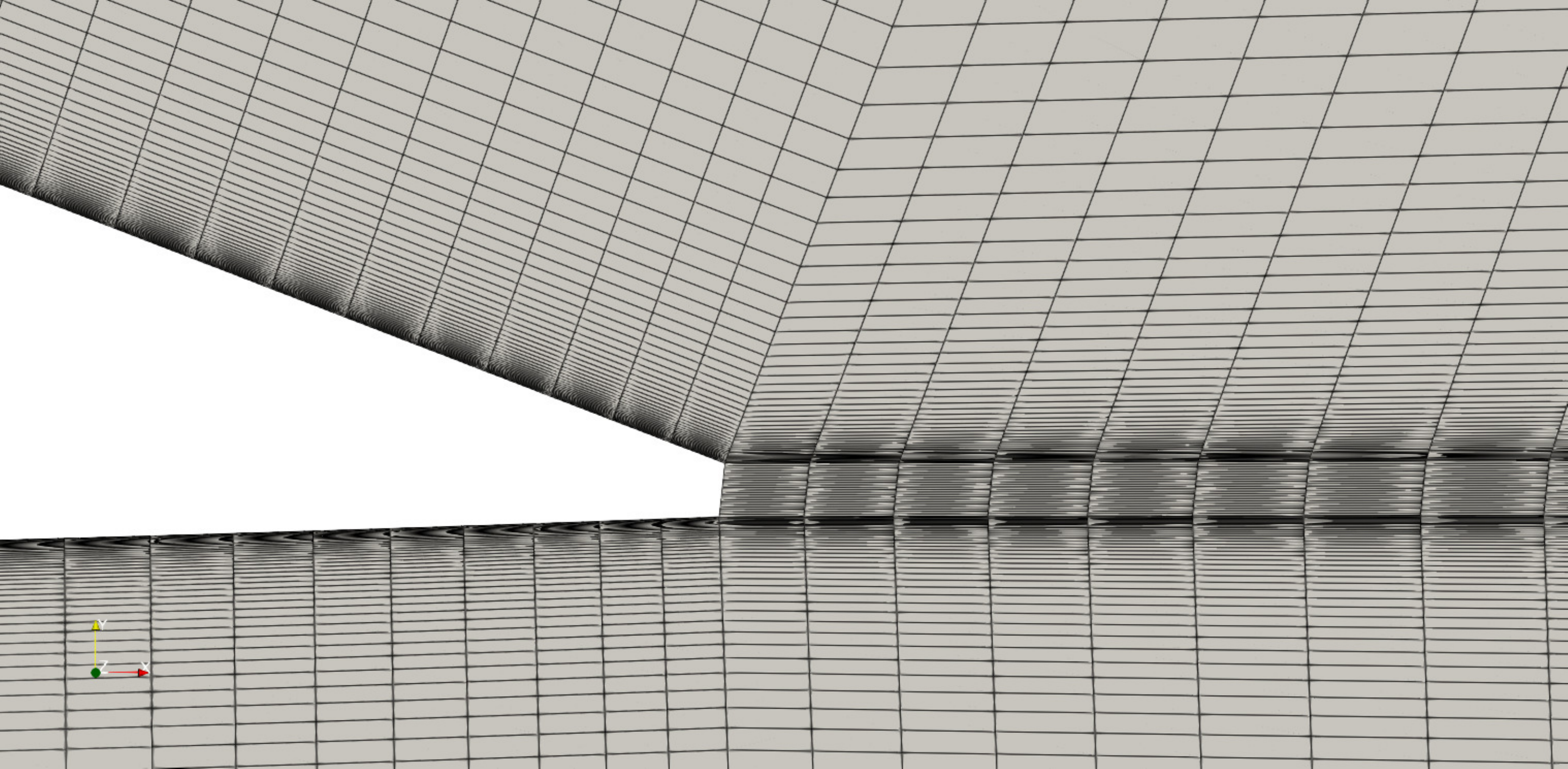}
		\caption{}
		\label{fig:MeshCrossTE}
	\end{subfigure}
	\newline
	\begin{subfigure}[h]{0.5\textwidth}
		\centering
		\includegraphics[width=1\textwidth]{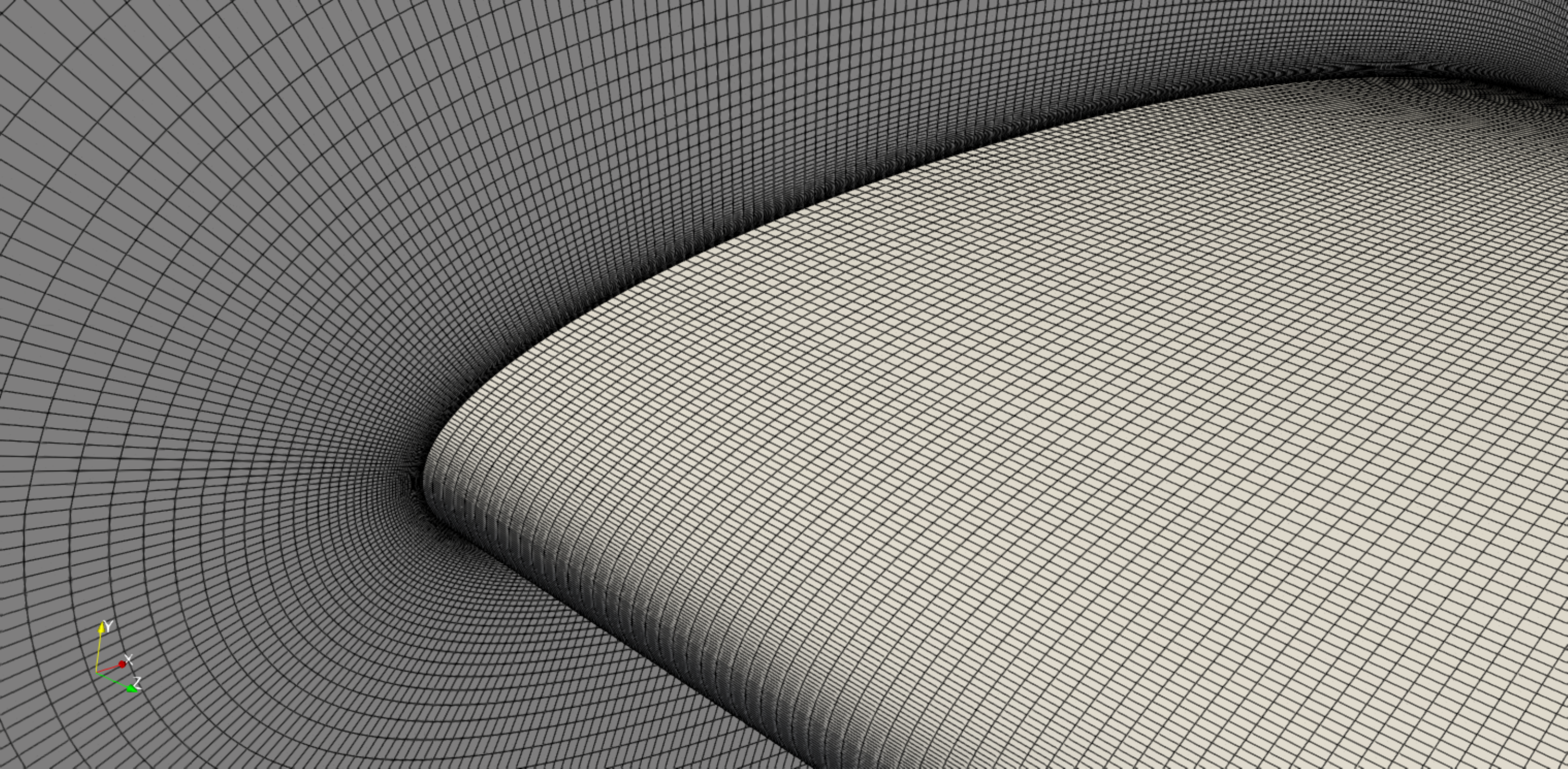}
		\caption{}
		\label{fig:MeshLE}
	\end{subfigure}
	\begin{subfigure}[h]{0.5\textwidth}
		\centering
		\includegraphics[width=1\textwidth]{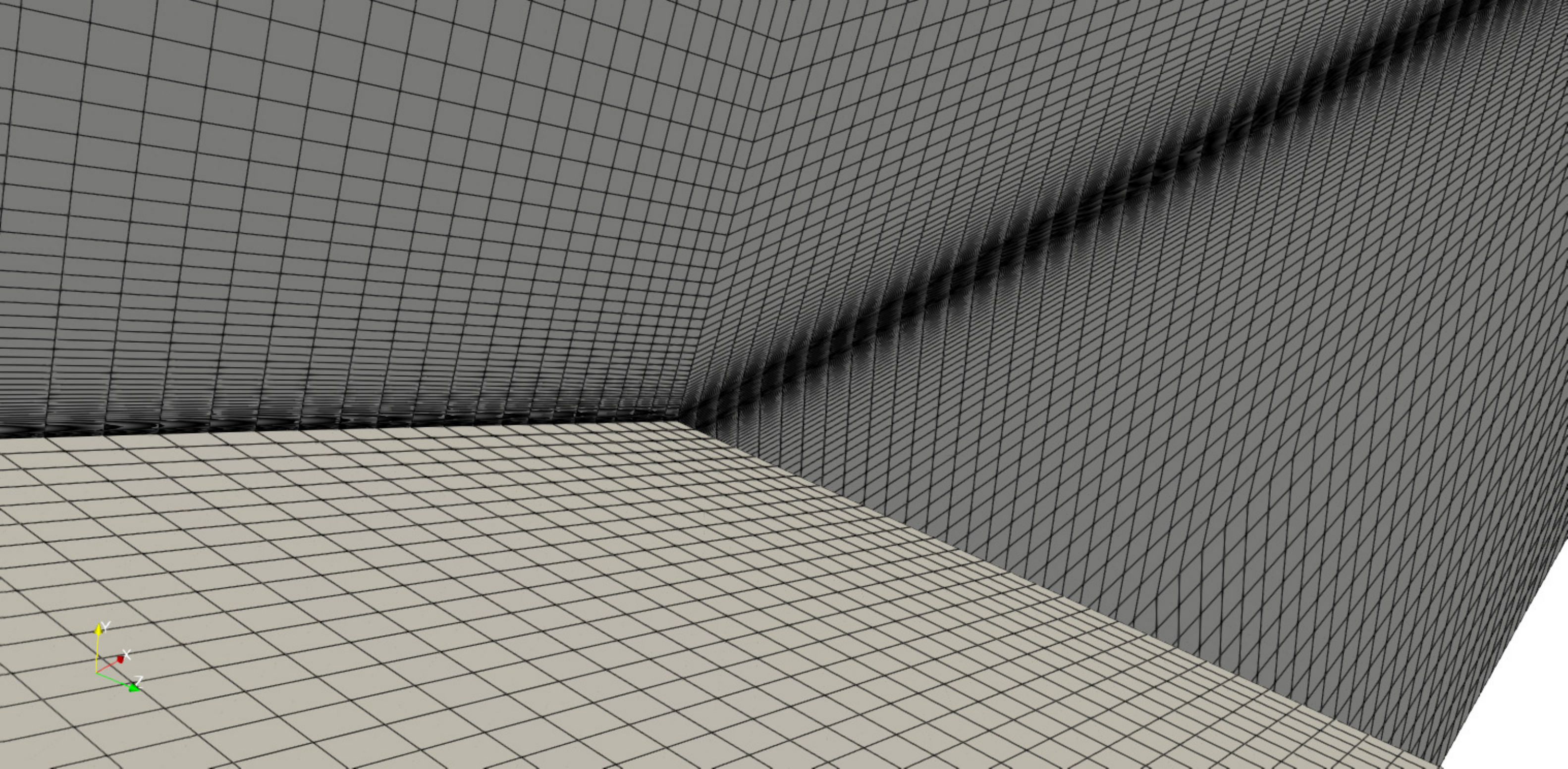}
		\caption{}
		\label{fig:MeshTE}
	\end{subfigure}
	\caption{Cross-sectional view of mesh M2 at (a) boundary layer and near wake, (b) close to the trailing edge. (c,d) Resolution of the hydrofoil span at the leading and trailing edges }
	\label{fig:Mesh_Res}
\end{figure}

\begin{figure}[!hb]
	\begin{subfigure}[h]{0.5\textwidth}
		\centering
		\includegraphics[width=1\textwidth]{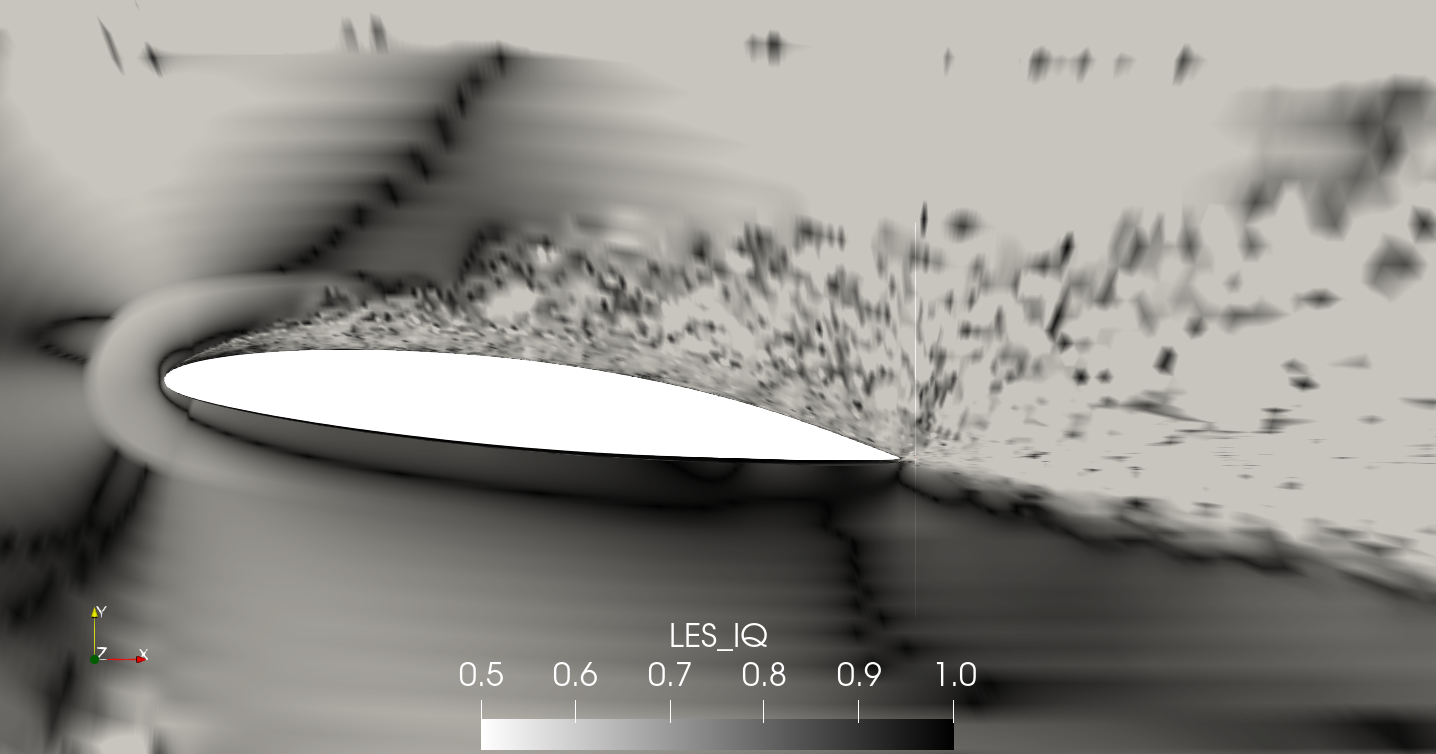}
		\caption{}
		\label{P0}
	\end{subfigure}
	\begin{subfigure}[h]{0.5\textwidth}
		\centering
		\includegraphics[width=1\textwidth]{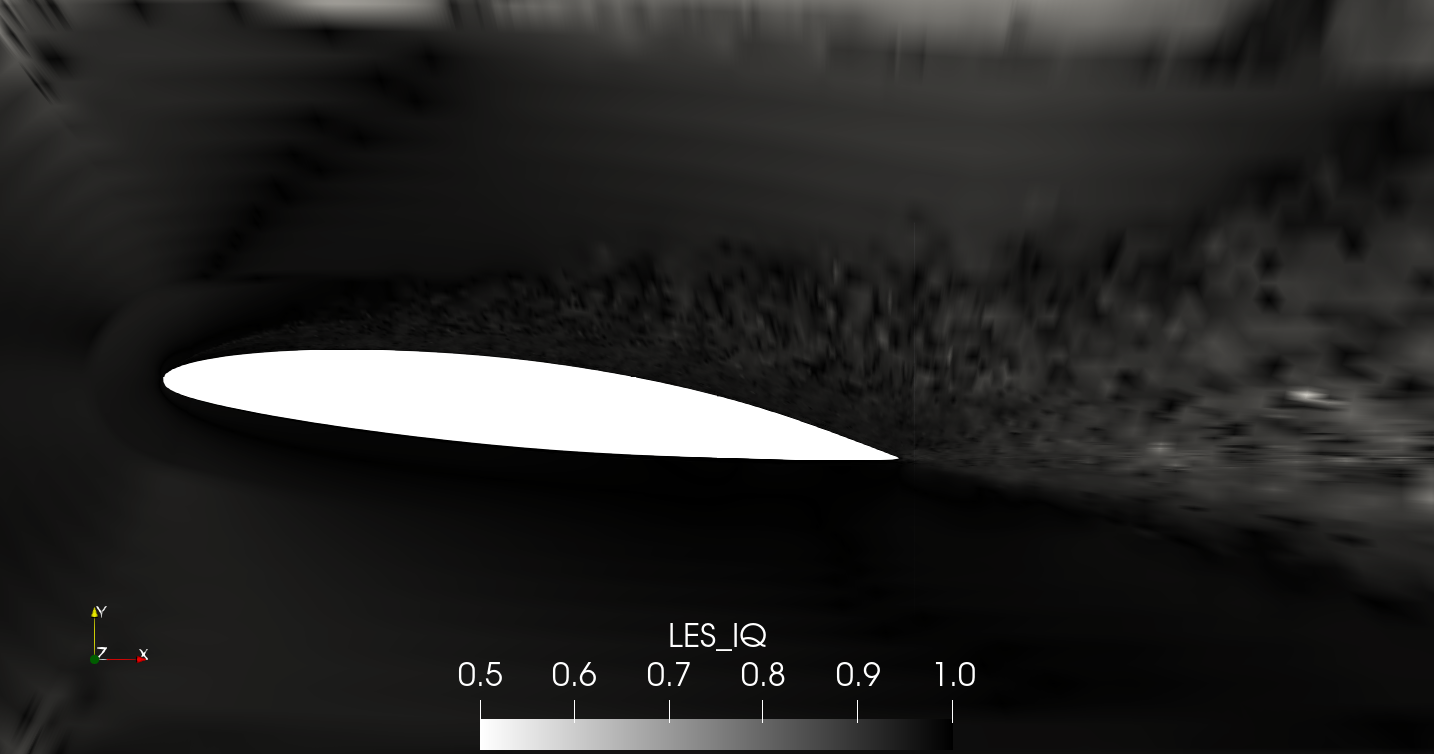}
		\caption{}
		\label{P1}
	\end{subfigure}
	\newline
	\begin{subfigure}[h]{0.5\textwidth}
		\centering
		\includegraphics[width=1\textwidth]{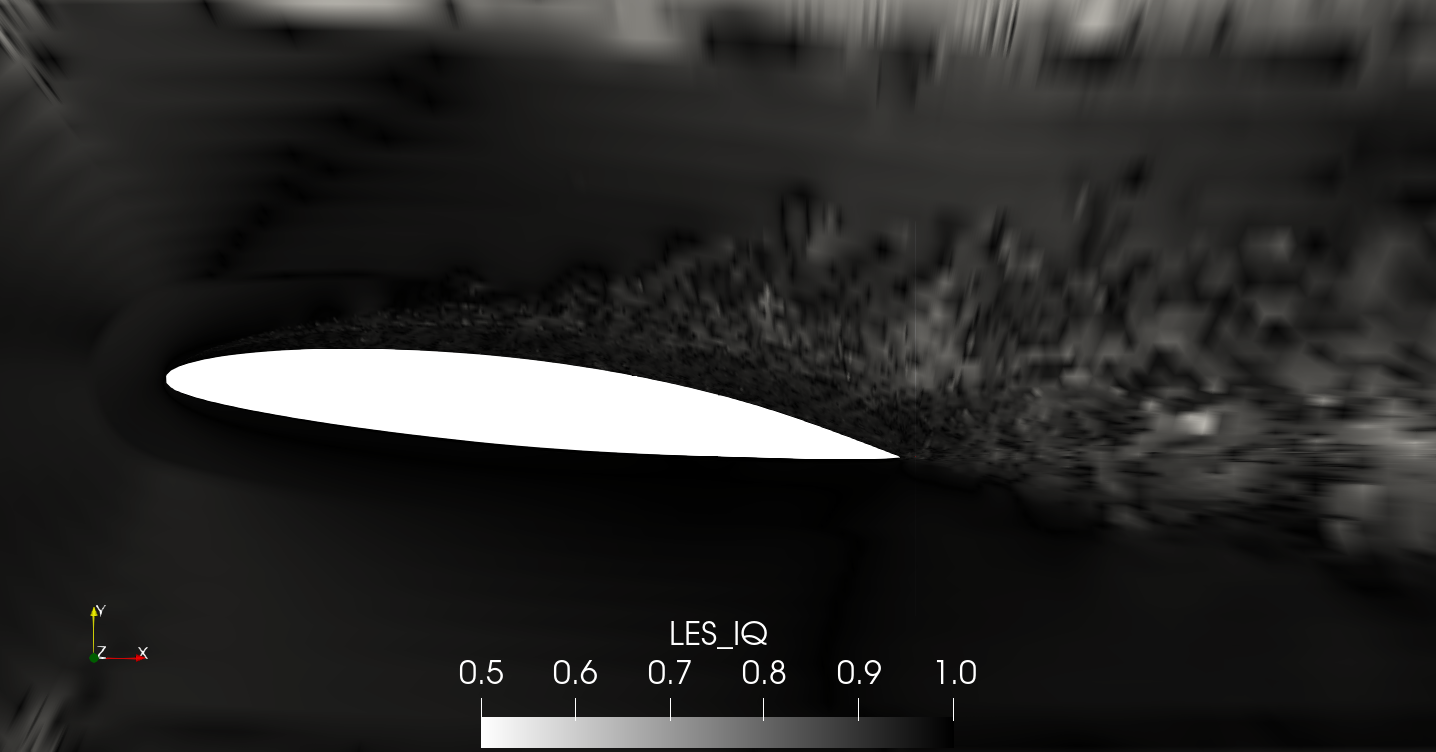}
		\caption{}
		\label{D0}
	\end{subfigure}
	\begin{subfigure}[h]{0.5\textwidth}
		\centering
		\includegraphics[width=1\textwidth]{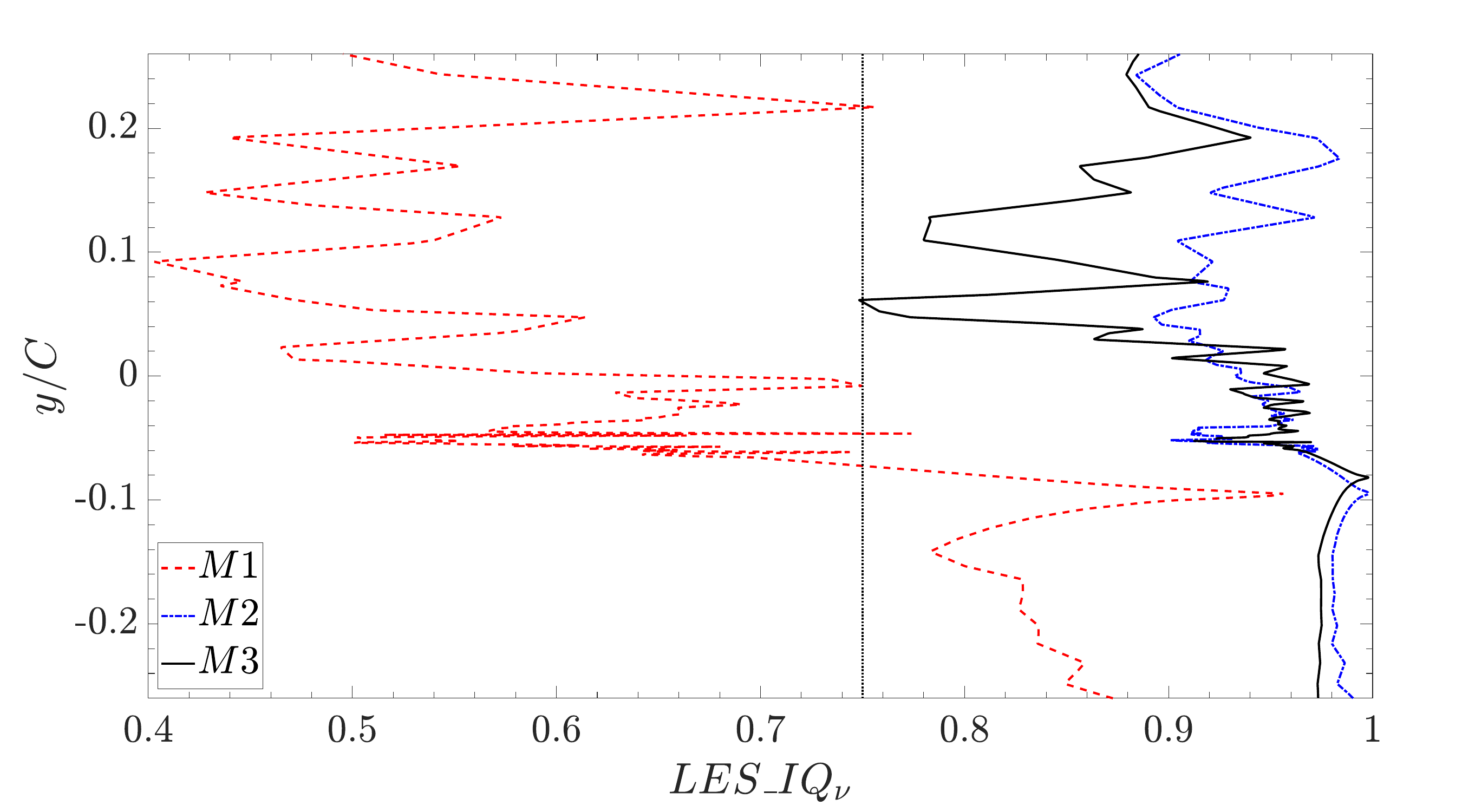}
		\caption{}
		\label{fig:compLESIQ}
	\end{subfigure}
	\caption{Index of quality for LES  in the region of interest around the hydrofoil for three meshes (a) M1,  (b) M2,  (c)  M3,  and (d) Comparison of $LES\_IQ_\nu$ at $(x,z) = (C,0.15C)$.}
	\label{fig:LES_IQ}
\end{figure}
To accurately capture cavity inception and the evolution of re-entrant jet dynamics in the laminar sub-layer, we set $y^+ < 1$ along the hydrofoil surface. In the case of cavitating flow, the fluctuation in $y^+$ can be substantial owing to regions covered in the vapor phase. Under unsteady partial cavitation conditions, the re-entrant jet velocity has been observed to be nearly $50\%$ of the freestream velocity, with a thickness smaller than $30\%$ of the cavity thickness \citep{callenaere2001CavInstab}. As a consequence, the re-entrant jet is associated with high-velocity gradients normal to the hydrofoil, and the mesh resolution can be established based on estimates of re-entrant jet momentum. To get a preliminary estimate for mesh size, we use the data obtained by \citet{suraj2021femcav} with respect to the re-entrant jet momentum for a rigid cavitating hydrofoil, and evaluate the magnitude of wall shear stress $\tau_w$ as follows: \begin{align}
\mathbf{T}_v &= \mu^l \left(\nabla \mathbf{u}^\mathrm{f} + \left(\nabla \mathbf{u}^\mathrm{f}\right)^T\right) \mathbf{n},\\
\mathbf{T}_{v\parallel} &= \mathbf{T}_v - \left(\mathbf{T}_v\cdot\mathbf{n}\right)\mathbf{n},\\
\tau_w &= ||\mathbf{T}_{v\parallel}||_2,
\end{align} where $\mathbf{T}_v$ represents the traction vector due to viscous stresses on the hydrofoil surface, and $\mathbf{T}_{v\parallel}$ is the shearing component of traction arising due to viscous stresses. We use the wall-shear stress to compute the friction-velocity $u_\tau$ in the re-entrant jet region as follows: \begin{equation}
u_\tau = \left(\frac{\tau_w}{\rho_l}\right)^\frac{1}{2}.
\end{equation}
This allows us to evaluate the wall-normal distance $y$, based on a target $y^+$ value as $y^+ = u_\tau y/\nu$. This method of estimation has been used to construct three successively refined meshes, labeled M1 (coarsest), M2, and M3 (finest). Further, LES requires the mesh to be isotropic in the wake region, in order to homogeneously resolve the turbulent structures. We refine the mesh by resolving the hydrofoil along chord-wise ($\Delta x^+$) and span-wise ($\Delta z^+$) directions, increasing the number of points in the boundary layer and wake regions in accordance with the guidelines prescribed by \citet{spalart2000turbstrat}. The meshes are constructed in Gmsh \citep{geuzaine2009gmsh}, and the parameters associated with the meshing process are presented in table \ref{tab:MeshParameters}. Fig.\ref{fig:Mesh_Res} presents the resolution offered by mesh M2 around the suction side and wake of the hydrofoil.

Owing to the filtering of governing equations based on the spatial resolution offered by the mesh, LES does not lend itself to a mesh convergence analysis. Theoretically, LES on a sufficiently refined mesh has the capability to resolve flow patterns down to the Kolmogorov length scale. However, taking into account the computational expense involved, it is necessary to identify the coarsest level of flow resolution acceptable for the problem under consideration. Along with the preliminary criteria established for meshing, LES quality has to be satisfactory on the constructed mesh. Some of the criteria proposed to evaluate LES quality are (a)$LES\_IQ_\eta$ based on the Kolmogorov length scale $\eta$ \citep{celik2005LESIQ}, (b)$LES\_IQ_\nu$ based on the turbulent eddy viscosity $\nu_t$ \citep{celik2005LESIQ} and (c) turbulent kinetic energy resolution \citep{pope2000TurbFlows}. For the current validation study, we adopt the turbulent eddy viscosity criteria as defined by \citet{celik2005LESIQ} as: \begin{equation}
LES\_IQ_\nu = \frac{1}{1 + \alpha_\nu\left(\frac{\left<\nu_{t,eff}\right>}{\nu}\right)^n}.
\end{equation} $\alpha_\nu$ and $n$ are parameters governing the Index of Quality $(IQ)$ derived based on the spatial resolution of LES relative to DNS, and are set to $\alpha_\nu = 0.526$ and $n = 0.52$. $\left<\nu_{t,eff}\right>$ is the time-averaged eddy viscosity, and $\nu$ is the kinematic viscosity. For LES to be considered adequate, we require $LES\_IQ_\nu$ to be greater than $0.75$ in regions of interest. In this regard, we consider the wall-normal distance of $0.2C$ over the hydrofoil's suction side, and the immediate wake $\left(x/C < 2.0\right)$ downstream of the hydrofoil to be our regions of interest. We evaluate $LES\_IQ_\nu$, with averaging conducted over two cavitation cycles (discussed in section \ref{sec:Hydrodynamic}). We compare the contours of $LES\_IQ_\nu$ in Fig. \ref{fig:LES_IQ}(a-c), for each of the meshes considered, with Fig.\ref{fig:compLESIQ} indicating the comparison at $x/C = 1$. We observe mesh M2 offers adequate resolution for verification purposes, and the corresponding meshing parameters will be employed for further validation studies.

\begin{figure}[!b]
	\begin{center}
	\includegraphics[width=0.6\textwidth,trim={0cm 1.25cm 9.8cm 1cm},clip]{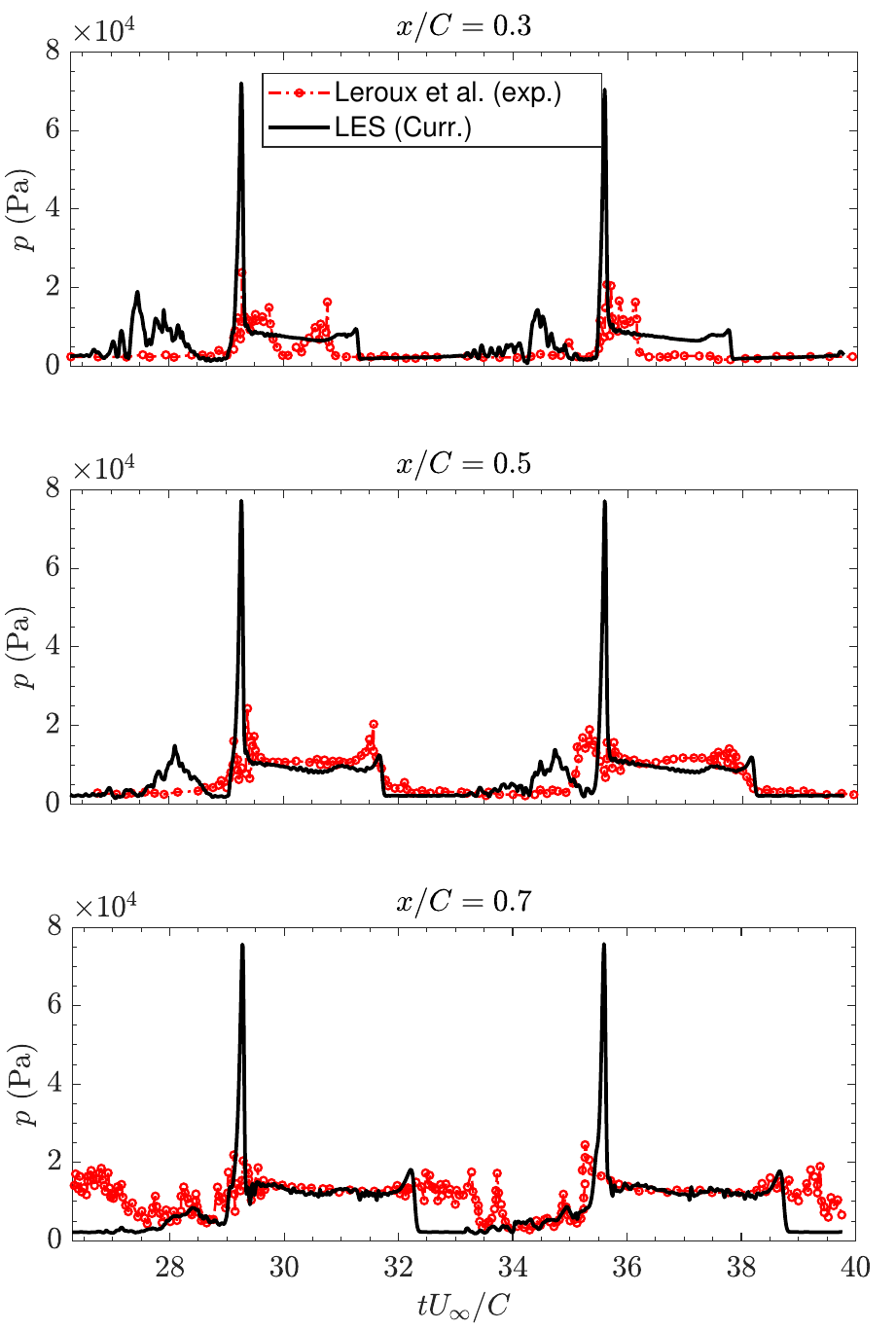}
	\end{center}
	\caption{Comparison of pressure at different chord-wise locations along the hydrofoil.}
	\label{fig:CompPres}
\end{figure}

\begin{figure}[!h]
	\begin{center}
	\includegraphics[width=0.7\textwidth]{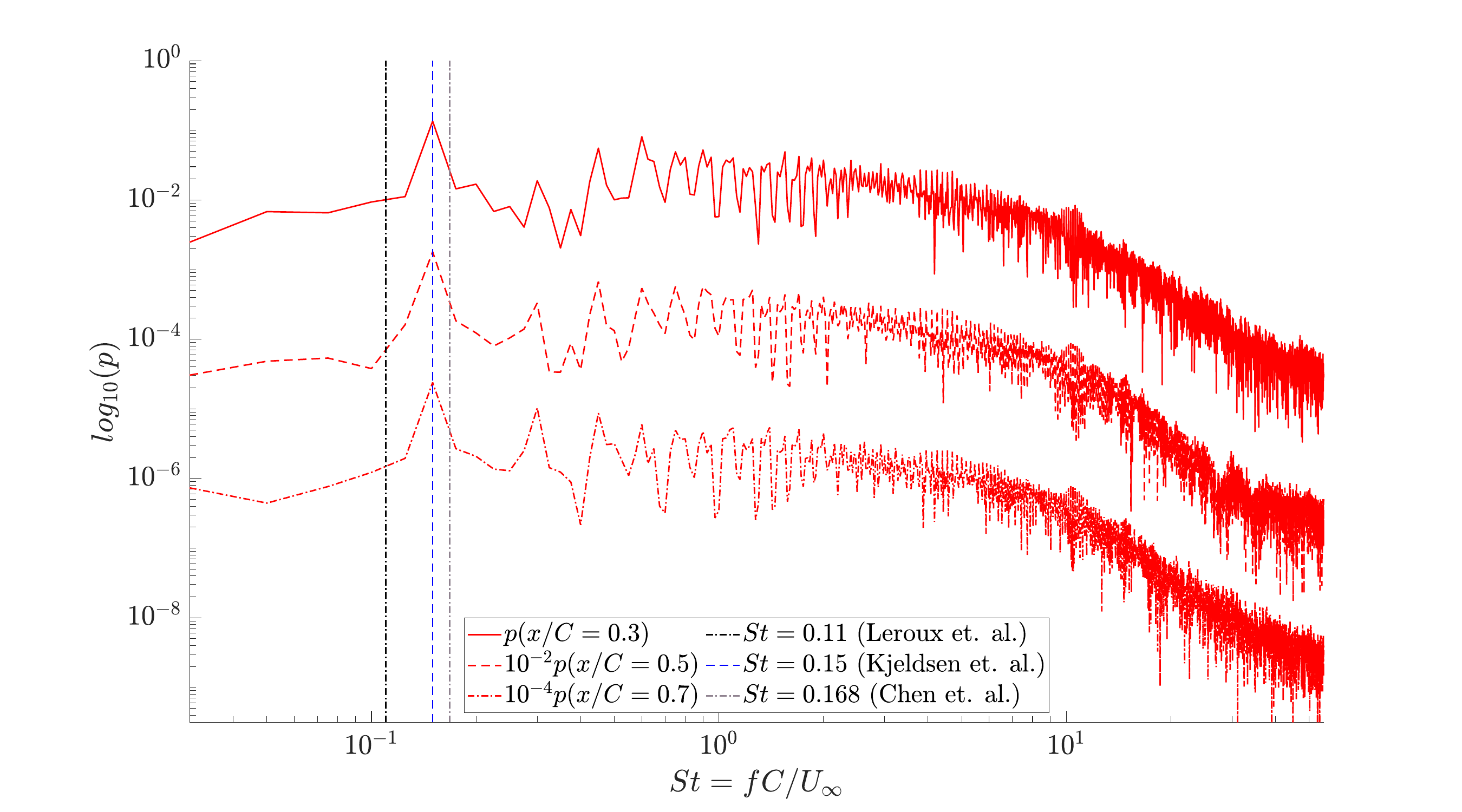}
	\end{center}
	\caption{Comparison of pressure frequency spectrum at chord-wise locations $x/C \in \{0.3, 0.5,0.7 \}$ with the dominant frequency modes observed in the studies of \citet{leroux2004experimental,kjeldsen2000sheetcloud,chen2019LES}.}
	\label{fig:CompFrq}
\end{figure}

\subsection{Hydrodynamic Coefficients and Cavitation Characteristics} \label{sec:Hydrodynamic}	
\noindent The hydrodynamic parameters obtained from the LES of cavitating flow past the NACA66 hydrofoil at $Re= 7.5\times10^5$ are compared against experimental and numerical results in Table \ref{tab:CLCD_Rigid}. We evaluate the fluid loads on the hydrofoil by integrating the surface traction over the first layer of elements in contact with the hydrofoil. In this context, we compute the instantaneous coefficients of lift and drag using \begin{align}
	C_L = \frac{2}{\rho^fU_{\infty}^2CS}\int_\Gamma \left(\mathbf{n}\cdot\boldsymbol{\sigma}\right)n_y d\Gamma \label{eq:lift},\\
	C_D = \frac{2}{\rho^fU_{\infty}^2CS}\int_\Gamma \left(\mathbf{n}\cdot\boldsymbol{\sigma}\right)n_x d\Gamma \label{eq:drag},
\end{align} where $\mathbf{n}$ is the unit normal to the hydrofoil surface, where $n_x$ and $n_y$ are its Cartesian components. $\boldsymbol{\sigma}$ is the fluid stress tensor and $\Gamma$ is the surface boundary of the hydrofoil. The pressure coefficient at a location in the domain is computed using \begin{equation}
	C_p = \frac{2\left(p - p_\infty\right)}{\rho_f U_\infty^2}, \label{eq:prescoeff}
\end{equation} where $p$ is the fluid pressure at the point of interest and $p_\infty$ is the far-field pressure. Owing to the multiphase nature of the flow, it is essential to characterize the cavity growth-collapse cycle and co-relate it with the pressure fluctuations observed close to the cavitating regions around the hydrofoil. In this regard, we integrate the cavitation fraction over the flow domain using \begin{equation}
	\int_{\Omega^f} (1-\phi_h)d\Omega, \label{eq:cavVolume}
\end{equation} to obtain the instantaneous global vapor volume. To evaluate the spectral characteristics in the current study, we define the non-dimensional frequency $f^*$, and the chord-based Strouhal number $St$ as
\begin{align}
	f^* = \frac{fC}{U_\infty},~ St = \frac{f^\mathrm{cav}C}{U_\infty},
\end{align} where $f$ denotes a general frequency observed over the course of the analysis and $f^\mathrm{cav}$ is the dominant cavity shedding frequency.
\begin{figure}[!ht]
	\centering
	\includegraphics[width=0.7\textwidth]{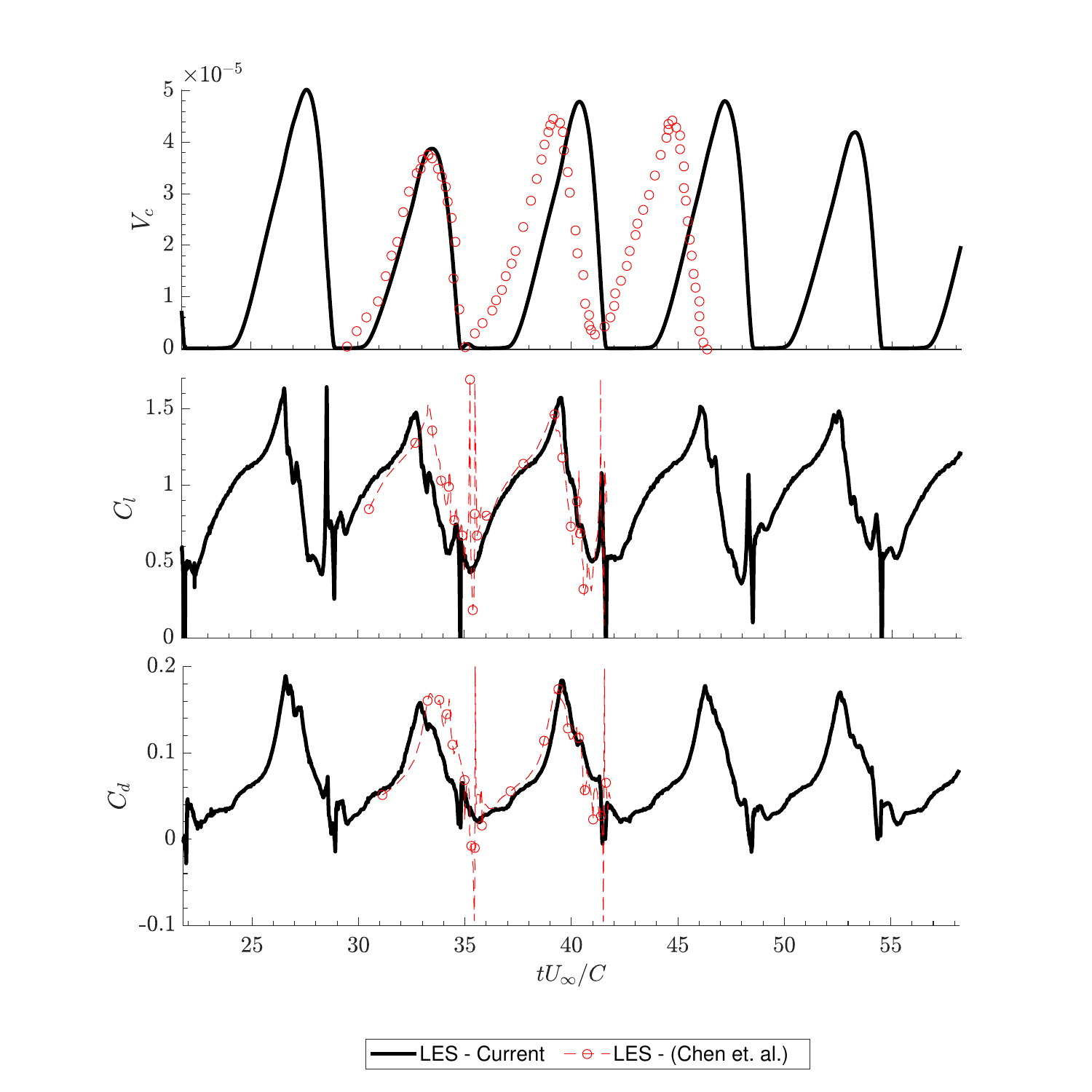}
	\caption{Comparison of cavity volume and force coefficients computed in the current study with the LES studies of \citet{chen2019LES}.}
	\label{fig:CompVcCLCDRigid}
\end{figure}

Fig. \ref{fig:CompPres} presents the comparison of pressure at different chord-wise locations on the hydrofoil with experimental data of \citet{leroux2004experimental}. We observe a good agreement in the observed pressure spike associated with the cloud cavity's collapse in the hydrofoil's wake. Over the next cavitation cycle, the sheet cavity's growth and cloud cavity's separation show reasonable agreement in the form of a uniform pressure field at the probe locations. Note that instants of pressure recovery vary from $x/C =0.3$ to $x/C=0.7$, owing to the convection of the separated cloud cavity. Subsequently, we observe the cloud cavity's destabilization where-in minor pressure spikes are observed relative to experimental measurements. While the current LES simulations capture the instantaneous pressure development over the cavity cycle well, we observe an over-prediction of the pressure spikes relative to experimental studies.

With respect to the periodicity of pressure oscillations, we observe the frequency spectra of the wall pressure at the chord-wise locations $x/C \in \{0.3, 0.5, 0.7\}$ in Fig.\ref{fig:CompFrq}. In the current study, we observe the dominant frequency of the pressure oscillation to be $f^\mathrm{cav} = 5.33$ Hz, which is equivalent to $St = 0.15$. This result is in agreement with the experimental studies of \citet{kjeldsen2000sheetcloud}, wherein the authors observed $St = 0.15$ with respect to the cavitation cycles of a NACA0015 hydrofoil. While \citet{leroux2004experimental} observed $St\approx0.11$ for the cavitation cycles of a NACA66 hydrofoil, they indicate an agreement of their results with the earlier work of \citet{kjeldsen2000sheetcloud}.

As illustrated by \citet{callenaere2001CavInstab}, the extent of confinement alters the cavity's periodicity with respect to growth, shedding and collapse. In this regard, the nature of boundary conditions imposed in a numerical study can alter the periodicity observed in cavitation dynamics, leading to a shorter cycle when the lateral boundaries are no-slip walls. In the case of symmetric boundary conditions at the lateral boundaries, reasonable agreement with $St=0.11$ is observed \citep{ji2015les,suraj2023cavviv}, whereas the use of no-slip walls at the lateral boundaries indicates agreement with $St=0.15$ \citep{chen2019LES}. The effects of confinement and their role in capturing cavitation dynamics require further investigation.
\begin{table}[!h]
	\centering
	\begin{tabular}{|c|c c c c c|}
		\hline
		&$\overline{C}_L$ & $C^{rms}_L$ & $\overline{C}_D$ & $C^{rms}_D$ & $St$\\
		\hline
		Current & $0.950$ & $0.324$ & $0.072$ & $0.046$ & $0.15$\\
		\hline
		Num.-4\citep{chen2019LES} & $0.953$ & $0.311$ & $0.074$ & $0.041$ & $0.168$\\
		\hline
	\end{tabular}
	\caption{Summary of the hydrodynamic force coefficients, and the corresponding  Strouhal number observed during LES of cavitating flow.}
	\label{tab:CLCD_Rigid}
\end{table}

\begin{figure}[!ht]
\centering
\includegraphics[width=0.7\textwidth]{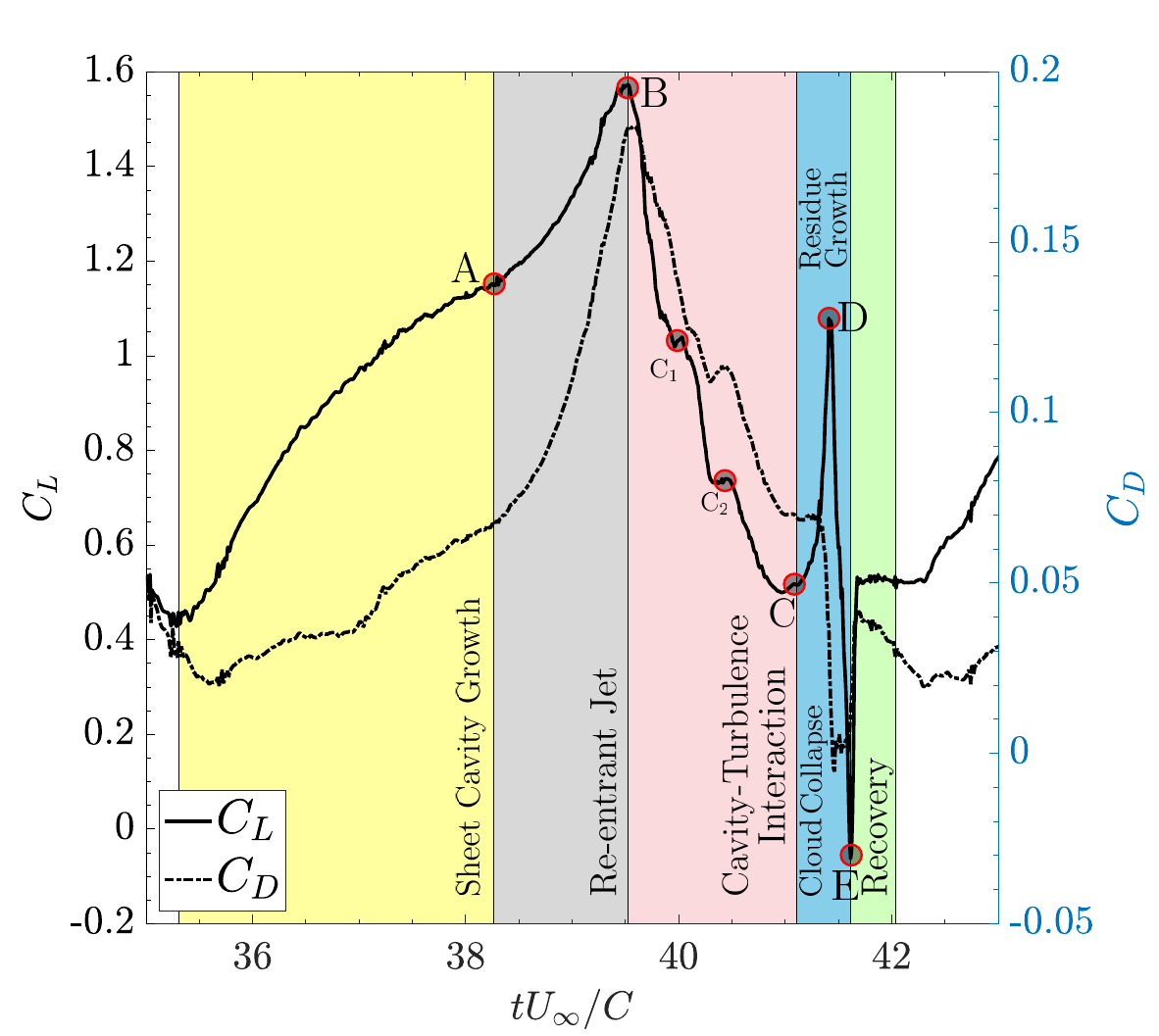}
\caption{Variation in $C_L$ and $C_D$ over one cycle of cavitation illustrating the duration of sheet-cavity growth, cloud-cavity separation and collapse, residual cavity growth and collapse. The points of change in flow behavior are labeled as (A) Instantiation of the re-entrant jet, (B) Separation of cloud cavity,  (C) Shedding of Trailing Edge Vortex Cavities. $\mathrm{C}_1$ and $\mathrm{C}_2$ are vortex cavities that are shed from the trailing edge and contribute to cloud cavity collapse (D) Complete collapse of cloud cavity (E) Collapse of residual sheet cavity. }
\label{fig:CLCDCycle}
\end{figure}
\noindent Further, in Fig. \ref{fig:CompVcCLCDRigid} we compare the time-series of cavity volume, the lift and drag coefficients over two cavitation cycles with the LES data of \citet{chen2019LES}.The mean and rms lift $(\overline{C}_L,C^{rms}_L)$ and drag $(\overline{C}_D,C^{rms}_D)$ coefficients are summarized along with the Strouhal number $(St)$ in table \ref{tab:CLCD_Rigid}. We observe the current study shows good agreement with the observed lift and drag coefficients in previous numerical studies \citep{chen2019LES}. The variation of lift and drag coefficients over a cavitation cycle is presented in Fig.\ref{fig:CLCDCycle}. Assuming the cavitation cycle to begin at the non-dimensional time $t_0^{\mathrm{cav}}$, and the cavity cycle's period to be $T^\mathrm{cav}$, we define a fractional cavitation cycle time as \begin{equation}
t_f^\mathrm{cav} = \frac{t^* - t_0^\mathrm{cav}}{T^\mathrm{cav}}.
\end{equation} Correspondingly, for $t_f^\mathrm{cav} \in [0, 0.62]$, we observe the sheet cavity's growth. As the cavity grows, an increasing fraction of the suction side experiences pressures close to the vapor pressure. Furthermore, the sheet cavity poses an obstruction to the incoming liquid phase, resulting in a re-direction of the liquid over the cavity. The combination of these effects leads to the hydrofoil experiencing higher lift and drag respectively.
\begin{table}[!h]
\centering
\begin{tabular}{|c|c c c c|}
\hline
$t_f^\mathrm{cav} = \frac{t^* - t_0^\mathrm{cav}}{T^\mathrm{cav}}$ & $l^\mathrm{cav}/C$ & $x^\mathrm{cav}/C$ & $100\delta_j/C$ & $V_j/U_\infty$ \\
\hline
$0.16$ & $0.021$ & $0.021$ & $-$ & $-$\\
$0.21$ & $0.061$ & $0.061$ & $-$ & $-$\\
$0.26$ & $0.080$ & $0.080$ & $-$ & $-$\\
$0.32$ & $0.240$ & $0.219$ & $-$ & $-$\\
$0.37$ & $0.421$ & $0.372$ & $0.122$ & $0.543$\\
$0.42$ & $0.584$ & $0.489$ & $0.307$ & $0.489$\\
$0.48$ & $0.715$ & $0.566$ & $0.426$ & $0.328$\\
$0.53$ & $0.820$ & $0.573$ & $0.576$ & $0.315$\\
$0.58$ & $0.909$ & $0.581$ & $0.933$ & $0.267$\\
\hline
\end{tabular}
\caption{Evolution of the sheet cavity and re-entrant jet over a cavitation cycle. We measure the cavity length $l^\mathrm{cav}$ and location of cavity closure $x^\mathrm{cav}$ from the leading edge, whereas the re-entrant jet thickness $\delta_j$ is measured normal to the hydrofoil at the location of cavity closure. The mean re-entrant jet velocity $V_j$ is estimated based on the Strouhal number of the flow \citep{callenaere2001CavInstab,pham1999sheetcloud}. The vapor-liquid interface is constructed using $\phi = 0.585$, and the characteristics are evaluated at $z/S = 0.5$.}
\label{tab:Re-entrantJet}
\end{table}

\begin{figure}[!h]
	\centering
	\includegraphics[scale=0.65]{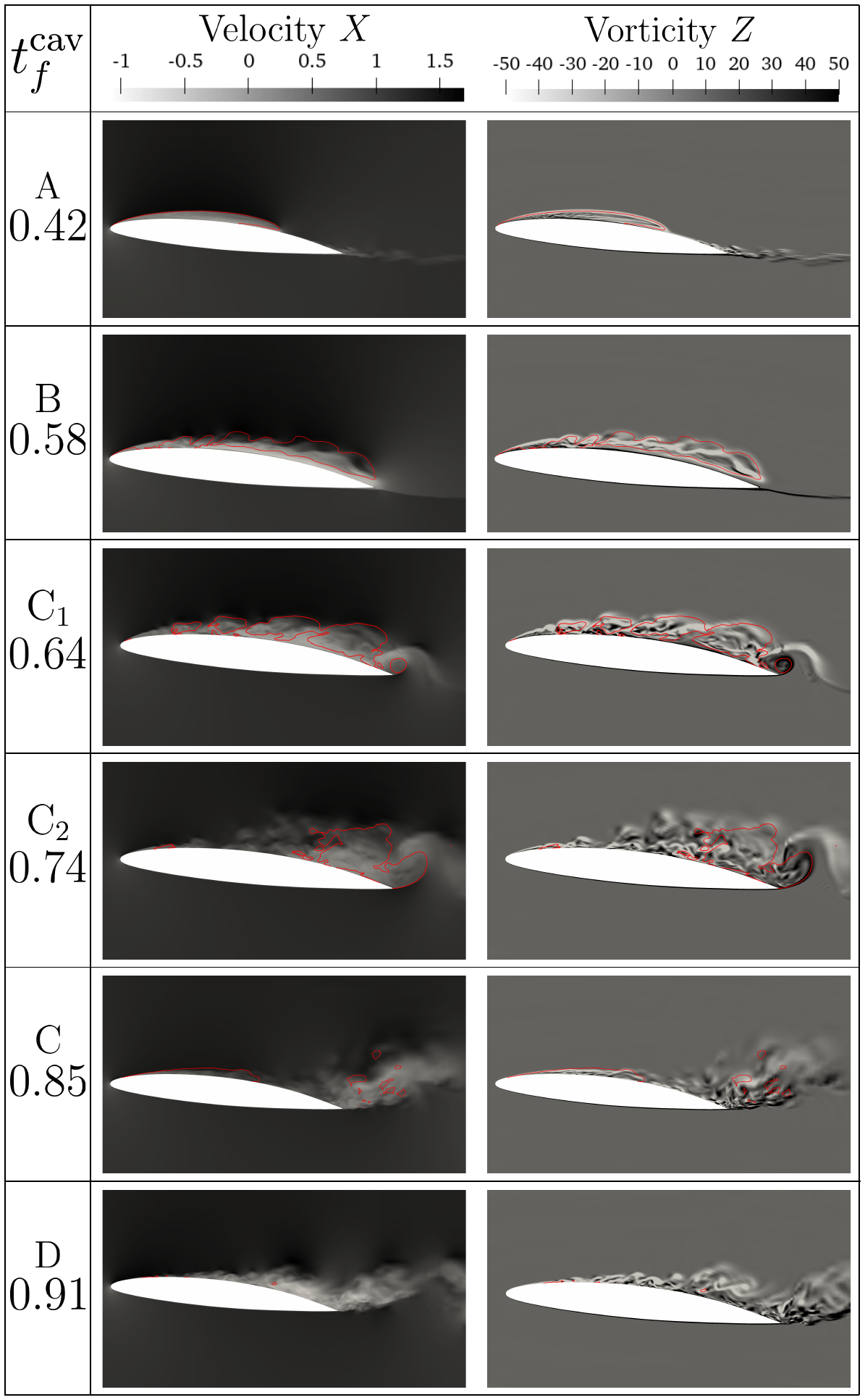}
	\caption{Evolution of the cavity cycle at locations highlighted in Fig.\ref{fig:CLCDCycle}. The non-dimensional x-velocity and z-vorticity $(\omega_z^* = \omega_z C/U_{\infty})$ components are presented at $Z/S = 0.5$. The red lines correspond to iso-contours of $\phi = 0.585$ to indicate the cavity structure(s) at different instances of the cycle.}
	\label{fig:cloudCav}
\end{figure}

Owing to the re-entrant jet instability, the sheet cavity experiences a breakdown, leading to the shedding of a cloud cavity at $t_f^\mathrm{cav} = 0.62$ and the formation of a residual cavity on the hydrofoil surface. The cavity shedding process results in the liquid phase impinging back on the hydrofoil surface, while simultaneously interacting with the cloud cavity. Furthermore, the obstruction effect imposed by the smaller residual cavity reduces, leading to a recovery in the drag force experienced by the hydrofoil. For $t_f^\mathrm{cav} \in [0.62,0.86]$, the cloud cavity convects downstream and leads to the emergence of a highly unsteady and turbulent flow field over the suction side of the hydrofoil. The cloud cavity collapses into smaller cavities, culminating with the formation of vortex cavities at the trailing edge of the hydrofoil. The unsteady flow field drives the vortex cavities downstream, and the residual cavity grows rapidly in the interval $t_f \in [0.86,0.94]$. The vortex cavities can no longer sustain themselves in the high-pressure region in the hydrofoil's wake, and experience a sudden collapse close to the instant $t_f \approx 0.91$, leading to an impulsive load on the hydrofoil. The pressure spike associated with this collapse drives the complete collapse of the residual sheet cavity over the hydrofoil surface, leading to an impulsive negative lift. For $t_f^\mathrm{cav} \in [0.94,1]$, the flow around the hydrofoil adjusts to create conditions conducive to the re-emergence of cavitation. The flow patterns observed over the course of a cavitation cycle are illustrated in Fig.\ref{fig:cloudCav}.

The observations presented here are consistent with the works of \citet{leroux2004experimental} and \citet{pham1999sheetcloud}, where the latter study indicated the emergence of interfacial perturbations over the cavity surface. \citet{chen2019LES} propose that the re-entrant jet does not possess adequate momentum to drive the cavity breakdown process and emphasize the role of turbulent fluctuations induced on the cavity surface in driving its breakdown. As highlighted by table \ref{tab:Re-entrantJet}, the mean velocity of the re-entrant jet $V_j$ decreases as the jet reaches closer to the leading edge. However, we observe an increase in the re-entrant jet thickness, indicating the continuous entrainment of the liquid phase into the re-entrant jet \citep{callenaere2001CavInstab}. While we observe the emergence and growth of instabilities on the cavity surface, the role of these fluctuations in influencing the cavity shedding process cannot be conclusively demarcated owing to the absence of a sharp vapor-liquid interface.

\begin{figure}[!h]
	\centering
	\includegraphics[scale=0.095,trim={5cm 0 0 1cm},clip]{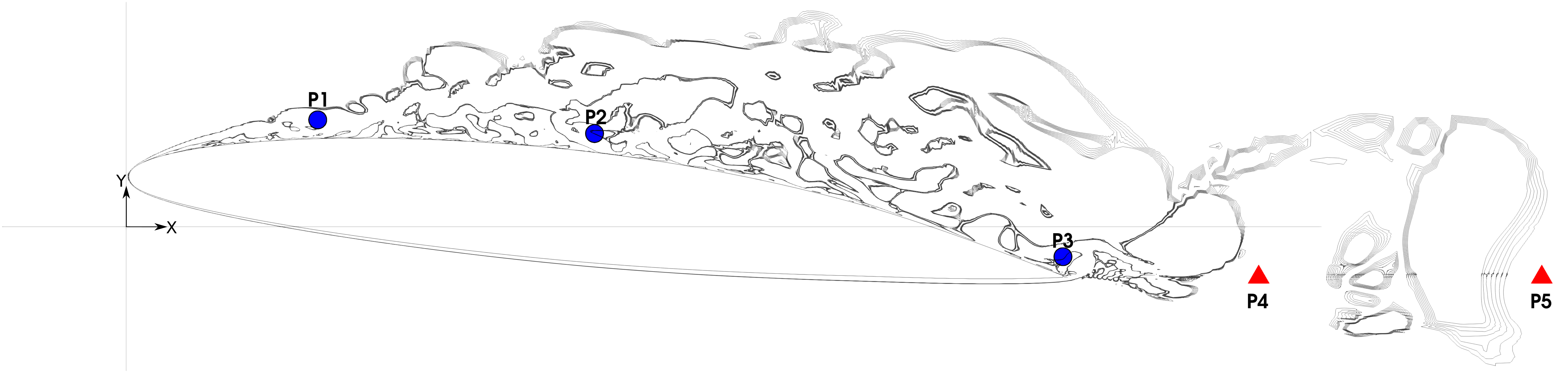}
	\caption{Arrangement of probes to assess turbulent flow characteristics: Probes P1-P3 are placed at a distance of $0.02C$ from the hydrofoil, at $x/C \in \left\{0.2, 0.5, 1.0\right\}$ respectively. Probes P4-P5 are placed downstream of the trailing edge at $x/C \in \{1.25, 1.5\}$. The origin is located at the leading edge of the hydrofoil, with the angle of attack created by rotating the hydrofoil about mid-chord point. Contour lines of Z-vorticity $\omega_z^* = \omega_z C/U_{\infty} \in \left[-5,-2\right]\cup\left[2,5\right]$ are presented to indicate unsteady flow-structures in the vicinity of the hydrofoil.}
	\label{fig:TurbPts}	
\end{figure}

\subsection{Turbulence-Cavity interactions}\label{sec:tCint}
\begin{figure}[!h]
	\begin{subfigure}[h]{0.5\textwidth}
		\centering
		\includegraphics[width=\textwidth,trim={0 4.5cm 0 0},clip]{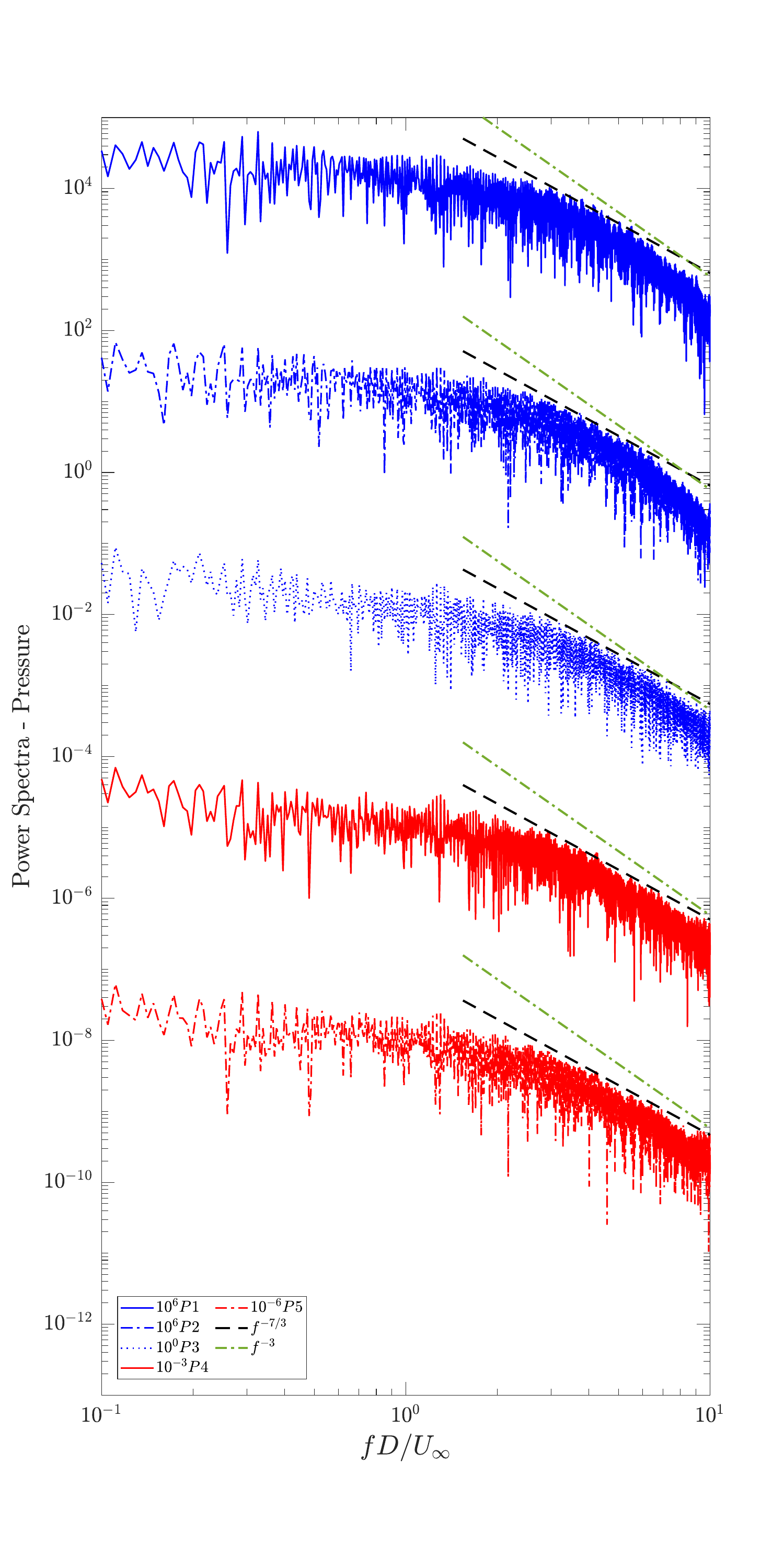}
		\caption{}
		\label{fig:freqPres}
	\end{subfigure}
	\begin{subfigure}[h]{0.5\textwidth}
		\centering		
		\includegraphics[width=\textwidth,trim={0 4.5cm 0 0},clip]{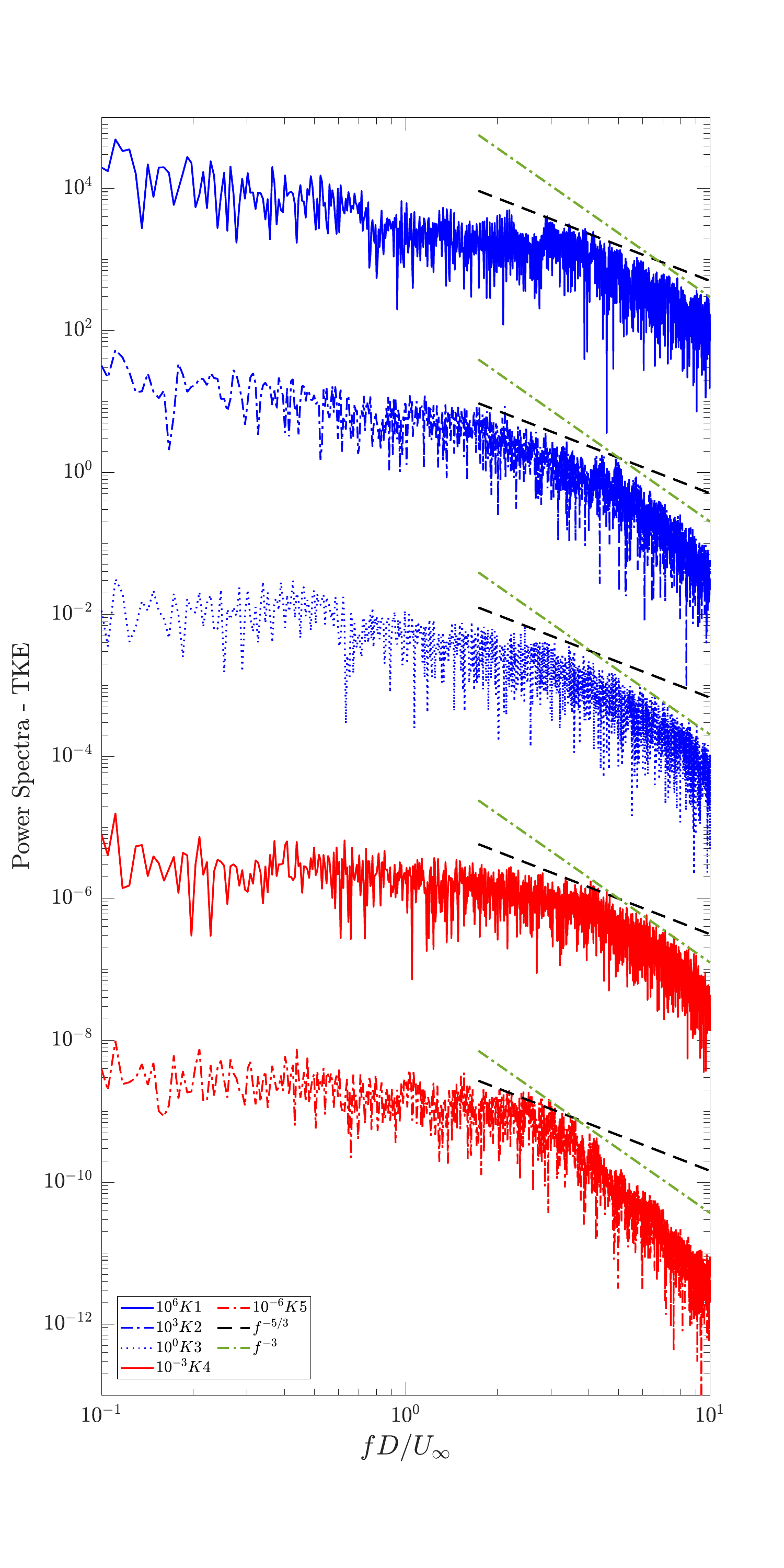}
		\caption{}
		\label{fig:freqTurb}
	\end{subfigure}
	\caption{ FFT Spectra of (a) pressure fluctuations and (b)turbulent kinetic energy at locations (see Fig. \ref{fig:TurbPts}) in the cavitating region and near wake of the hydrofoil. For sake of clarity, spectra of points from the same group are shifted, and the magnitude of shift is indicated in the legend.}
	\label{fig:frqSpectra}
\end{figure}

\noindent In this section, we discuss the evolution of turbulent structures as a consequence of cavitation over the suction surface around the hydrofoil. We initially present the turbulence power spectra in the cavity developing region, and near wake of the hydrofoil, with the probes of assessment shown in Fig.\ref{fig:TurbPts} to establish the adequacy of mesh M2 in capturing the turbulent features of the flow. Subsequently, we briefly discuss the cavity topology and vorticity dynamics developing around the hydrofoil in the interval $t_f^\mathrm{cav} \in [0.62,0.86]$.

\subsubsection{Energy spectra surrounding the hydrofoil}\label{sec:rigTurbSpect}
In this section, we assess the turbulence power spectra at the points illustrated in Fig. \ref{fig:TurbPts} to verify the mesh resolution offered by mesh M2. For locally homogeneous turbulence, Kolmogorov's $-5/3$ power law relates the turbulent kinetic energy $(E \equiv u_i'u_i'/2)$ to the wave number of integral scale ($\kappa$) eddies as $E(\kappa) \propto \kappa^{-5/3}$ in the inertial range. \citet{taylor1938Turb} in their frozen turbulence hypothesis indicated $-5/3$ decay of the kinetic energy versus frequency is a reasonable indicator of the turbulence spectrum. While anisotropic and highly unsteady mono-phase turbulent flows show good agreement with Kolmogorov's turbulent kinetic energy decay law, multi-phase flows have been observed to deviate significantly. \citet{lance1991TurbLiqBubbly} studied the turbulence exhibited by the liquid phase in a turbulent bubbly flow field and observed the high-frequency decay to observe a $-8/3$ power law. Further works investigating the turbulence decay characteristics of spectra \citep{prakash2016TurbBubly,martinez2010TurbBubly} have depicted agreement with the deviation of turbulence in multi-phase flows from the $-5/3$ power law.

In the current study, we expect the integral scale eddies to lie in the frequency band of $0.1\leq fC/U_{\infty}\leq 10$. In this range, Fig. \ref{fig:freqTurb} indicates the spectrum of turbulent kinetic energy decay to be approximately parallel to a $-3$ slope in the high-frequency region, and parallel to a $-5/3$ slope in the mid-range frequency region. Furthermore, we observe a variation in the decay spectra with the spatial location at which turbulent characteristics are being monitored. The non-local effects of phase transitions and their contributions to turbulence are observed in the form of TKE spectra, where the decay is observed to deviate beyond the $-3$ slope. From these observations, and their validation with the studies of turbulence spectra in multi-phase flows, we can confirm the current SGS turbulence model is adequate for capturing cavitating flows at high Reynolds numbers. 

\begin{figure}[!h]
	\centering
	\includegraphics[width=\textwidth]{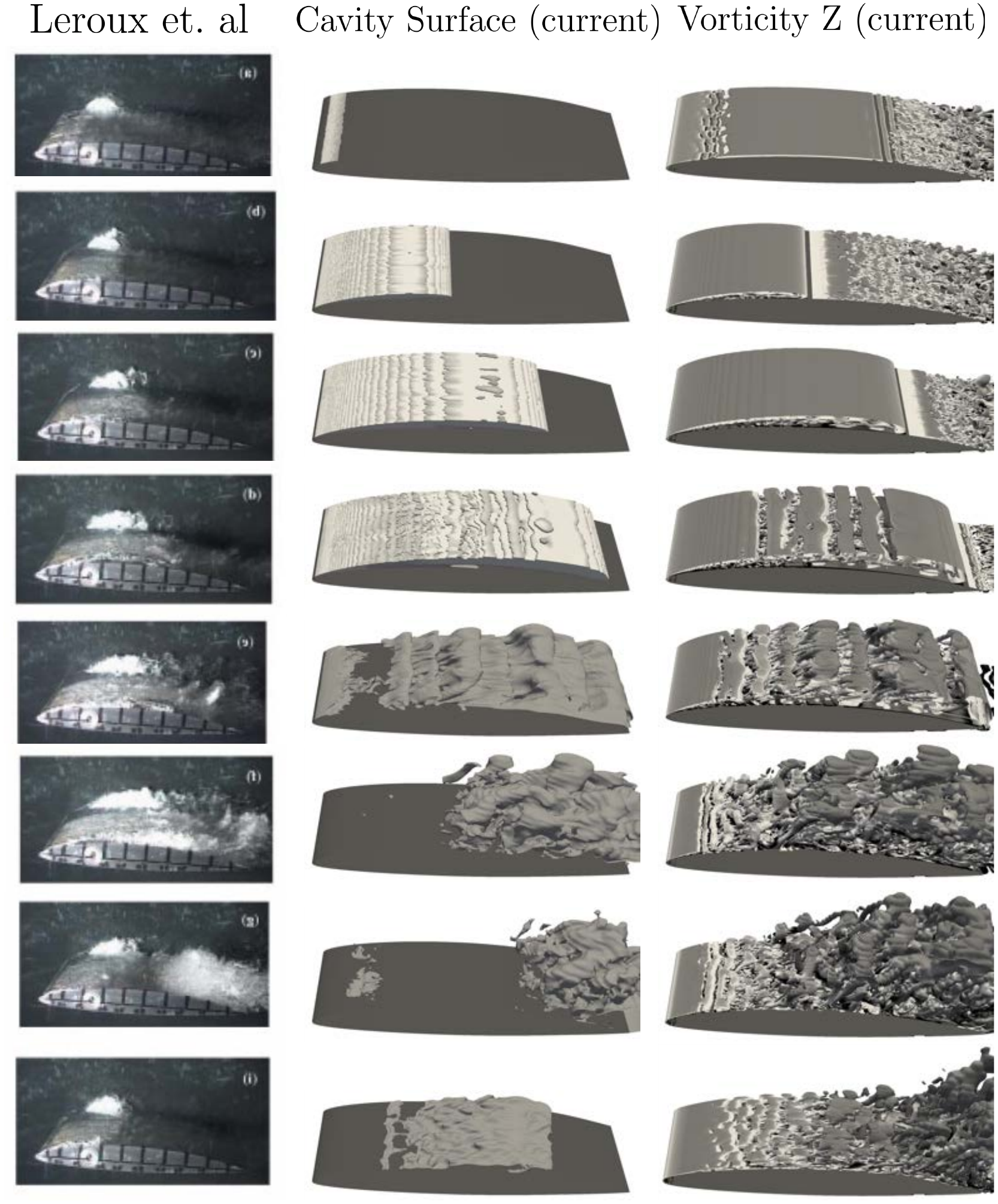}
	\caption{Evolution of the cavity and vorticity over a single cycle. The first column illustrates the experimental results obtained by \citet{leroux2004experimental}. The second column depicts the cavity developed in the current numerical framework. The third column shows the  contours of Q-criterion ($Q^*= 8$) colored by $Z$-vorticity.}
	\label{fig:CavityVortCycle}
\end{figure}

\subsubsection{Cloud Cavity-Vorticity Interaction over Rigid Hydrofoil} \label{sec:cloudcavVort}
The evolution of the cloud cavity interplays closely with the onset and growth of turbulent structures on the cavity surface. As illustrated in sec.\ref{sec:Hydrodynamic}, turbulent fluctuations evolve over the sheet cavity and continue to be present over the cloud cavity surface. Furthermore, the streamlined shape of the hydrofoil ensures a high-pressure region is recovered in its immediate wake. In conjunction with the unsteady flow patterns, the wake pressure drives the collapse of the cloud cavity, leading to the formation of smaller vapor pockets and vortex cavities. This interaction primarily occurs during $t_f^\mathrm{cav} \in [0.62,0.86]$.

Upon separation of the cloud cavity at $t_f^\mathrm{cav} = 0.62$, the low inertia and irregular shape of the cavity induce velocity fluctuations in the liquid phase. Consequently, turbulent vortical structures form in the cloud cavity's wake, while the flow upstream of the cloud cavity is highly perturbed. The suction-side vortex structures close to the trailing edge create a high-pressure differential across the hydrofoil, forcing a vortex roll-up at the trailing edge with some vaporization of the liquid. The shedding of this vortex cavity has been labeled as $\mathrm{C}_1$ in Fig.\ref{fig:CLCDCycle} and the corresponding velocity and vorticity fields around the hydrofoil are illustrated in Fig.\ref{fig:cloudCav}.

The breakdown of $\mathrm{C}_1$ leads to the emergence of smaller vortices in the wake, while the larger cloud cavity continues to progress downstream. Within this period, cavity growth close to the leading edge subsides completely, with flow up to $x/C = 0.5$ being highly unsteady and populated with multiple small-scale vortices \citep{franc1988oschydrofoil}. Once the cloud cavity encounters the trailing edge, the pressure differential across the hydrofoil forces the roll-up of a second vortex around a fragment of the cloud cavity. This event is labeled $\mathrm{C}_2$ in Fig.\ref{fig:CLCDCycle}, and is depicted in Fig.\ref{fig:cloudCav}. A consequence of this fragmentation is the formation of multiple vapor clouds of smaller sizes. This breakdown also allows for the oncoming flow to stabilize again, leading to the formation of a residual sheet cavity on the hydrofoil surface.

The collapse of these vapor clouds at $t_f^\mathrm{cav} = 0.86$ leads to the growth of highly unsteady fluctuations upstream of the hydrofoil. This results in the collapse of the residual sheet cavity, resulting in the impulsive loads experienced by the hydrofoil. Further investigation of $\mathrm{C}_1$, $\mathrm{C}_2$ and $\mathrm{C}$ events in the cavitation cycle are necessary to elucidate the key cloud collapse characteristics. The current framework lends itself to such a study, owing to the capture of unsteady features present in the liquid phase during the cloud-cavity breakdown process. To conclude, we present a comparison of the cavitation structures observed during a cavitation cycle between the current study and the work of \citet{leroux2004experimental} in Fig.\ref{fig:CavityVortCycle}.

\section{LES of Cavitating Flow past Cantilevered Flexible Hydrofoil} \label{sec:flexhydrofoil}
\noindent Using the current framework, we next study a cantilevered flexible NACA66 hydrofoil's response characteristics under and cavitating conditions. The flow is allowed to develop at $Re =  7.5\times 10^5$, $\sigma = 1.4$ and $\alpha_0 = 8^\circ$, where $\alpha_0$ denotes the hydrofoil's static angle of attack. The flexibility of the hydrofoil is characterized by Young's modulus $E^\mathrm{s} = 3$ GPa, Poisson's ratio $\nu^\mathrm{s} = 0.35$ and density $\rho^\mathrm{s} = 1480~\mathrm{kg}/\mathrm{m^3}$. The two key non-dimensional parameters of the fluid-structure coupling are
\begin{align}
	m^* = \frac{\rho^\mathrm{s}}{\rho^\mathrm{l}} = 1.48,\nonumber\\
	U^* = \frac{U_\infty}{f_1C} = 0.678,
\end{align} where $m^*$ and $U^*$ denote the mass ratio and reduced velocity respectively.

Since the current study is conducted in an unsteady multiphase flow, the effects of added mass are time varying, owing to the phase of the fluid interacting with the hydrofoil. In this regard, the natural frequencies of the hydrofoil are derived for still-water based on the studies of \citet{benaouicha2012addedmass} and \citet{delatorre2013AddedMass}. We obtain the mass-normalized mode-shapes along with the modal frequencies. The dominant modal frequencies along with the effective modal mass ratio in the bending ($y$) and twisting ($\theta_z$) directions are summarized in table\ref{tab:naturalFrequencies}. In the current study, the clockwise direction is taken to be positive, with the pitching moment computed about the quarter-chord point. From the extracted modes, we observe that the first mode of response involves a dominant twisting component relative to the second mode. This indicates the presence of bend-twist coupling for the hydrofoil under consideration.
\begin{table}[!h]
	\centering
	\begin{tabular}{|c|c c c c|}
		\hline
		Mode $(n)$ & $f_n$ (Hz) & $m^n_y/m$ & $I^n_z/I_z$ & $St_n = f_nC/U_\infty$ \\
		\hline
		$1$ & $55$ & $5.98\times10^{-1}$ & $4.71\times10^{-1}$ & $1.48$\\
		$2$ & $177$ & $7.71\times10^{-6}$ & $1.55\times10^{-1}$ & $4.70$\\
		$3$ & $325$ & $1.90\times10^{-1}$ & $1.45\times10^{-1}$ & $8.62$\\
		$4$ & $353$ & $1.15\times10^{-2}$ & $2.01\times10^{-2}$ & $9.36$\\
		$5$ & $558$ & $1.29\times10^{-4}$ & $1.76\times10^{-2}$ & $14.8$\\
		$6$ & $802$ & $5.12\times10^{-2}$ & $4.44\times10^{-2}$ & $21.3$\\
		$7$ & $921$ & $5.96\times10^{-3}$ & $7.91\times10^{-3}$ & $24.4$\\	
		$8$ & $1022$ & $3.88\times10^{-3}$ & $1.34\times10^{-3}$ & $27.1$\\	
		\hline	
	\end{tabular}
	\caption{ Modal frequencies and effective modal mass ratios obtained from the solution to the eigenmode problem corresponding to the linear elastic structural model. $m^n_y$ denotes the effective modal mass along the bending direction, and $I^n_z$ denotes the effective mass-moment of Inertia in the twisting direction. $(m,I_z)$ denote the mass and mass-moment of inertia along the $z$-axis respectively. For the current study $m = 1.56\mathrm{kg}$ and $I_z = 0.0093\mathrm{kg}/\mathrm{m}^2$. $St_n$ denotes the non-dimensional structural frequency.}
	\label{tab:naturalFrequencies}
\end{table}
Furthermore, as highlighted in the studies of \citet{benaouicha2012addedmass}, the hydrofoil in a cavitating flow experiences both positive and negative damping effects, which can lead to instabilities in the dynamical system. In order to allow fluid flow instabilities to trigger structural instabilities, we do not model structural damping in the hydrofoil system.

\subsection{Mesh Characteristics}
\begin{figure}[!h]	
	\centering
	\includegraphics[width=1\textwidth]{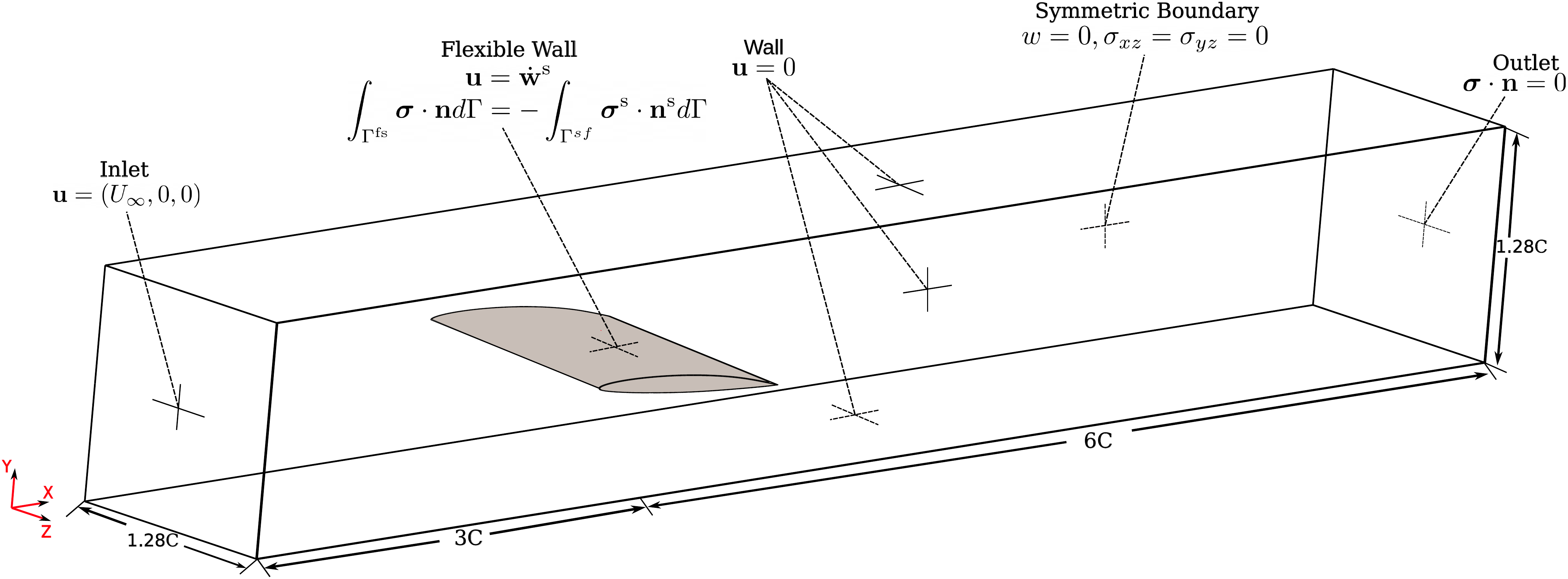}      \caption{A schematic of the flow past flexible cantilevered NACA66 hydrofoil at $Re = 750000$. The computational setup and boundary conditions are shown for the filtered Navier-Stokes equations. }
	\label{Domain_0}
\end{figure}

\begin{figure}[!h]
	\centering
	\includegraphics[width=0.7\textwidth]{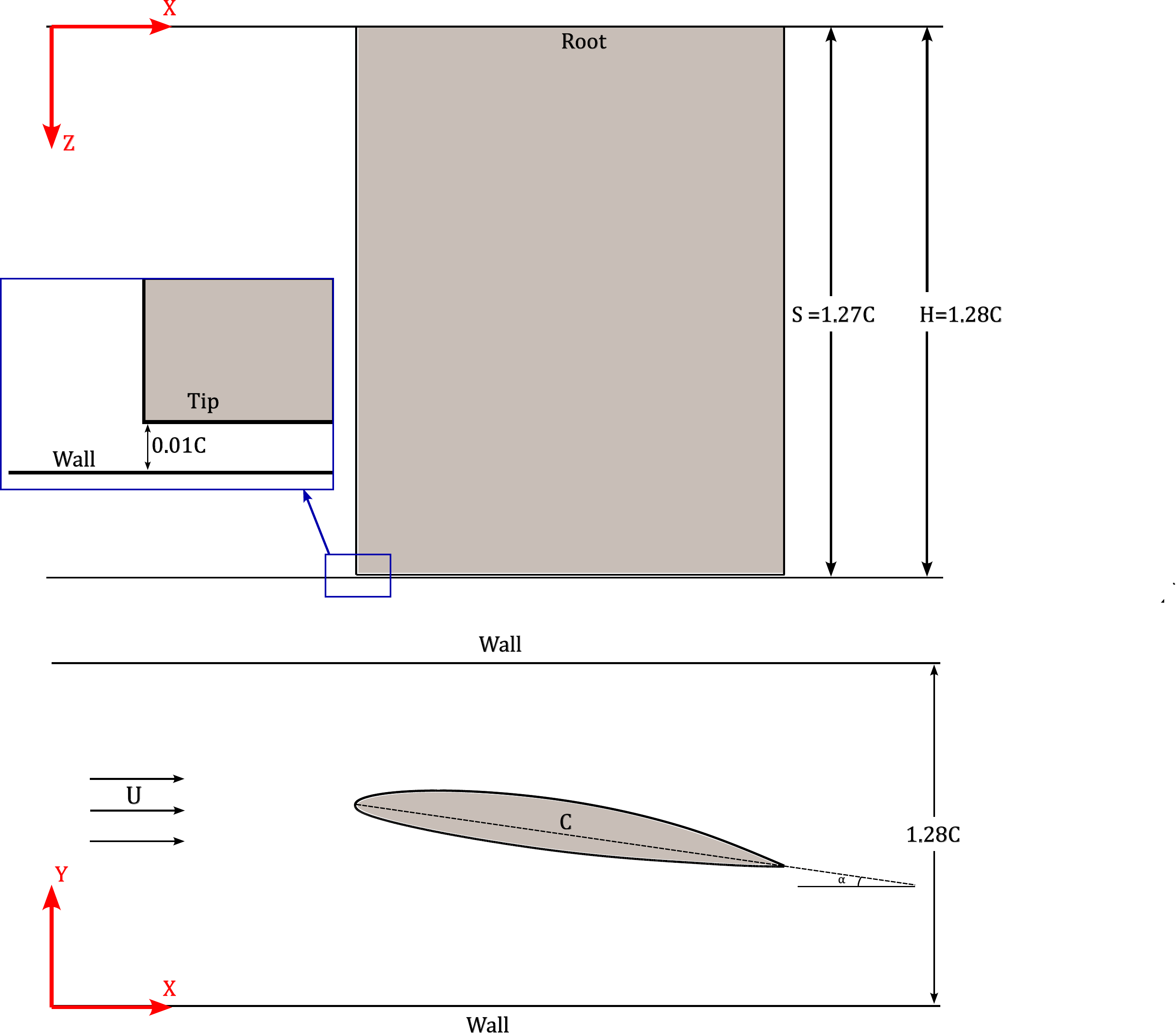}      \caption{Top and side views of the flexible hydrofoil, illustrating the angle of attack and the gap between the hydrofoil tip and front wall.}
	\label{Domain_1}
\end{figure}

The domain for studying the flow past the cantilevered hydrofoil is presented in Figs. \ref{Domain_0} and \ref{Domain_1}. The inlet and outlet boundaries are positioned $3C$ and $6C$ upstream and downstream of the leading edge (LE) respectively. The channel's width is set to $1.28C$, with side walls equidistant from the half-chord point of the hydrofoil, with no-slip boundary conditions. The span of the cantilevered hydrofoil $S = 1.27C$, with symmetric boundary conditions imposed at the hydrofoil root. At the hydrofoil tip, the domain is extended to give the channel width of $1.28C$ along the spanwise ($Z$) direction, to give a gap-to-span ratio of $0.00523$. A no-slip boundary condition is imposed at the tip wall of the channel. A uniform flow field is imposed at the inlet and is set as the initial condition for the non-cavitating flow. For conducting LES of the cantilevered hydrofoil, we consider the meshing requirements established by $M2$ in section \ref{sec:rigidmesh}. Owing to the large aspect ratio of the flexible hydrofoil, a uniform mesh-sizing of $\Delta z+ = 60$, requires $495$ mesh layers in the span-wise direction, leading to a mesh with $32M$ nodes. In order to manage the computational expense, we decrease mesh-sizing along the hydrofoil span, maintaining a ratio of $r_z = 0.988$, from root to tip, leading to a mesh with $12.12M$ nodes. The mesh information is summarized in table \ref{tab:MeshParametersFlex}. A uniform time-step of $\Delta t^* = \Delta tU_\infty/C = 9.434\times10^{-3}$ is employed for the rigid hydrofoil study.
\begin{table}[h]	
	\centering
	\begin{tabular}{|c c c|}
		\hline
		Property & M21-Flex & M22-Flex \\
		\hline
		$y^+$ & $0.3$ & $0.3$ \\
		$\left(\Delta x^+,\Delta z^+\right)$ & $(65,120)$ & $(48,60)$ \\
		$N_{chord}$ & $200$ & $280$ \\
		$\left(N_{BL}, r_{BL}\right)$ & $(60,1.15)$ & $(60,1.15)$ \\
		$\left(N_{span}, r_{span}\right)$ & $(240, 0.988)$ & $(495, 1.0)$\\
		$N^{nodes}$ & $12.12 M$ & $32.29M$ \\
		$N^{elem.}$ & $15.81 M$ & $42.33M$ \\
		\hline
	\end{tabular}
	\caption{Mesh parameters for the LES validation of flow past flexible cantilevered NACA66 hydrofoil. A mixed mesh is employed, with 8 node tri-linear hexahedral elements used in the regions of interest surrounding the hydrofoil and near-wake. 6-node prismatic elements are used in the far-field.}
	\label{tab:MeshParametersFlex}
\end{table}

\subsection{Fully-wetted response} \label{sec:fullywetted}

\begin{figure}[!h]
	\begin{subfigure}[h]{0.5\textwidth}
		\centering
		\includegraphics[width=\textwidth,trim={0 2.7cm 0 0},clip]{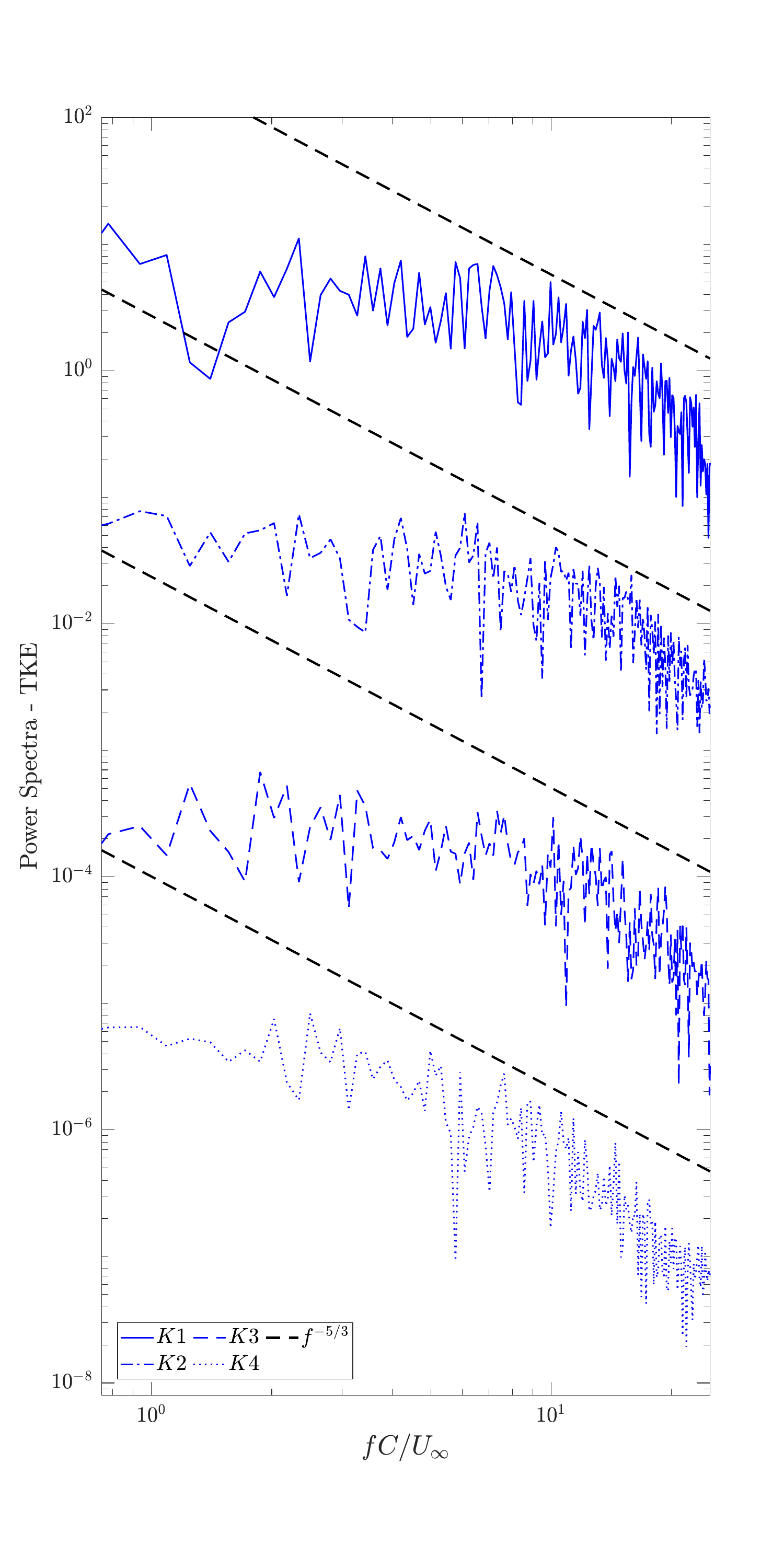}
		\caption{}
		\label{fig:frqK1flex}
	\end{subfigure}
	\begin{subfigure}[h]{0.5\textwidth}
		\centering
		\includegraphics[width=\textwidth,trim={0 2.7cm 0 0},clip]{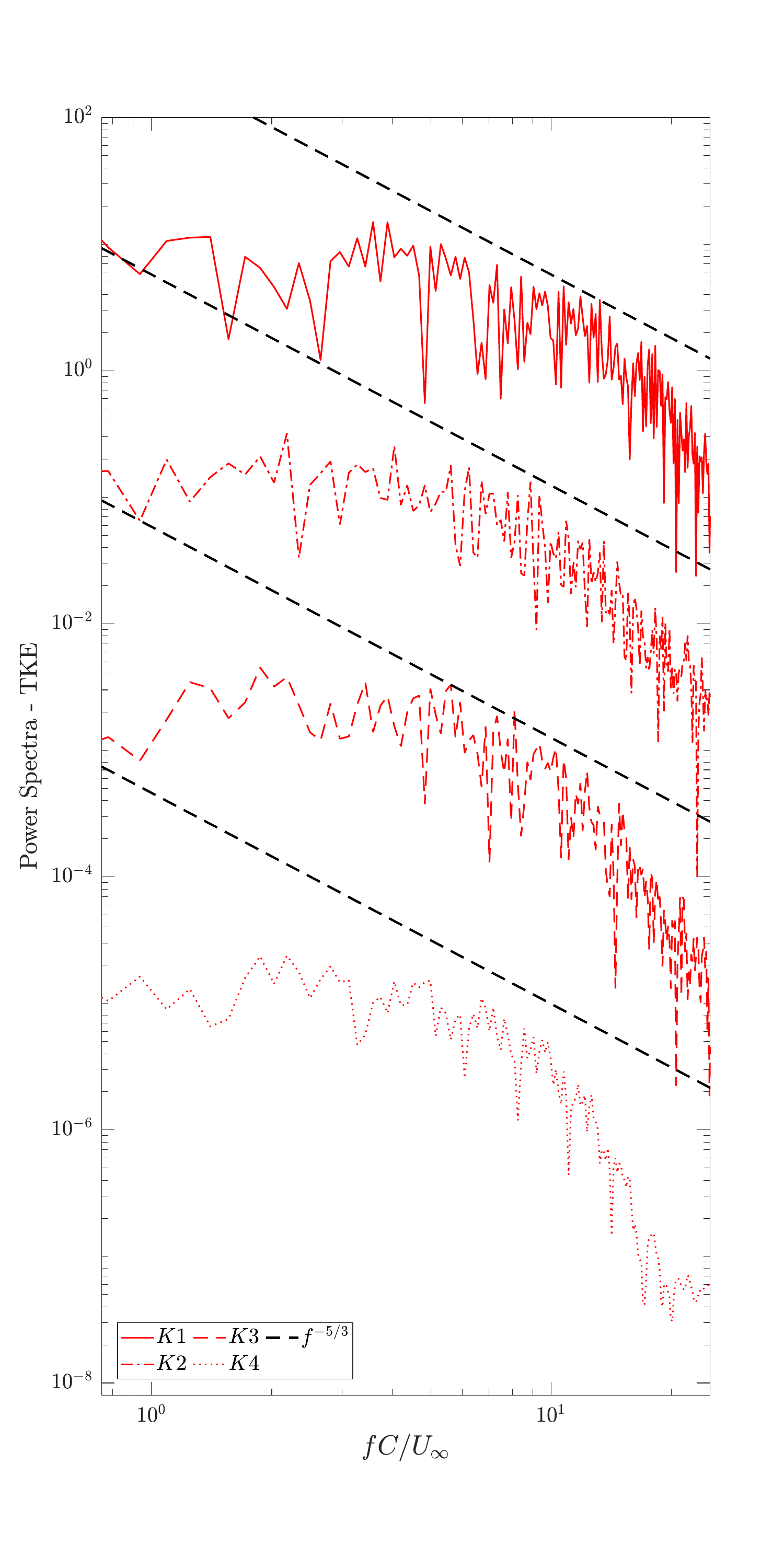}
		\caption{}
		\label{fig:frqK2flex}
	\end{subfigure}
	\caption{ FFT Spectra of (a) turbulent kinetic energy at the spanwise location $z/S = 0.5$ and (b)turbulent kinetic energy at the spanwise location $z/S = 1$ at the chordwise locations defined by Fig. \ref{fig:TurbPts} for the hydrofoil under fully-wetted conditions. For sake of clarity, spectra of points from the same group are shifted.}
	\label{fig:frqSpectraflex}
\end{figure}
In the current section, we illustrate some representative results for rigid and flexible hydrofoils in fully-wetted flow. Table \ref{tab:CLCDNonCav} presents the mean drag and lift coefficients of the rigid hydrofoil, along with comparisons against experimental data of \citet{leroux2004experimental}. Furthermore, we observe good agreement of the mean transverse displacement and tip-rotation of the hydrofoil with experimental data of \citet{ducoin2012expcav}. The performance of the dynamic SGS model is further evaluated through the wake characteristic data for the flexible hydrofoil as shown in Fig.\ref{fig:frqSpectraflex}. Based on the data presented in Table \ref{tab:CLCDNonCav}, the current LES simulations accurately predict the magnitude and spectral content of the forces, which in turn influence the fidelity of structural response.
\begin{table}[!h]
	\centering
	\begin{tabular}{|c| c c c c |c c c|}
		\hline	
		\multirow{2}{*}{Parameters} & 
		\multicolumn{4}{c|}{Rigid} & \multicolumn{3}{c|}{Flexible} \\
		& Exp.\citep{leroux2004experimental} & Num.-1 \citep{akcabay2014influence} & Num.-2\citep{harwood2014gapflow} & Current& Exp. \citep{ducoin2012expcav}& Num.-3\citep{akcabay2014influence} & Current\\
		\hline
		$C_L$ & $1.031$ & $1.19$ & $1.16$ & $1.122$ & $-$& $1.22$ & $1.112$ \\
		$C_D$ & $0.041$ & $0.022$ & $0.036$ & $0.044$ & $-$ & $0.022$ & $0.051$ \\
		$h_{max}/C$ & $-$ & $-$ & $-$ & $-$ & $0.024$& $0.01$ & $0.031$ \\
		$\theta_{max}~(deg)$ & $-$ & $-$ & $-$ & $-$ & $0.39$& $0.18$ & $0.27$ \\
		\hline
	\end{tabular}
	\caption{Comparison of hydrodynamic load coefficients and motion characteristics for fully-wetted conditions.}
	\label{tab:CLCDNonCav}
\end{table}

\begin{figure}[!h]
	\begin{subfigure}[h]{\textwidth}
		\centering
		\includegraphics[width=\textwidth,trim={0 3.8cm 0 3.8cm},clip]{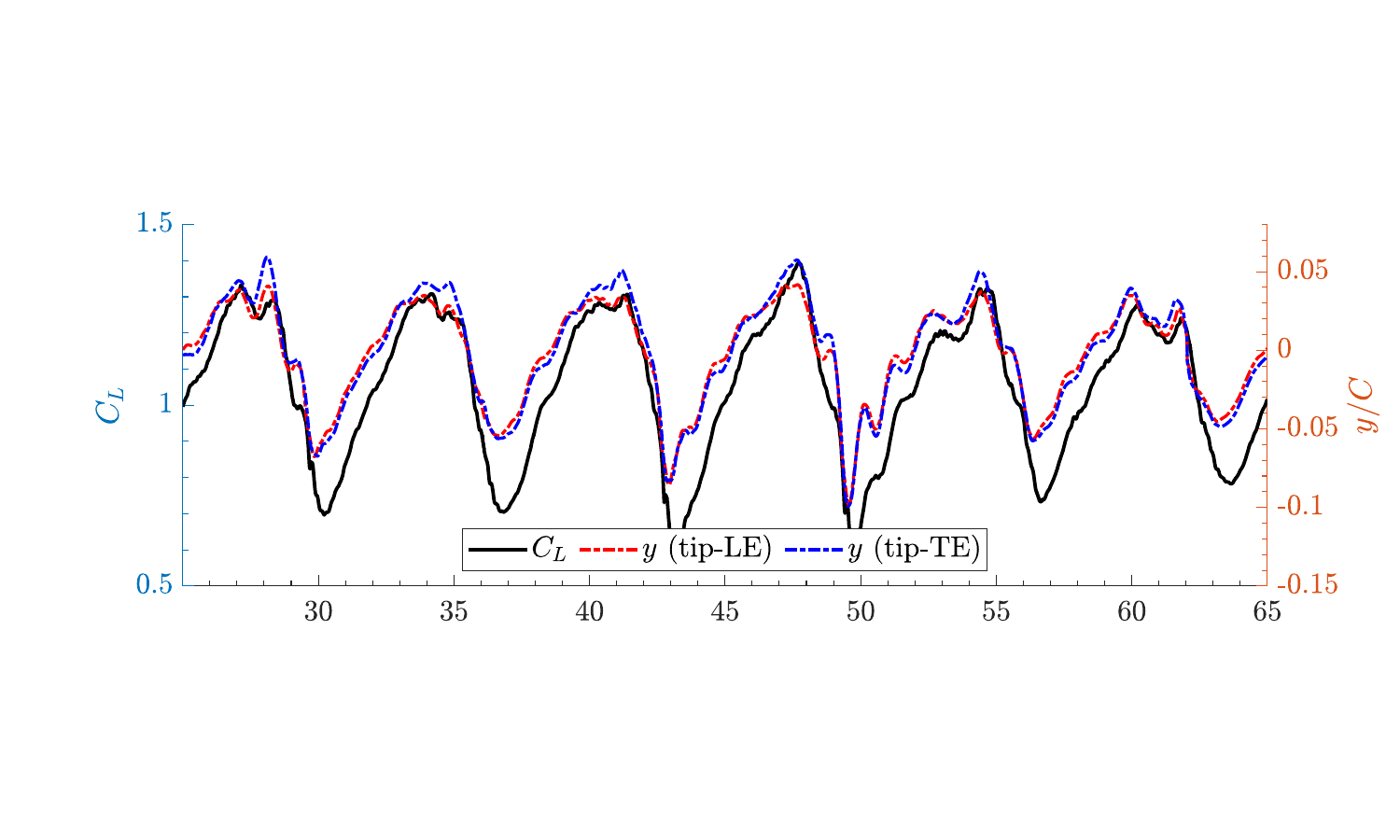}
		\caption{}
		\label{fig:yCL}
	\end{subfigure}
	\begin{subfigure}[h]{\textwidth}
		\centering
		\includegraphics[width=\textwidth,trim={0 4.0cm 0 4.8cm},clip]{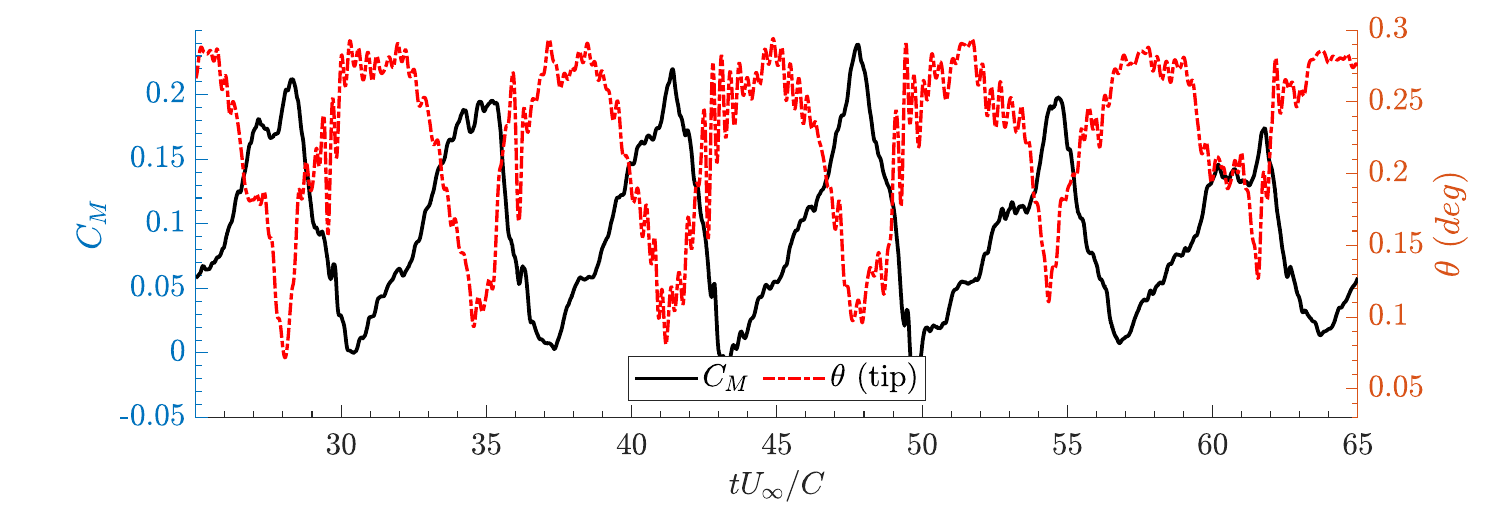}
		\caption{}
		\label{fig:thetaCM}
	\end{subfigure}
	\caption{ (a) Comparison of the bending displacement at the leading and trailing edges at the hydrofoil tip, with the Coefficient of Lift ($C_L$). (b) Comparison of the angular displacement at the hydrofoil tip, with the Coefficient of Moment ($C_M$) computed about the quarter-chord point of the static hydrofoil. }
	\label{fig:CLCM}
\end{figure}

\begin{figure}[!ht]
	\begin{subfigure}[h]{0.5\textwidth}
		\centering
		\includegraphics[width=\textwidth]{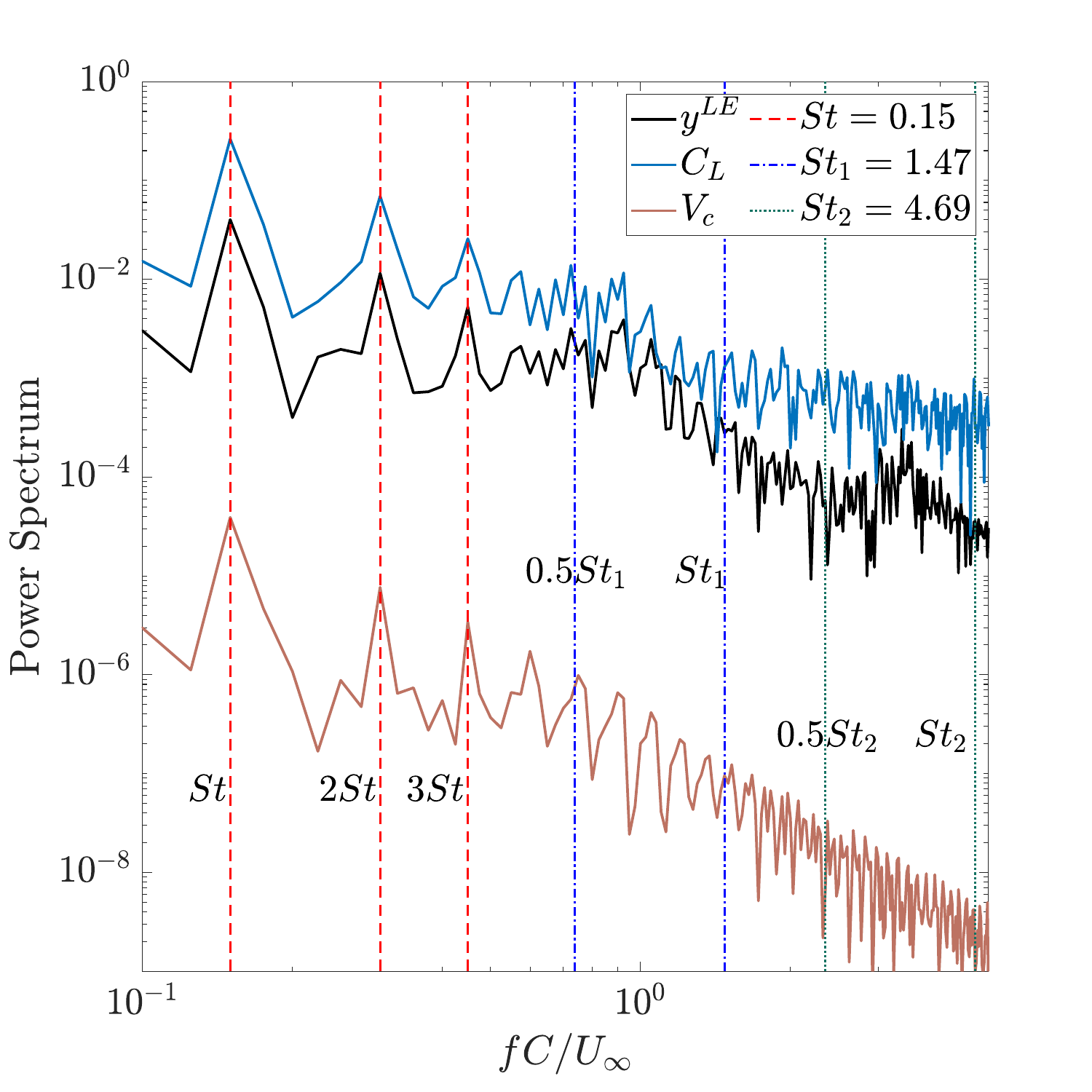}
		\caption{}
		\label{fig:yFFT}
	\end{subfigure}
	\begin{subfigure}[h]{0.5\textwidth}
		\centering
		\includegraphics[width=\textwidth]{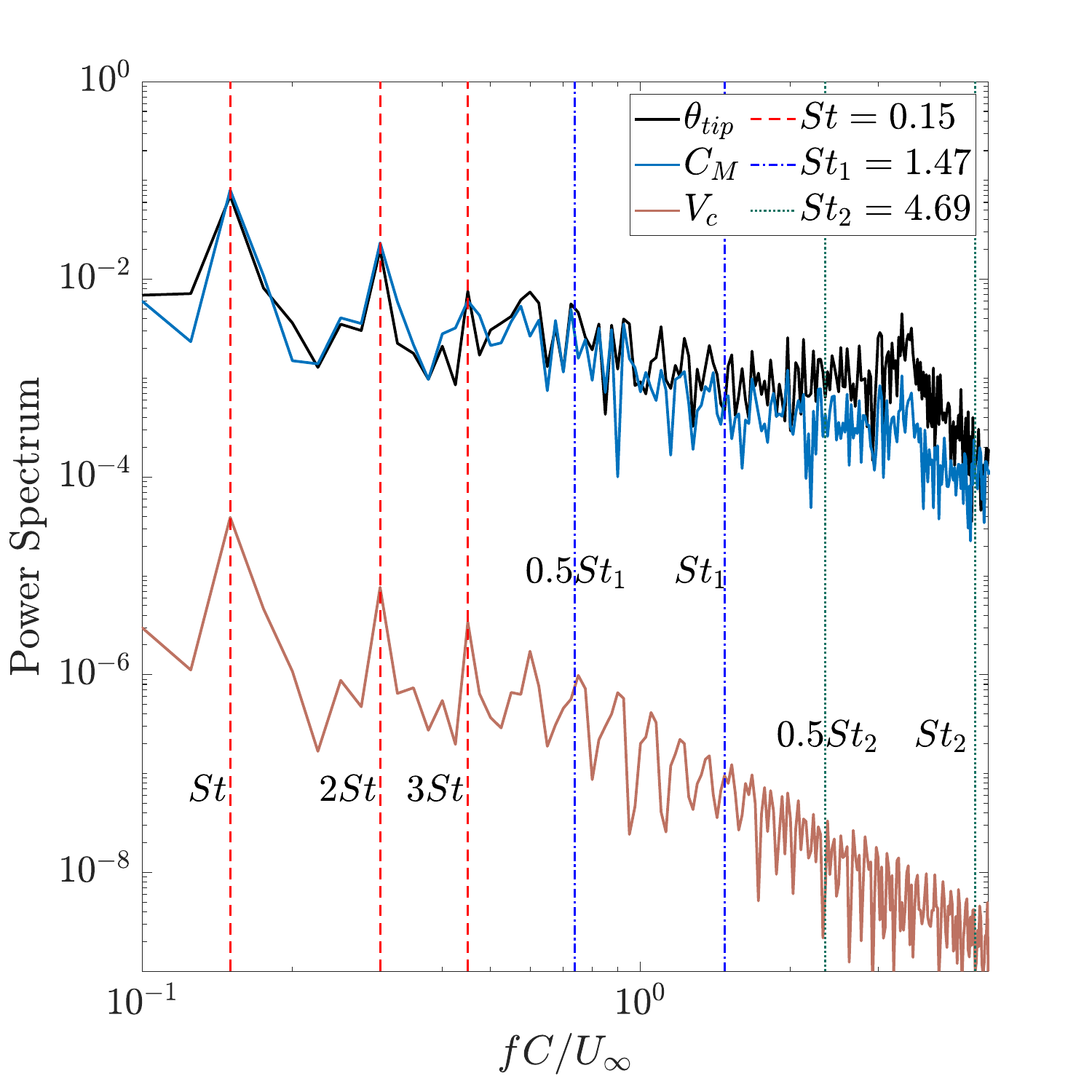}
		\caption{}
		\label{fig:thetaFFT}
	\end{subfigure}
	\caption{ FFT Spectra of tip (a) bending displacement ($y^{LE}$) and (b) angular displacement ($\theta_{tip}$) of the hydrofoil. While cavitation governs the structural response characteristics with a dominant frequency matching the Strouhal number of $0.15$, a broad frequency spectrum with a central peak corresponds to $f^*=3.35$. The central peak is not correlated with the structural frequencies or the Strouhal number.
 }
	\label{fig:frqSpectraDisp}
\end{figure}

\begin{figure}[!h]
	\centering
	\begin{subfigure}[h]{0.48\textwidth}
		\centering
		\includegraphics[width=\textwidth]{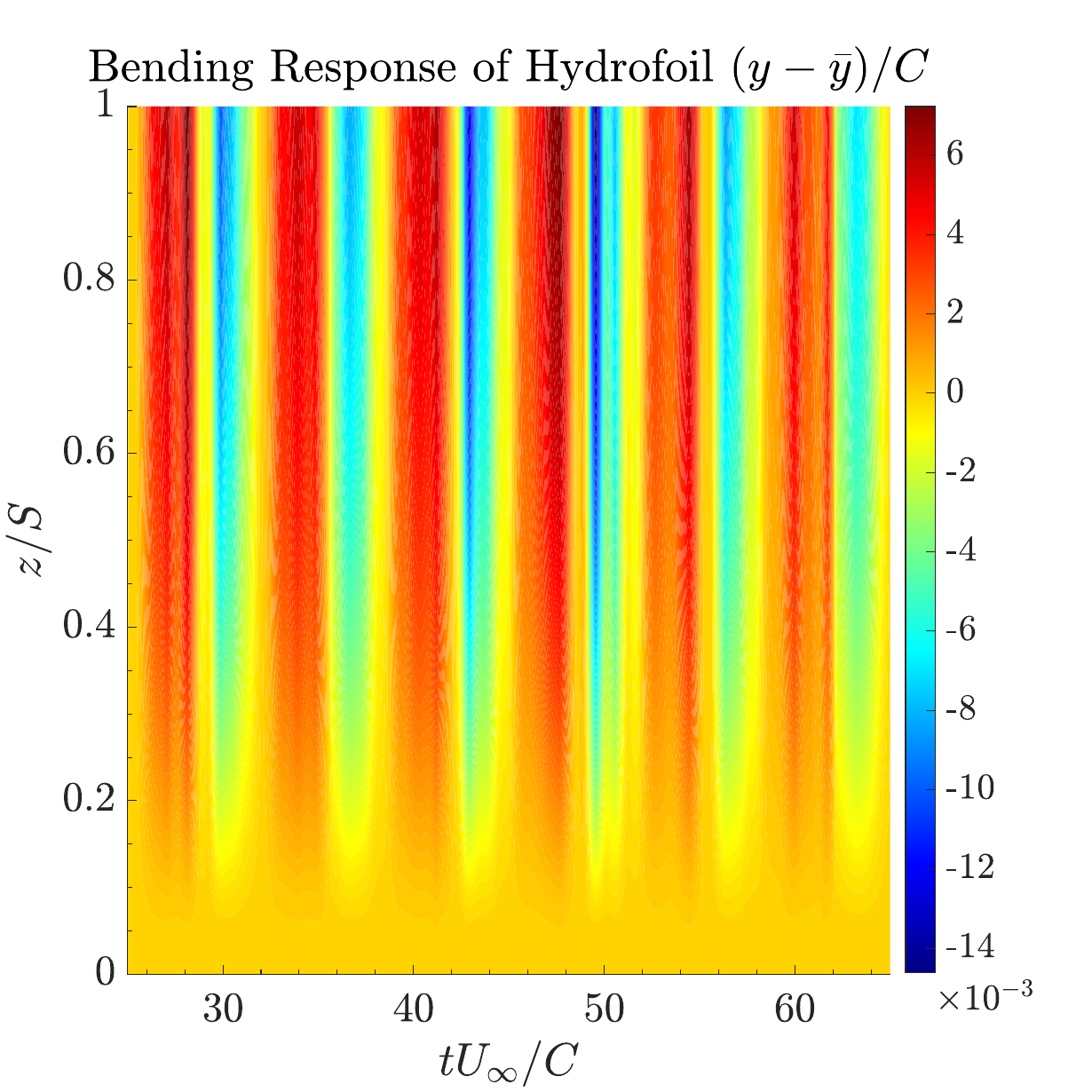}
		\caption{}
		\label{fig:bend}
	\end{subfigure}
	\begin{subfigure}[h]{0.48\textwidth}
		\centering
		\includegraphics[width=\textwidth]{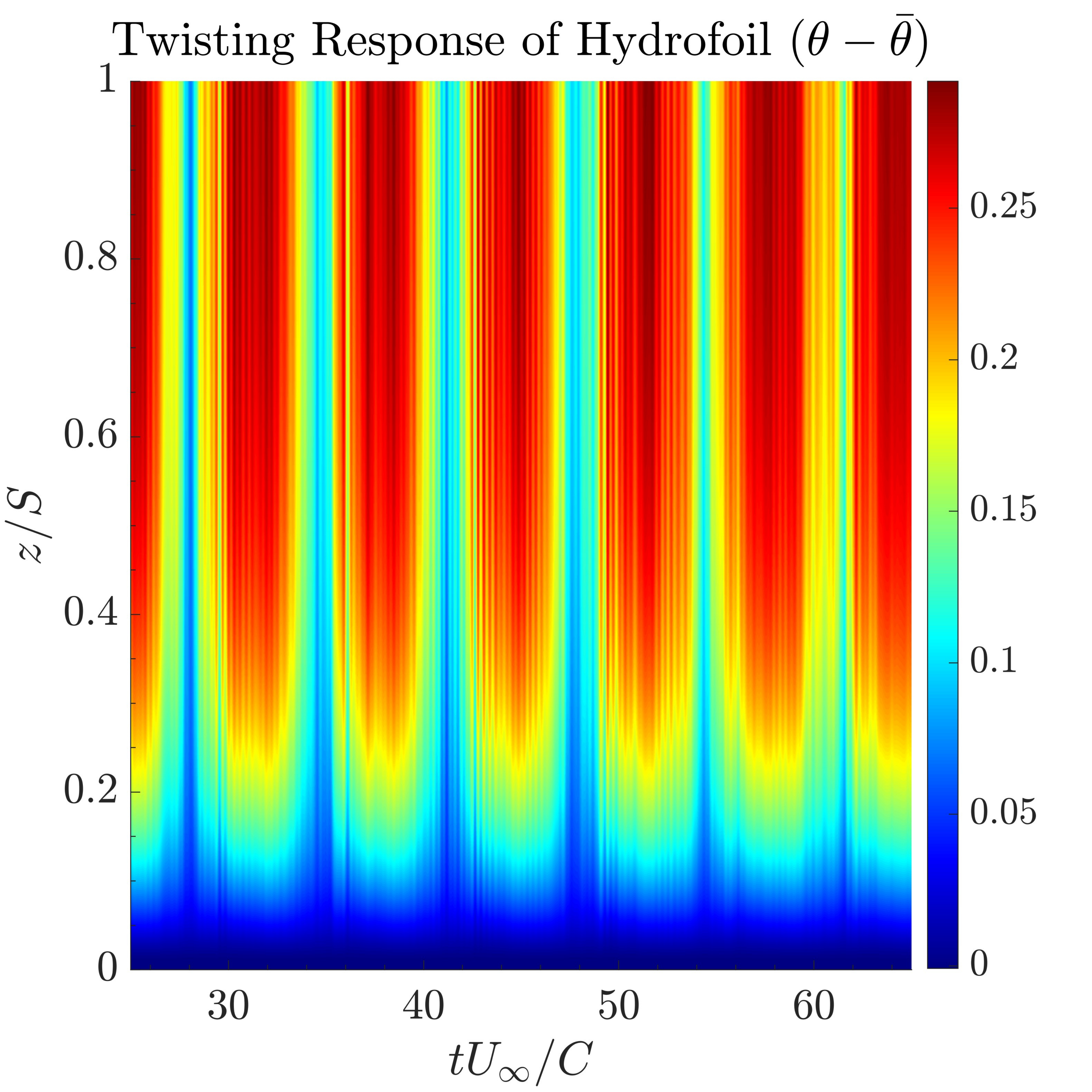}
		\caption{}
		\label{fig:twist}
	\end{subfigure}
	\caption{ Scalogram of (a) bending and (b) twisting response of the hydrofoil. We observe the superposition of several temporal harmonics, however a single spatial mode dominates the spanwise response. }
	\label{fig:bendtwist}
\end{figure}

\subsection{Cavitating Response at $\sigma= 1.4$}\label{sec:cavFlex}
\begin{figure}[!p]
	\begin{subfigure}[h]{0.45\textwidth}
		\centering
		\includegraphics[width=\textwidth]{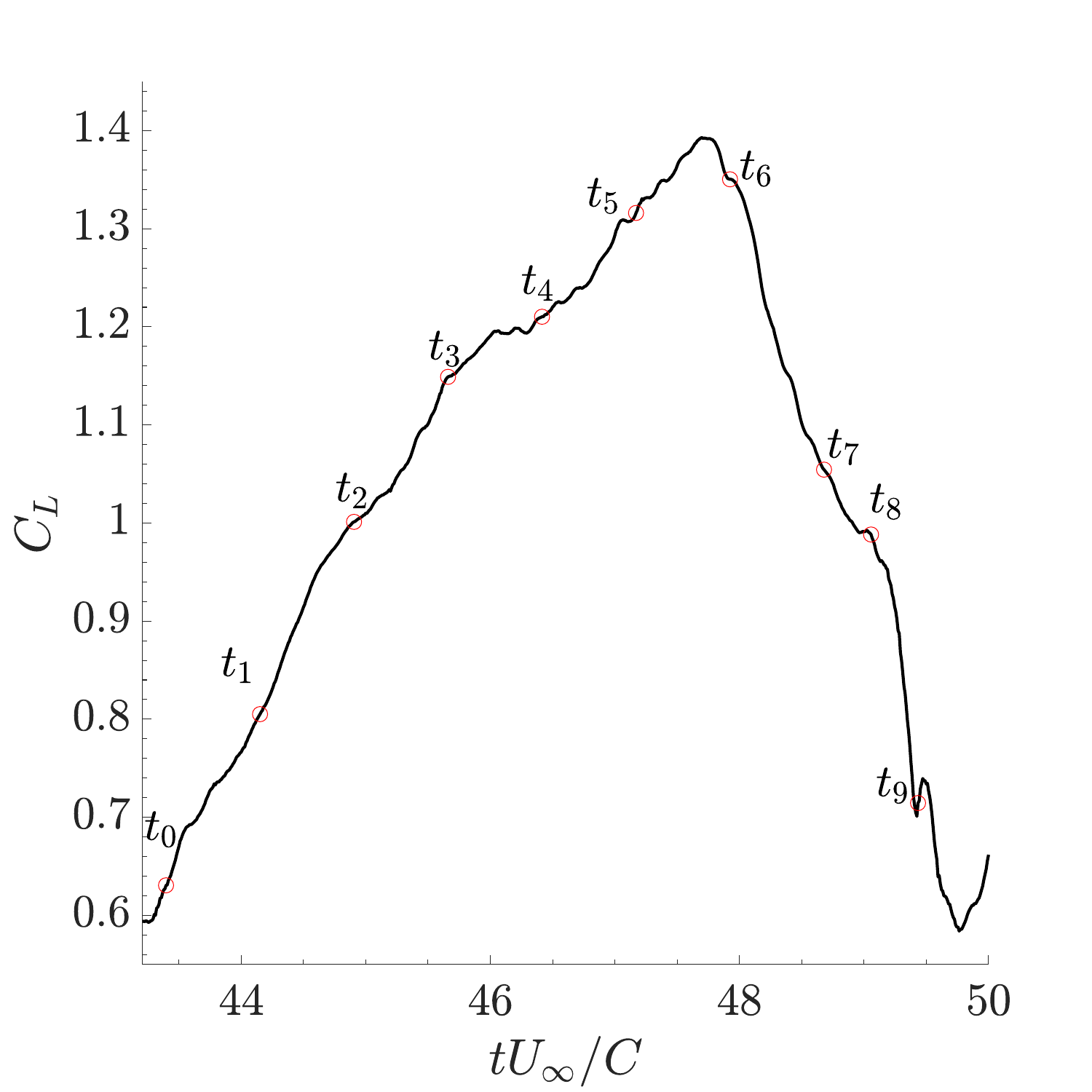}
		\caption{}
		\label{fig:CLcycle}
	\end{subfigure}
	\begin{subfigure}[h]{0.45\textwidth}
		\centering
		\includegraphics[width=\textwidth]{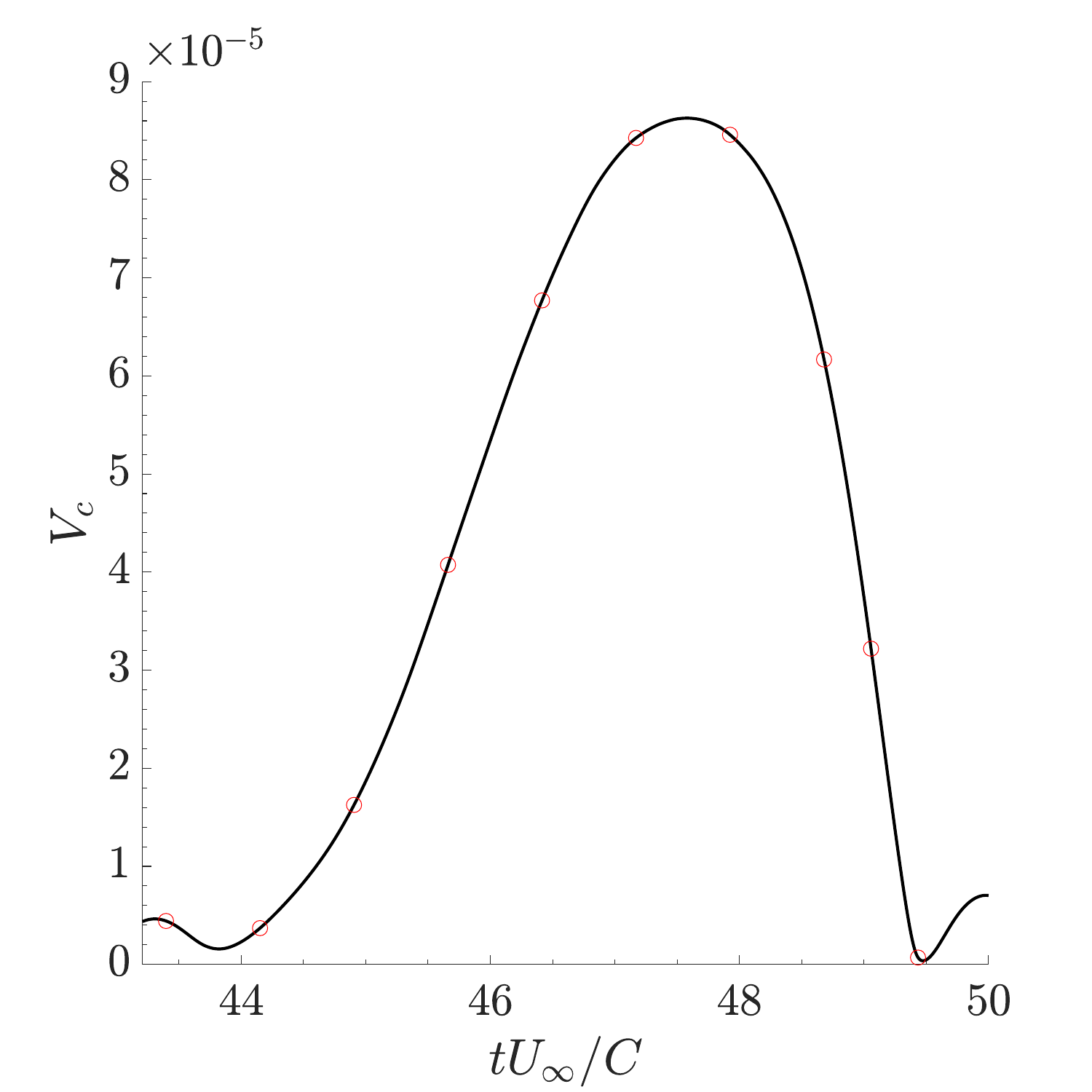}
		\caption{}		
		\label{fig:Vccycle}
	\end{subfigure}
	\begin{subfigure}[h]{\textwidth}
		\centering
		\includegraphics[width=0.8\textwidth]{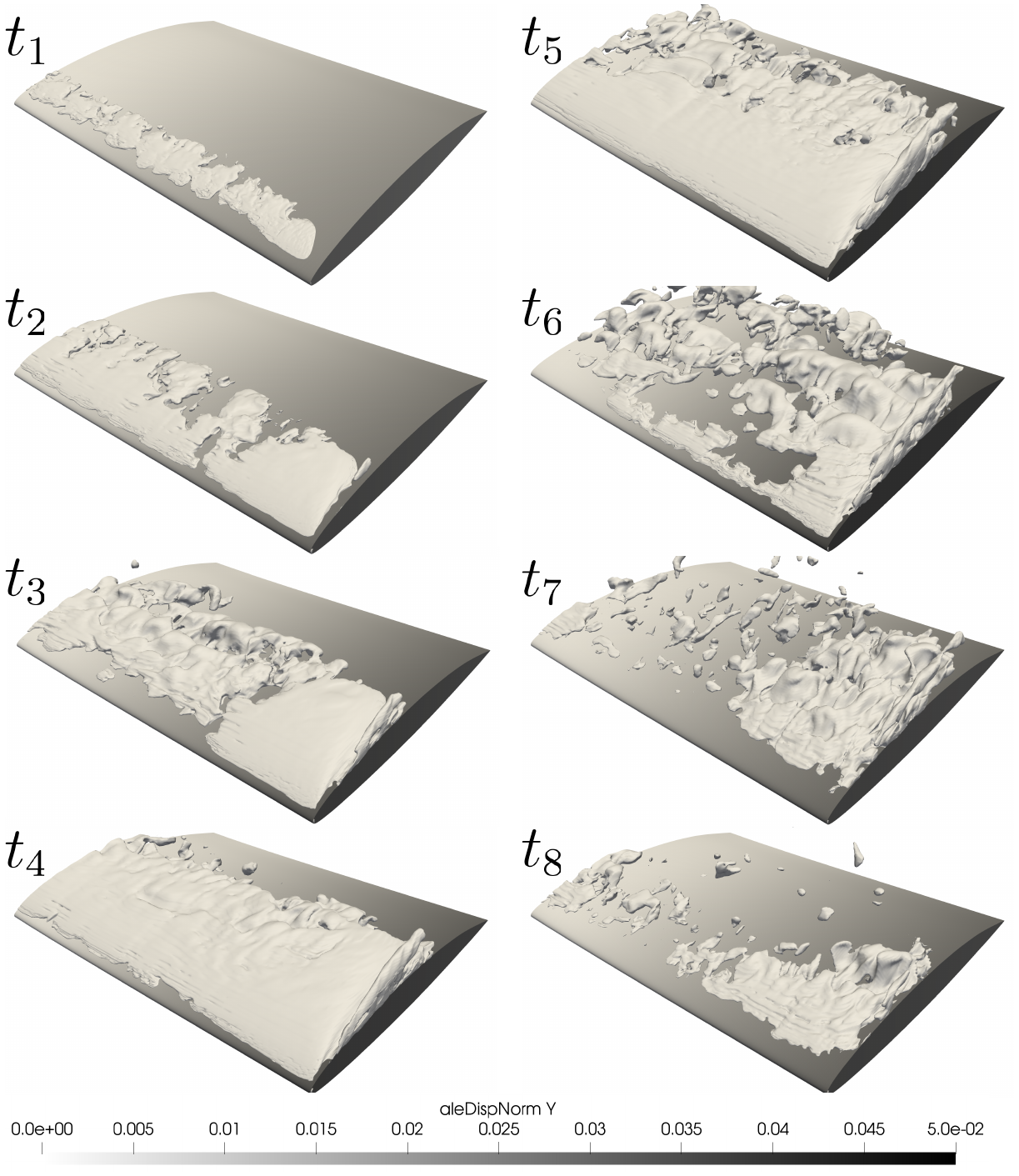}
		\caption{}
		\label{fig:CavCycle1}
	\end{subfigure}	
	\caption{ Evolution of (a) lift coefficient ($C_L$),  (b) cavity Volume ($V_c$), and (c) Cavity structures at the vapor-liqud interface ($\phi = 0.5$) observed over a cycle of cavitation. The hydrofoil is shaded according to the colour-map of bending deformation. }
	\label{fig:CycleFlex}
\end{figure}

\begin{table}[!h]
	\centering
	\begin{tabular}{|c|c c c|}
		\hline	
		\multirow{2}{*}{Parameters} & \multicolumn{3}{c|}{Flexible} \\
		&Exp. \citep{benaouicha2009Exp}& Num.-3\citep{akcabay2014influence} & Current\\
		\hline
		$h_{max}/C$ & $0.032$& $0.016$ & $0.041$ \\
		$\theta_{max}~(deg)$ & $-$& $0.39$ & $0.29$ \\
		$St = fC/U_\infty$ & $0.3054$ & $0.2391$ & $0.3132$\\
		\hline
	\end{tabular}
	\caption{Comparison of response characteristics for cavitating conditions at $\sigma = 1.4$.}
	\label{tab:CLCDCav}
\end{table}

We next evaluate the dynamic response of the hydrofoil under unsteady cavitating conditions at $\sigma = 1.4$. With regard to both the amplitude and frequency response of the hydrofoil, we observe good agreement in measurements relative to experimental and numerical studies. However, we observe some over-prediction in the bending response of the hydrofoil. We attribute this to the limited data available with regard to the added-mass coefficient to account for its time-varying nature within the modal response model. Added mass oscillations have been found to broaden the frequency response of the hydrofoil, owing to variations in the natural frequency of the structure \citep{akcabay2014influence}.

Figure \ref{fig:yCL} presents the hydrofoil's tip displacements at the leading and trailing edges along with the coefficient of lift. We observe the lift to be in phase with the tip displacement, with nearly identical spectral content. The presence of low amplitude, high-frequency fluctuations in the lift signal indicates the presence of cavity or vortex shedding events as was observed in the case of a rigid hydrofoil. As a result, we observe the absence of high-frequency impulsive loads observed in the case of a rigid hydrofoil (Fig.\ref{fig:CLCDCycle}) over the flexible hydrofoil, indicating the absence of abrupt cavity collapse events. Fig.\ref{fig:thetaCM} illustrates the tip angular rotation with the moment coefficient, wherein we observe the moment coefficient is out of phase with the angular rotation of the hydrofoil. The pitching moment acts to increase the hydrofoil's effective angle of attack close to the tip, resulting in higher lift coefficients relative to a rigid hydrofoil \citep{akcabay2014influence}.

\begin{table}[!h]	
	\centering
	\begin{tabular}{|c|c|}
		\hline	
		$f^* = fC/U_\infty$ (Current) & $f^* = fC/U_\infty$ (Experimental) \citep{akcabay2014influence,benaouicha2009Exp}\\
		\hline
		$0.15$ & $0.1729$\\		
		$0.3$ & $0.3054$\\		
		$0.45$ & $0.5261$\\
		$0.73$ & $0.7487$\\
		$0.93$ & $1.2969$\\
		\hline
	\end{tabular}
	\caption{Comparison of the dominant frequencies observed in the displacement response of the structure with experimental data.}
	\label{tab:frqCavComp}
\end{table}

With regard to the spectral characteristics exhibited by the flexible hydrofoil, we observe good agreement with experimental observations of \citet{benaouicha2009Exp} and \citet{akcabay2014influence}. The frequencies from the respective studies are non-dimensionalized w.r.t the time-scale $C/U_\infty$. Through Fig.\ref{fig:yFFT}, we illustrate the power spectrum of the bending displacement, lift coefficient and global cavity volume. We notice the emergence of peaks in the spectral response which are integral multiples of the Strouhal number $St = 0.15$, which indicates that for the current structural parameters, the cavity dynamics govern the hydrofoil's response. Therefore, we conclude the presence of a two-way lock-in between cavitation and displacement for $U^* = 0.678$. This observation is consistent with the results of \citet{suraj2023cavviv}, who observed the emergence of this Strouhal number at $U^* = 1$ for an elastically mounted cavitating hydrofoil. 

A similar power spectrum is presented for the torsional response of the hydrofoil in Fig.\ref{fig:thetaFFT}. While the dominant frequencies are the integral multiples of $St$, we also observe the emergence of a broadband frequency spectrum whose central peak occurs at $f^* = 3.35$, which is absent in the spectrum of the global vapor volume, and does not correlate with the evaluated natural frequencies of the structure or their integral sub-harmonics. We attribute the emergence of this frequency to unsteady flow characteristics induced as a consequence of the intricate interaction between cavitation, turbulence and structural deformation. This phenomenon requires further investigation with regard to its origin and is beyond the scope of the current work.

With the flow parameters considered, the flexible hydrofoil's deformation perturbs the flow within the boundary layer and inhibits the formation of a sheet cavity. This renders the developing cavity unstable, leading to the emergence of unsteady flow structures in the wake. Furthermore, as highlighted by Fig.\ref{fig:yCL}, there is a significant variation in the forces experienced by the hydrofoil from cycle to cycle. This indicates that, in certain cycles, the flow has the propensity to re-stabilize upstream of the leading edge, creating conditions conducive for sheet cavitation. However, in such cycles, the hydrofoil experiences large deformations owing to extensive pressure-side loading. This deformation leads to the perturbation of the re-entrant jet, resulting in early sheet cavity collapse. Consequently, only the region close to the hydrofoil tip retains conditions conducive to cavity formation. A representative cycle is presented in Fig.\ref{fig:CycleFlex}.

\section{Conclusion} \label{sec:conclusions}
A robust and accurate variational finite element framework for the numerical study of cavitating flows past flexible structures has been presented. An LES approach is adopted to capture the turbulent nature of the unsteady cavitating flow at high Reynolds numbers. In this regard, a re-entrant jet-based meshing criterion is proposed for flow resolution close to the hydrofoil. Through an extensive validation study of a rigid NACA66 hydrofoil, we establish the fidelity and efficacy of the LES-cavitation transport equation framework. We extend this framework for cavitating flows past flexible structures, wherein we model the flexible structure as a superposition of its dominant spatial modes obtained through the solution of a generalized eigenvalue problem. We incorporate the coupled partitioned NIFC scheme which provides a stable solution for the cavitating flow past a cantilevered flexible NACA66 hydrofoil. 

Within this framework, we have validated the response of a rigid hydrofoil in a turbulent cavitating flow. We observed good agreement of the spectral characteristics pertaining to cavitation, with cavity shedding occuring at a chord-based Strouhal number of $St = 0.15$. While pressure spikes associated with cavity collapse are overpredicted, good agreement has been observed with respect to the forces computed on the structure, which is critical for the flexible structure framework. To illustrate the physical fidelity of the framework, a brief investigation of the cavitation cycle is presented and compared to previous experimental and numerical studies, and two prominent trailing edge vortical structures have been identified which contribute to cloud cavity collapse. 

Subsequently, we have numerically validated the turbulent cavitating flow past a flexible hydrofoil. In this regard, the response characteristics of a flexible hydrofoil in non-cavitating flow have been evaluated with respect to a range of experimental and numerical benchmarks to illustrate the adequacy of the linear-eigenmode model adopted for capturing structural dynamics. Good agreement is observed with respect to the bending and twisting deformations experienced by the hydrofoil. Within the highly unsteady cavitating flow regime, the hydrofoil is found to conform closely to the response characteristics observed in experimental studies, and performs significantly better in the context of physical fidelity relative to numerical studies of cavitating flows involving elastically-mounted hydrofoils. 

Based on the power spectra of the moment, angular displacement and global cavity volume, the cavity dynamics if found to govern the hydrofoil's structural response, with a dominant frequency occuring at $St = 0.15$ and its integral multiples. A broadband response is observed in the $Z$-moment coefficient and torsional deformation, with the central peak in the spectrum occuring at $f^* = 3.35$. Over the course of a cavitation cycle, multiple vortex/cavity shedding events are detected and correlate with the emergence of low-amplitude high frequency fluctuations in the fluid flow.  The structural motion leads to the disruption of the re-entrant jet up to the mid-span of the hydrofoil, leading to an induction of velocity fluctuations on the hydrofoil surface. As a result, cloud cavity shedding is disrupted resulting in a damping of high-frequency force fluctuations on the hydrofoil. A prominent tip-leakage vortex is observed near the hydrofoil tip, leading to the extension of sheet cavity into the resulting low pressure region.

The developed framework depicts high fidelity within the scope of the homogeneous mixture-theory model for predicting cavitation. The proposed framework has successfully provided insight into turbulence-cavity interactions during cloud cavity collapse and investigations into the energy transfer mechanism associated with the excitation of unsteady dynamics of the flexible hydrofoil. The scope of the current framework can be extended through the deployment of phase-field cavitation modeling and nonlinear structural dynamics. 

\section*{Acknowledgements}
The authors would like to acknowledge the Natural Sciences and Engineering Research Council of Canada (NSERC) and Seaspan Shipyards for the funding. This research was enabled in part through computational resources and services  provided by (WestGrid) (https://www.westgrid.ca/), Compute Canada (www.computecanada.ca) and the Advanced Research Computing facility at the University of British Columbia. 
	
	\bibliography{mybibfile}   

\end{document}